\documentclass{article}
\usepackage{graphicx}  % standard LaTeX graphics tool;
\usepackage{amsmath}   % For getting proper math eqns
\usepackage{amssymb}   % Used for getting various symbols eg. gtrsim, lesssim etc.
\usepackage{bm} % For bold math (esp of lower greek letters)
\usepackage{dcolumn}% Align table columns on decimal point
\usepackage{color}
\usepackage{colortbl}
\usepackage{colordvi}
\usepackage{mathrsfs}
\usepackage{amsfonts}
\usepackage{varioref}
\usepackage{float}
\usepackage{graphicx}

\usepackage[compress]{cite}

\RequirePackage[colorlinks,citecolor=blue,urlcolor=magenta,linkcolor=blue]{hyperref}
\input epsf
\usepackage{tablefootnote}
\usepackage{multirow}
\usepackage[T1]{fontenc}
\usepackage[utf8]{inputenc}
\usepackage{amsmath, amsfonts, amssymb, mathrsfs}
\usepackage{subcaption}
\usepackage{graphicx}
\graphicspath{{./Report/}}
\usepackage{enumitem}
\usepackage{comment}

\usepackage[noabbrev]{cleveref}
\crefname{subsection}{section}{sections}
\crefname{subfigure}{figure}{figures}
\usepackage[utf8]{inputenc}
\usepackage{amsmath, amsfonts, amssymb}
\usepackage{subcaption}
\usepackage{float}
\usepackage{geometry}

\title{ Deciphering signatures of Kerr-Sen black holes in presence of plasma from the
Event Horizon Telescope data
}

\author{Siddharth Kumar Sahoo \footnote{521ph1007@nitrkl.ac.in}~$^{1}$ and Indrani Banerjee \footnote{banerjeein@nitrkl.ac.in}~$^{1}$\\
{\small{$^{1}$Department of Physics and Astronomy, National Institute of Technology, Rourkela, Odisha-769008, India}}}

\date{ }

\begin{document}

\maketitle
\begin{abstract}
The present work explores the role of the dilaton charge $r_2$ and the plasma environment in explaining the observed images of M87* and Sgr A*. Dilaton charges are associated with Kerr-Sen black holes, the stationary, axi-symmetric black hole solution in the Einstein-Maxwell-dilaton-axion (EMDA) gravity which arise in the low energy effective action of superstring theories. We investigate the impact of the background spacetime (here dilaton charge and spin) and the plasma environment in modifying the shape and size of the black hole shadow. The theoretically derived shadow is compared with the observed images of M87* and Sgr A* which enable us to constrain the background spacetime in presence of the plasma environment. {  Our analysis reveals that the shadow of M87* favors the Kerr scenario and rules out $r_2>0.48$, while the shadow of Sgr A* exhibits a marginal preference towards the Kerr-Sen scenario (although GR is allowed within 1-$\sigma$) and rules out $r_2>1$. Thus, large values of dilaton charge are disfavored for M87* and Sgr A* and this result holds good irrespective of the inhomogeneous plasma environment. Moreover, the shadows of M87* and Sgr A* rule out very dense inhomogeneous plasma environments surrounding these objects but the plasma density is further constrained from the electron number density and accretion rate estimates. As a consequence, with the current level of precision of the shadow related data we cannot distinguish between the Kerr and mildly charged Kerr-Sen black holes.  
}

\end{abstract}
%\tableofcontents
\section{Introduction\label{sec:1}}

The detection of gravitational waves\cite{LIGOScientific:2016aoc, LIGOScientific:2017vwq, LIGOScientific:2018mvr} and the release of images of M87* \cite{EventHorizonTelescope:2019dse, EventHorizonTelescope:2019ggy, EventHorizonTelescope:2019jan, EventHorizonTelescope:2019pgp, EventHorizonTelescope:2019ths, EventHorizonTelescope:2019uob} and Sgr A* \cite{EventHorizonTelescope:2021dqv, EventHorizonTelescope:2019uob, EventHorizonTelescope:2022apq, EventHorizonTelescope:2022exc, EventHorizonTelescope:2022urf, EventHorizonTelescope:2022wkp, EventHorizonTelescope:2022wok, EventHorizonTelescope:2022xqj} have demonstrated the success of general relativity (GR) in the strong field regime. These observations have also opened a window for strong field tests of GR (particularly with enhanced sensitivity in near future), where we expect to observe deviations from predictions of GR\cite{Will:2014kxa, Wright:2024mco, LIGOScientific:2016lio, Berti:2015itd}.  Detecting deviations from GR is important because,  even if GR has passed many experimental tests,  it still falls short in providing a complete understanding of gravity.  
 The cosmological constant problem \cite{Martin:2012bt, Weinberg:1988cp},  the accelerated expansion of the universe\cite{SupernovaSearchTeam:1998fmf, SupernovaCosmologyProject:1998vns},  the presence of singularities in the theory\cite{Penrose:1964wq, Hawking:1970zqf, Hawking:1966vg} and its inconsistency with quantum theory \cite{Page:2004xp} highlight the inadequacies of GR and the necessity for a more complete  theory of gravity which can potentially address the above issues.  
 
Black holes (BHs) are the most compact objects in the Universe and their extreme gravity makes them one of the ideal laboraboties to test GR and alternative theories of gravity using astrophysical observations \cite{Bambi:2017khi, Berti:2015itd}. Light from different sources, which also include radiation from the accretion disk surrounding a BH, are affected by its strong gravity before reaching the observer. Not all rays of light are able to escape the strong gravity of the black hole and reach the observer at infinity, thus the observer in general sees a dark region surrounded by a bright ring which corresponds to light rays which just succeed to escape the gravitational potential of the BH \cite{Gralla:2019xty} .
This dark region is called the black hole shadow and the bright ring is called the photon ring \cite{Gralla:2019xty}.   In many of the cases the light geodesics form spherical photon orbits \cite{Teo:2003ltt} and the projection of these orbits on the observer's sky is called the \textit{critical curve}/\textit{boundary curve} \cite{Gralla:2019xty, perlick2000ray, Perlick:2004tq} or more generally the shadow. The shadow of a Schwarzschild BH was first calculated by Synge \cite{Synge:1966okc} while Bardeen investigated the shadow of a Kerr BH for the first time \cite{1974IAUS...64..132B}. Later, Synge's work was extended by incorporating a static and spherically symmetric plasma distribution \cite{Perlick:2015vta} which followed studies on stationary, axi-symmetric plasma distributions in the Kerr spacetime \cite{Perlick:2017fio}.

In the present work we investigate the interplay between the string inspired Einstein-Maxwell- dilaton-axion  (EMDA) gravity \cite{Rogatko:2002qe, Sen:1992ua} and the surrounding plasma environment in explaining the observed images of M87* and Sgr A*.
EMDA gravity arises in the low energy effective action of heterotic string theory compactified on a $6-d$  torus, $T^6$,  giving rise to a pure $N=4$,  $d=4$ supergravity coupled to $N=4$ super Yang-Mills theory, which results in a pure supergravity theory after appropriate truncation. The bosonic sector of this supergravity theory coupled to the U(1) gauge field corresponds to the EMDA gravity \cite{Rogatko:2002qe}. The Kerr-Sen solution in EMDA gravity  represents the stationary axi-symmetric space time around  a charged rotating black hole (BH) \cite{Sen:1992ua} which  is characterized uniquely by three quantities ,  i. e,  $M\text{ (mass of BH),  }a\text{ (spin of BH),  and }r_2\text{ (dilaton charge)}$. The dilaton charge stems from the electric charge and the dilaton field while the axion field imparts rotation to the Kerr-Sen BHs.

Astrophysical implications of Kerr-Sen BHs have been explored in the context of strong gravitational lensing and black hole shadows \cite{Gyulchev:2006zg,An:2017hby,Younsi:2016azx,Hioki:2008zw,Narang:2020bgo,Jana:2023sil,Sahoo:2023czj}, continuum and reflection spectrum of black holes \cite{Banerjee:2020qmi, Tripathi:2021rwb}, quasi-periodic oscillations \cite{Dasgupta:2025fuh} and jet power and radiative efficiency of microquasars \cite{Banerjee:2020ubc}. In our previous work \cite{Sahoo:2023czj},  we had obtained constraints on the dilaton charge using the observations of M87* and Sgr A* by EHT collaboration\cite{EventHorizonTelescope:2019dse, EventHorizonTelescope:2022apq}. The present work is a continuation of our previous work \cite{Sahoo:2023czj} where we revisit the constrains on the dilaton charge of M87* and Sgr A* by taking into account the effect of the surrounding plasma environment. This is important because astrophysical black holes are surrounded by an accretion disk containing plasma \cite{Abramowicz:2011xu, Perlick:2017fio}, hence considering the effect of plasma on the shadow outline may provide more reasonable constraints on the background metric. Plasma is a dispersive medium and it affects light rays of different frequencies differently. While this effect may be insignificant for optical and higher frequencies, in the radio frequency domain its effect is expected to be more pronounced, particlularly if the plasma density is high. Since, the EHT observes the shadows of M87* and Sgr A* in the radio frequency range \cite{EventHorizonTelescope:2019dse, EventHorizonTelescope:2019ggy, EventHorizonTelescope:2019jan, EventHorizonTelescope:2019pgp, EventHorizonTelescope:2019ths, EventHorizonTelescope:2019uob, EventHorizonTelescope:2021dqv, EventHorizonTelescope:2022apq, EventHorizonTelescope:2022exc, EventHorizonTelescope:2022urf, EventHorizonTelescope:2022wkp, EventHorizonTelescope:2022wok, EventHorizonTelescope:2022xqj}, investigating the role of plasma on the observed shadows is important.

Investigating the impact of pressureless, non-magnetized plasma on radio signals began since the 1960s with the study of deflection of radio signals near the solar corona which can be approximated by a non-magnetized, pressureless plasma \cite{PhysRevLett.17.455,Muhleman:1970zz}.
Since BHs are surrounded by an accretion disk, it is important to investigate the impact of plasma on the radio signals reaching the earth from the vicinity of BHs. This motivated the study of light deflection in the Schwarzschild  and Kerr spacetime in pressureless, non-magnetized plasma environments \cite{2000rofp.book.....P} following which  gravitational lensing was investigated with different methods in the presence of plasma \cite{Bisnovatyi-Kogan:2010flt,Tsupko:2013cqa,2013ApSS.346..513M,Crisnejo:2018uyn,Crisnejo:2018ppm,Crisnejo:2019xtp,Crisnejo:2019ril}. This followed studies on the
implications of plasma in the strong-bending regime (e.g. multiple imaging properties) \cite{Tsupko:2013cqa,Liu:2016eju,2000rofp.book.....P} and other astrophysical observations \cite{Rogers:2015dla,Rogers:2016xcc,Er:2013efa,Tsukamoto:2014tja,Tsukamoto:2017fxq}. Recent years have witnessed increasingly more interest in studies related to gravitational lensing \cite{Kumar:2022zky,Crisnejo:2022qlv,Bisnovatyi-Kogan:2022yzj,Tsupko:2019axo,Sun:2022ujt,Er_2020,Kala:2024fvg,Kala:2022uog} and shadows \cite{Yan:2019etp,Chowdhuri:2020ipb,Badia:2021kpk,Badia:2022phg,Bezdekova:2022gib,Briozzo:2022mgg} in the presence of plasma. 

In the present work we investigate the trajectories of light rays in the vicinity of Kerr-Sen BHs surrounded by a pressureless, non-magentized plasma. We consider plasma distributions which ensure separability of the Hamilton-Jacobi equations leading to the presence of a generalized Carter constant \cite{Perlick:2017fio,Bezdekova:2022gib}. This in turn leads to first order geodesic equations for all the four coordinates and enables us to analytically obtain the shadow of Kerr-Sen BHs in the presence of plasma. These calculations hold good for any observer position and inclination \cite{Perlick:2017fio,Perlick:2023znh}. The theoretically obtained shadows are compared with the observed images of M87* and Sgr A* which enables us to establish constrains on the dilaton charge of the these BHs and the surrounding plasma environment. The present work thus provides a framework to constrain the deviations from the Kerr scenario in the presence of plasma.

\paragraph{Paper outline:}
In \Cref{sec:2} we give a brief overview of the EMDA gravity and discuss briefly about the Kerr Sen black hole. \Cref{sec:3} summarizes the propagation of light rays in a stationary, axisymetric spacetime in presence of a pressureless, non-magnetized plasma environenment. \Cref{sec:3.1}
discusses the first order geodesic equations for photons moving in the Kerr-Sen spacetime in presence of non-magnetised, pressureless plasma while in \Cref{sec:3.2} the expression of shadow outline considering an observer at a finite distance from the black hole is derived. In \Cref{sec:4}, we report the variation in the shadow of Kerr-Sen BHs with variation in the dilaton charge, spin, inclination and plasma environments. \Cref{eht} outlines the methodology used to constrain the dilaton charge parameter in the presence of plasma from the EHT data and reports the constrains for M87* (\Cref{Constraining dilaton charge and plasma environment from M87* shadow}) and Sgr A* (\Cref{Constraining dilaton charge and plasma environment from Sgr A* shadow}). We summarize the main findings and implications of our work in \Cref{conclusion} and discuss some avenues which can be explored in future.
We use geometrized units ($G=c=1$) and the metric signature is chosen to be $(-, +, +, +)$.  However, during comparison with observations we covert back to SI units.

\section{Einstein-Maxwell-dilaton-axion gravity}\label{sec:2}
{Einstein-Maxwell-dilaton-axion gravity is a string theory based alternate gravity model which roughly speaking results from the compactification of heterotic string theory to 4 dimensions and taking low energy limit of the effective action ${S}$ \cite{Sen:1992ua, Rogatko:2002qe, EMDAGarfinkle, EMDAGarfinkle2}.  Along with the metric tensor and Maxwell field,  the theory also contains dilaton and axion fields which are related to string theory.  One of the interesting feature of  EMDA gravity is,  classical solutions in this theory can be used to investigate signatures of string theory as it still retains S and T dualities of string theory\cite{Rogatko:2002qe} .  The action $S$ of EMDA gravity is given as,   }

%\begin{align}
%S=\frac{1}{16\pi} \int \sqrt{-g}\ d^{4}x(R-2\partial_{\mu} \xi \partial^{\mu} \xi-\frac{1}{3} \mathbf{H}_{\rho \sigma \delta}\mathbf{H}^{\rho \sigma \delta}+ e^{-2 \xi} F_{\alpha \beta} F^{\alpha \beta})
%\label{eqn:1}
%\end{align}

%The Einstein-Maxwell dilaton-axion (EMDA) gravity \cite{Sen:1992ua, Rogatko:2002qe} is an alternate gravity theory obtained from string theory.   It results from the compactification of he terotic string theory to 4 dimensions and taking low energy limit of the effective action $\mathcal{S}$ .  The action of EMDA gravity is given as
%results from the compactification of ten dimensional heterotic string theory on a six dimensional torus $T^6$.  %In EMDA gravity,  $N=4$,  $d=4$ supergravity is coupled to $N=4$ super Yang-Mills theory which can be suitably truncated to a pure supergravity theory exhibiting $S$ and $T$ dualities.  The bosonic sector of this supergravity theory when coupled to the $U(1)$ gauge field is known as the Einstein-Maxwell dilaton-axion (EMDA) gravity \cite{Rogatko:2002qe} which provides a simple framework to study classical solutions.  The four dimensional effective action for EMDA gravity consists of a generalization of the Einstein-Maxwell action such that the metric $g_{\mu\nu}$ is coupled to the dilaton field $\xi$,  the $U(1)$ gauge field $A_\mu$ and the Kalb-Ramond field strength tensor ${H}_{\alpha\beta\gamma}$.  The action corresponding to EMDA gravity assumes the form, 
\begin{align}
S=\frac{1}{16\pi} \int \sqrt{-g}d^{4}x(R-2\partial_{\mu} \xi \partial^{\mu} \xi-\frac{1}{3}W_{\rho \sigma \delta}W^{\rho \sigma \delta}+ e^{-2 \xi} F_{\alpha \beta} F^{\alpha \beta})
\label{eqn:1}
\end{align}
In \Cref{eqn:1}  $g$ is the determinant of the metric tensor and $R$ the Ricci scalar associated with the 4-dimensional metric tensor $g_{\mu\nu}$,  
\(\xi\) represents the dilatonic field,  \(F_{\mu \nu}=\nabla_{\mu} \mathcal{A}_{\nu}-\nabla_{\nu}\mathcal{ A}_{\mu}\) is Maxwell field strength tensor and \(W_{\rho \sigma \delta}\)  is  the Kalb-Ramond field 
strength tensor%related to  \(\mathcal{A}_{\mu}\) the vector potential 
\cite{Rogatko:2002qe, Sen:1992ua}.  In four dimensions the Kalb-Ramond field strength tensor \(W_{\rho \sigma \delta}\) can be written in terms of the pseudo-scalar axion field $\Lambda$\cite{Rogatko:2002qe, Ganguly:2014pwa},  such that, 
\begin{align}
\label{eqn:2}
{W}_{\alpha\beta\delta} = \frac{1}{2}e^{4\xi}\epsilon_{\alpha\beta\delta\gamma}\partial^{\gamma}\Lambda
\end{align}
 The action in \Cref{eqn:1}  written in terms of the axion field assumes \cite{ Rogatko:2002qe} the form, 
\begin{align}
\label{eqn:3}
S = \frac{1}{16\pi}\int\sqrt{-g}~d^{4}x\bigg{[}{R} - 2\partial_{\nu}\xi\partial^{\nu}\xi - \frac{1}{2}e^{4\xi}\partial_{\nu}\Lambda\partial^{\nu}\Lambda + e^{-2\xi}{F}_{\rho\sigma}{F}^{\rho\sigma}  + \Lambda{F}_{\rho\sigma}\tilde{{F}}^{\rho\sigma}\bigg{]} 
\end{align}
The equations of motion for the axion $\Lambda$,  dilaton $\xi$ and the vector potential $\mathcal{A}_{\mu}$ can be obtained by  varying the action $S$ with respect to the corresponding fields.   %The solutions of these equations for axion $\Lambda$ ,  dilaton $\xi$ and vector   potential $\mathcal{A}_\mu$ have been worked out and can be found in \cite{Sen:1992ua, Rogatko:2002qe, Ganguly:2014pwa} . 
%Solving the aforesaid equations one obtains solutions for the dilaton,  axion and the Maxwell field,  respectively \cite{Ganguly:2014pwa, Sen:1992ua, Rogatko:2002qe}, 
%where $M$ is the mass,  $a$ is the spin and $q$ is the charge of the black hole.  In \Cref{S2-9} $r_2$ is associated with the dilaton charge parameter and is given by $r_{2} = \frac{q^{2}}{{M}}e^{2\xi_{0}}$ where $\xi_{0}$ represents the asymptotic value of the dilatonic field.  The dilaton charge parameter also depends on the electric charge of the black hole,  which owes its origin from the axion-photon coupling and not from the in-falling charged particles.  This is because the axion and dilaton field strengths vanish if the electric charge $q=0$ (see \Cref{S2-10} and \Cref{S2-9}).  It is further important to note that the axion field renders a non-zero spin to the black hole since the field strength corresponding to the axion field vanishes if the black hole is non-rotating (\Cref{S2-10}). 
The field equations for gravity in EMDA theory can be obtained by varying the action in \Cref{eqn:3}  with  respect to $
g_{\mu\nu}$.   The modified Einstein equations take the form 
\begin{align}
\label{eqn:4}
{G}_{\mu\nu} = \mathcal{T}_{\mu\nu}({F}, \xi, \Lambda) 
\end{align}

and
\begin{align}
\label{eqn:5}
\mathcal{T}_{\mu\nu}({F}, \xi, \Lambda)& = e^{2\xi}(4{F}_{\mu\rho}{F}_{\nu}^{\rho} - g_{\mu\nu}{F}^{2}) - g_{\mu\nu}(2\partial_{\gamma}\xi\partial^{\gamma}\xi + \frac{1}{2}e^{4\xi}\partial_{\gamma}\Lambda\partial^{\gamma}\Lambda) 
\nonumber \\
&+ \partial_{\mu}\xi\partial_{\nu}\xi + e^{4\xi}
\partial_{\mu}\Lambda\partial_{\nu}\Lambda
\end{align}
In \Cref{eqn:4} and \Cref{eqn:5},   $\mathcal{T}_{\mu\nu}$ is the energy-momentum tensor and $G_{\mu\nu}$ is the Einstein tensor. 
The stationary,  axisymmetric,  charged, rotating black hole solution of \Cref{eqn:4} in EMDA gravity is the Kerr Sen solution\cite{Sen:1992ua, Ganguly:2014pwa}.  The form of the metric in Boyer-Lindquist coordinates $(t, r, \theta, \phi)$ \cite{Boyer:1966qh}  is given by,

%\section{The Kerr Sen Black hole\label{sec:3}}
%The Kerr Sen metric represents a stationary and axis symmetric space time around a charged rotating black hole in Einstein Maxwell dilaton-axion gravity\cite{Sen:1992ua}.  For a Kerr Sen black hole of mass $M$ and charge $q$ and rotation parameter \(a\),  the metric of the space time in Boyer-Lindquist coordinates\((t, r, \theta, \phi)\)\cite{Boyer:1966qh, Garcia:1995qz} is written as:
\begin{multline}
\label{eqn:6}
    ds^2=-\left(1-\frac{2Mr}{\rho}\right)dt^2+\frac{\rho}{\Delta}dr^2+\rho d\theta^2+\frac{\sin^2{\theta}}{\rho}\left[((r+r_2) r+a^2)^2-\Delta a^2 \sin^2\theta\right]d\phi^2-\frac{4aMr}{\rho}dtd\phi
\end{multline}
where
\begin{gather}
\label{eqn:7}
\rho=r(r+r_2
)+a^2 \cos^2\theta\\
\Delta=r(r+r_2)+a^2-2 M r    
\end{gather}
The dilaton charge parameter \(r_2=\frac{q^2 e^{2 \xi_{0}}}{M}\)   is related to the electric charge  of the black hole $q$,  and the asymptotic value of the dilaton field $\xi$.  The expression of $r_2$ {indicates} $r_2\geq 0$.  The event horizon of the Kerr Sen black hole can be obtained by solving the equation $g^{rr}=0$,  {which} gives us
    \begin{gather}
    \label{eqn:8}
        \frac{r_{h_{\pm}}}{M}=\left(1-\frac{r_2}{2 M}\pm
   \sqrt{\left(1-\frac{r_2}{2
   M}\right)^2-\frac{a^2}{M^2}}\right)
    \end{gather}
    In above equation $r_{h_-}$ and $r_{h_+}$ represent the inner and outer  horizons of the black hole respectively.  Imposing the condition of black hole having real, positive event horizon,  we obtain two restrictions.  First,  the upper bound of dilaton charge parameter $\frac{r_{2max}}{M}= 2$.  Second,  the maximum spin $a_{max}$ for a Kerr Sen black hole is $\frac{a_{max}}{M}=\left(1-\frac{r_2}{2M}\right)$ {for a given $r_2$}.   In what follows we will scale all distances with $M$ such that $r\equiv\frac{r}{M}$ and $r_2\equiv\frac{r_2}{M}$ .  

{  It is important to note that in Equation (\ref{eqn:6}), the mass parameter $M$ which appears in the Kerr-Sen metric corresponds to the ADM mass of the Kerr Sen black hole (BH).
This can be verified from the prescription discussed in \cite{Pal:2021nul,Piotr}, which states that a spacetime metric with spatial part given by,
\begin{align}
ds_{spatial}^2=\lambda(r) dr^2 + r^2\chi(r) d\Omega^2
\label{1}
\end{align}
has an ADM mass given by,
\begin{align}
M_{ADM}=\lim_{r\to \infty}\frac{1}{2}\Big[-r^2 \chi'(r) +r \lbrace \lambda(r) -\chi(r)\rbrace \Big]
\label{eqn:2-1}
\end{align}
It turns out that for the Kerr-Sen metric,
\begin{align}
\lambda(r)=\frac{1+\frac{r_2}{r}+\frac{a^2\cos^2\theta}{r^2}}{1+\frac{r_2}{r}-\frac{2M}{r}+\frac{a^2}{r^2}} ~~~~~{ \rm{and}}~~~~\chi(r)=1+\frac{r_2}{r}+\frac{a^2\cos^2\theta}{r^2} ~~~{ \rm{(in ~the~ large~ r ~limit)}}
\label{eqn:3-1}
\end{align}
Using Equation (\ref{eqn:2-1}) and Equation (\ref{eqn:3-1}), it can be shown that the ADM mass of the Kerr-Sen BH $M_{\rm ADM,KS}=M$, which appears in the Kerr-Sen metric (Equation \ref{eqn:6}).\\
For the Kerr BH,
\begin{align}
\lambda(r)=\frac{1+\frac{a^2\cos^2\theta}{r^2}}{1-\frac{2M}{r}+\frac{a^2}{r^2}} ~~~~~{ \rm{and}}~~~~\chi(r)=1+\frac{a^2\cos^2\theta}{r^2} ~~~{ \rm{(in ~the~ large~ r ~limit)}}
\label{eqn:4-1}
\end{align}
which when used in Equation (\ref{eqn:2-1}) yields, $M_{\rm ADM,Kerr}=M$. Thus, the mass parameter $M$ which appears in the Kerr and the Kerr-Sen metrics correspond to their ADM masses. This is important because, later we will be determining the theoretical angular diameter of the shadow of M87* and Sgr A* assuming them to be Kerr-Sen BHs. In order to determine the theoretical angular diameter we use previously determined masses of M87* and Sgr A*. Since the mass of Sgr A* and M87* are determined using stellar dynamics studies \cite{Gebhardt:2011yw,Abuter:2018uum}, where the closest distance of the star from the central BH can be $\sim$ few thousands of the gravitational radii (e.g., the point of closest approach of S0-2 star from Sgr A* is 120 A.U. which $\sim 2800 ~R_g$), considering the observationally determined mass of the BH to be its ADM mass seems reasonable \cite{KumarWalia:2024omf,Vagnozzi:2022moj}.

However, in this context, the active gravitational mass (AGM) of the BH may also be relevant. The AGM of a spacetime metric can be obtained by expanding the $g_{tt}$ component of the metric in powers of $1/r$ such that,
\begin{align}
g_{tt}=-\Big[1-\frac{2  M}{r} + \mathcal{O}\Big(\frac{1}{r^2}\Big)\Big] ~~~{ \rm{(in ~the~ large~ r ~limit)}}
\label{eqn:5-1}
\end{align}
where the coefficient of the $2/r$ term corresponds to the AGM, provided $\chi(r)\sim 1+ \mathcal{O}(\frac{1}{r})$. For the Kerr and the Kerr-Sen spacetime the $\chi(r)$ satisfies the aforesaid criteria. Thus, if we expand the $g_{tt}$ component of the Kerr-Sen metric in powers of $1/r$, we get,
\begin{align}
g_{tt}&=-\Big[1-\frac{2Mr}{r(r+r_2)+a^2\cos^2\theta}\Big]\\
 & \approx -\Big[1-\frac{2M}{r}\Big(1-\frac{r_2}{r}-\frac{a^2\cos^2\theta}{r^2} \Big)\Big]~~~{ \rm{(in ~the~ large~ r ~limit)}}
 \label{eqn:7-1}
\end{align}
In Equation (\ref{eqn:7-1}), if we put $r_2=0$ we get the corresponding result for the Kerr metric. Thus, for both the Kerr and the Kerr-Sen spacetimes
$M_{\rm ADM}=M_{\rm AGM}=M$, where $M$ is the mass parameter that appears in the two metrics. 
%Therefore, even if one assumes that the previously estimated masses of Sgr A* and M87* \cite{Gebhardt:2011yw,Abuter:2018uum} correspond to their AGM, then also our results would not change, as in our case $M_{\rm ADM}=M_{\rm AGM}=M$.
}
%It is important to note that for Kerr and Kerr-Sen spacetimes, the BHs' spins play no significant role in determination of their ADM mass, as eventually one has to take the large $r$ limit where the effect of spin is negligible. Hence, the above calculation can as well be done using the non-rotating versions of the Kerr and the Kerr-Sen metric, i.e. Schwarzschild and the dilaton BH, respectively, which would yield the same value for the ADM masses. }

    \section{Overview of light rays in a plasma in stationary axisymmetric spacetime \label{sec:3}}
As discussed in the previous section,  the Kerr-Sen spacetime is a stationary, axisymmetric spacetime admitting Killing vectors $\partial_t$ and $\partial_\phi$.  %Therefore,  it is important to discuss  basic ideas and terminologies of a plasma in a general stationary and axisymmetric spacetime $ds^2=g_{\mu\nu}(r, \theta)dx^\mu dx^\nu$ .  
For purposes specific to our work,  we consider plasma in a stationary, axisymmetric spacetime with metric tensor $g_{\mu\nu}(r, \theta)$.  The plasma frequency $\omega_{P}(x^\mu)$ is  related to the electron number density $\mathcal{N}(x^\mu)$  by  the relation 
\begin{gather}
\label{eqn:9}
    \omega^2_{P}(x^\mu) =\frac{4\pi e^2}{m_e} \mathcal{N}(x^\mu)
\end{gather}
In above equation $e \text{ and } m_e$ are the charge and mass of electron,  respectively.  The path of the light rays in the plasma can be determined using Hamiltonian formalism.   For our work we consider pressureless,  non-magnetised plasma where the Hamiltonian $\mathcal{H}$ for light ray is given by\cite{breuer1980propagation, breuer1981propagation, perlick2000ray}  
\begin{gather}
\label{eqn:11}
    \mathcal{H}(x^\mu, p^\mu)=\frac{1}{2}\left(g^{\mu \nu} p_{\mu} p_{\nu} +\omega_{P}(x^\mu)^2\right)
\end{gather}
and relations for obtaining $\dot{x}^\mu$ and  $p^\mu$ from $\mathcal{H}$ are given by 
\begin{gather}
\label{eqn:12}
    \dot{x}_\mu=\frac{\partial \mathcal{H}}{\partial p^\mu}
\end{gather}
\begin{gather}
\label{eqn:13}
    \dot{p}_\mu=-\frac{\partial \mathcal{H}}{\partial x^\mu}
\end{gather}
The dot on $x^\mu\text{ and }p^\mu$ represents derivative with respect to some curve parameter $\lambda$.   This implies $p_t\text{ and }p_\phi$ are the constants of motion.  %We note that the quantity $p_t=-\omega_0$ is related to the asymptotic frequency of the light {measured by an observer at rest at infinity}.  
%{In our study we consider pressureless non-magnetised plasma environments which depend only on $(r, \theta)$.   
Let $\omega(x)\equiv\omega(r, \theta)$ represent the frequency of light at $(r, \theta)$.  We decompose the momentum $p^\mu$ of the light ray in a direction parallel and orthogonal to a timelike observer with normalized four-velocity $u^\mu$,  i. e ,  $u_\mu u^\mu=-1$ in the curved spacetime.  Thus we decompose $p^\mu$ as 
\begin{equation}
\label{eqn:14}
    p^\mu=\omega(x) u^\mu+k^\mu
\end{equation}
In the above equation the first and second terms represent components of $p^\mu$ parallel and orthogonal to $u^\mu$ respectively.  The frequency of the light ray can be written as
\begin{equation}
\label{eqn:15}
    \omega(x)=-p_\mu u^\mu
\end{equation}
%{note $k^\mu$  also in general a function of $(r, \theta)$ is }
while
\begin{equation}
    \label{eqn:16}
    k^\mu=p^\mu+p_\nu u^\nu u^\mu
\end{equation}
{Substituting  \Cref{eqn:14} in   \Cref{eqn:11}  and using the property  $\mathcal{H}=0$ for photons or light rays,  we get }
\begin{equation}
    \label{eqn:17}
    \omega(x)^2=k^\mu k_\mu+\omega^{2}_P(x)
\end{equation}
{Due to spacelike nature of $k^\mu$,  it follows}
\begin{equation}
    \label{eqn:18}
    \omega^2(x)\geq \omega^2_P(x)
\end{equation}
{The above inequality is the condition for light propagation in a plasma.  This means light propagation in a region of plasma is possible as long as the inequality \Cref{eqn:18}  is satisfied. }
{The above condition \Cref{eqn:18} can also be interpreted in terms of the refractive index $\mathbf{\mathfrak{n}}(r, \theta)$.  We first define the phase velocity of the light wave $v_p$ and the index of refraction $\mathfrak{n}$ as, }
\begin{equation}
    \label{eqn:19}
    v_p(r, \theta)=\left(\frac{\omega^2(r, \theta)}{k_\mu(r, \theta) k^\mu(r, \theta)}\right)^\frac{1}{2}
\end{equation}
\begin{equation}
    \label{eqn:20}
    \mathfrak{n}(r, \theta)=\frac{1}{v_p(r, \theta)}
\end{equation}
since $c=1$. {Using \Cref{eqn:17}  and \Cref{eqn:19}  in \Cref{eqn:20} we get }
\begin{equation}
    \label{eqn:21}
    \mathfrak{n}=\sqrt{1-\frac{\omega^2_P(x)}{\omega^2(x)}}
\end{equation}
{Thus light propagation condition in  \Cref{eqn:18} translates to  $\mathfrak{n}(r, \theta)\geq 0$}\cite{synge1960relativity, Perlick:2017fio}.    
{ For a static timelike observer $u^\mu=\frac{\delta^{\mu}_t}{\sqrt{-g_{tt}}}$, }  the frequency of light  $\omega(r, \theta)\equiv\omega(x)$ measured at  location $(r, \theta)$   and $\omega_0$  are related by\cite{Perlick:2017fio}
\begin{gather}
\label{eqn:22}
    \omega(x)=\frac{\omega_0}{\sqrt{-g_{tt}}}
\end{gather}
%here $\omega(x)$ is frequency measured by an { non-moving} observer at location $(r, \theta)$.   
Thus,  $\omega_0$ corresponds to the frequency of light measured by an observer at rest at infinity.

\subsection{Motion of photons in plasma in Kerr Sen spacetime\label{sec:3.1}}%Light ray geodesics of Kerr Sen black hole surrounded by plasma\label{sec:3. }}
In this section,  we will obtain the equations of geodesics traced by light rays in presence of plasma around a Kerr-Sen black hole.   {We consider the %effect of
plasma as a dispersive medium,  i.e,  the presence of plasma modifies the refractive index of the surrounding space time,  through which the light rays travel,  hence affecting the path of the light rays. }
%The plasma environment around the Kerr Sen black hole is characterized by the plasma frequency function $\omega_{P}(x^\mu)$.  The plasma frequency $\omega_{P}(x^\mu)$ is related to electron number density $\mathcal{N}(x^\mu)$ function of plasma.  The relation between $\omega_{P}(x^\mu)$ and $\mathcal{N}(x^\mu)$ is given by, 
%\begin{gather}
\label{eq:5}
   % \omega_{P}(x^\mu)^2 =\frac{4\pi e^2}{m_e} \mathcal{N}(x^\mu)
%\end{gather}
%in the above equation $e \text{ and } m_e$ are the electronic charge and mass of electron,  respectively.  The Hamiltonian $\mathcal{H}$ for light ray in the spacetime was worked out  by\cite{breuer1980propagation, breuer1981propagation, perlick2000ray}  
%\begin{gather}
\label{eq:6}
   % \mathcal{H}(x^\mu, p^\mu)=\frac{1}{2}\left(g^{\mu \nu} p_{\mu} p_{\nu} +\omega_{P}(x^\mu)^2\right)
%\end{gather}
%In our study we consider plasma environments which depend only on $(r, \theta)$.  

%The relations for obtaining $\dot{x}^\mu$ and  $p^\mu$ from $\mathcal{H}$ are given by 

{Substituting the components of Kerr Sen metric in the Hamiltonian \Cref{eqn:11}  and simplifying we get }
\begin{gather}
    \label{eqn:23}
    \mathcal{H}=\frac{1}{2\rho} \left(\left( \frac{p_\phi}{\sin{\theta}}+a \sin{\theta}\ p_t\right)^2-\frac{1}{\Delta}\left(ap_\phi+p_t(a^2+r(r+r_2))\right)^2+p^2_\theta+\Delta p^2_r+\rho \omega^2_P\right)=0
\end{gather}
From the  Hamilton-Jacobi equation we have 
\begin{gather}
\label{eqn:24}
    \mathcal{H}+\frac{\partial \mathcal{S}(t, r, \theta, \phi)}{\partial \lambda}=0
\end{gather}
where \begin{gather}
\label{eqn:hjp}
p_\mu= \frac{\partial \mathcal{S}}{\partial x^\mu}
\end{gather} 
We note that for light rays in plasma in \Cref{eqn:11}  $\mathcal{H}=0$ which yields $\frac{\partial \mathcal{S}}{\partial \lambda}=0$. 
Taking action $\mathcal{S}$ of the form
\\\begin{equation}
\label{eqn:25}
    \mathcal{S}= p_t t+ S_r(r)+ S_\theta(\theta) +p_\phi \phi
\end{equation}
and substituting in \Cref{eqn:hjp}  and \Cref{eqn:25} in \Cref{eqn:23} we get, 
\begin{multline}
\label{eqn:26}
    \left( \frac{p_\phi}{\sin{\theta}}+a \sin{\theta}\ p_t\right)^2-\frac{1}{\Delta}\left(ap_\phi+p_t(a^2+r(r+r_2))\right)^2\\+\Delta \left(\frac{d \mathcal{S}_r(r)}{d r}\right)^2+\left(\frac{d \mathcal{S}_\theta(\theta)}{d \theta}\right)^2+\rho\ \omega_{P}(r, \theta)^2=0
\end{multline}
    In the presence of plasma,  the separability of the $r$ and $\theta$ dependent part of the above equation holds true \textit{iff}  the quantity $\omega_{P}(r, \theta)^2$ has a form like \begin{gather}
        \label{eqn:omegap_form}
   \omega_{P}(r, \theta)^2=\frac{f(r)+g(\theta)}{\rho}
\end{gather}
 where,  $f(r)$ and $g(\theta)$ are functions of $r$ and $\theta$ respectively.  The separability condition for $r_2=0$ (Kerr metric case) was first obtained in\cite{Perlick:2017fio} ,   for a more general result refer\cite{Bezdekova:2022gib}.  From the condition of light propagation in plasma ($\omega(r, \theta)^2\geq \omega_{P}(r, \theta)^2$)  we infer that,  in order for light to propagate from the source to observer in the plasma medium,  the plasma frequency function can not be unbounded.  \\ 

 Thus,   Hamiltonian   \Cref{eqn:23}  can be written in the form
 \begin{multline}
 \label{eqn:hjseparation}
       \left( \frac{p_\phi}{\sin{\theta}}+a \sin{\theta}\ p_t\right)^2+g(\theta)+\left(\frac{d \mathcal{S}_\theta(\theta)}{d \theta}\right)^2=\\ \frac{1}{\Delta}\left(ap_\phi+p_t(a^2+r(r+r_2))\right)^2-\Delta \left(\frac{d \mathcal{S}_r(r)}{d r}\right)^2-\Delta f(r)
 \end{multline}
{As the left hand side of the equation depends on $\theta$ and the right hand side depends on r,   thus,  both expressions must be equal to a constant.  We represent the constant by the symbol $\mathcal{C}$ . }
The constant of separability $\mathcal{C}$ is called the generalised Carter constant\cite{Perlick:2017fio, Carter:1968rr}.  As $p_r=\frac{d \mathcal{S}_r(r)}{d r}$ and $p_\theta=\frac{d \mathcal{S}_\theta(\theta)}{d \theta}$,  we obtain
 \begin{gather}
 \label{eqn:29}
     \Delta^2 p^2_{r}=\left(ap_\phi+p_t(a^2+r(r+r_2))\right)^2-\Delta f(r)-\mathcal{C}\Delta
     \end{gather}
     \begin{gather}
     \label{eqn:30}
      p^2_{\theta}= \mathcal{C}-\left( \frac{p_\phi}{\sin{\theta}}+a \sin{\theta}\ p_t\right)^2-g(\theta)
 \end{gather}
 The equations of geodesics of light and  refractive index in presence of plasma  in the Kerr Sen space-time are as follows:
 \begin{gather}
    \label{eqn:31}
    \frac{\rho^2\dot{r}^2}{\omega^2_{0}} \ =\left(-a\eta+(a^2+r(r+r_2))\right)^2-\Delta f(r)-\mathcal{Q}\ \Delta=K(r)
    \end{gather}
    \begin{gather}
        \label{eqn:32}
     \frac{\rho^2\dot{\theta}^2}{\omega^2_{0}}\ =\mathcal{Q}-\left(-a  \sin{\theta}+\frac{\eta}{\sin{\theta}}\right)^2-g(\theta)=J(\theta) 
    \end{gather}
       \begin{gather}
       \label{eqn:33}
       \frac{\rho \dot{\phi}}{\omega_{0}}  =  \frac{(\rho-2 r) \eta-2 a r \sin^2\theta}{\Delta\sin^2{\theta}}
       \end{gather}
       \begin{gather}
       \label{eqn:34}
       \frac{\rho \dot{t}}{\omega_{0}} = \frac{-(\Delta a^2 \sin^2{\theta}-((r+r_2) r+a^2)^2)-2 r a \eta}{\Delta}    
       \end{gather}
       The variation of refractive index $\mathfrak{n}$ of plasma  now becomes
\begin{gather}
    \label{eqn:ri}
    \mathfrak{n}^2(r, \theta)=1-\frac{f(r)+g(\theta)}{\omega(r, \theta)^2\rho}
\end{gather}
        In above equations $\mathcal{Q}=\frac{\mathcal{C}}{\omega_{0}^{2}}$ and $\eta=\frac{p_\phi}{\omega_0}$ (where $\omega_0=-p_t$).  \Cref{eqn:31},  \Cref{eqn:32},  \Cref{eqn:33}  and \Cref{eqn:34}  when solved,  gives the path of the light rays traveled in a plasma medium.  %(satisfying Perlik-Tsupko separability condition). 
{The effect of plasma on light ray geodesics is taken into account with presence of $f(r)$ and $g(\theta)$ in the geodesic equations. }

   Similar to Kerr black hole,  light rays can form spherical photon orbits in the Kerr-Sen spacetime as well\cite{Teo:2003ltt, Sahoo:2023czj, Banerjee:2021nza},   the projection of these spherical photon orbits on the observer's sky gives the $boundary\ curve$ or $critical\ curve$ (also generally referred as shadow outline).  {We want to take into account the effect of plasma on shadow outline which will be a more general and astrophysically relevant study. }%{ However,  we consider only the dispersive nature of the plasma and consider plasma distributions which satisfy the separability condition which keep calculations analytical}. 
   In order to find the equation of the shadow outline which is the projection of the spherical photon orbits,  we have to first obtain the condition of spherical photon orbits, which corresponds to $\dot{r}=0\text{ and }\ddot{r}=0$,  which means $K(r_p)=0 \text{ and }K'(r_p)=0$ ,  where $r_p$ represents the radius of the spherical photon orbit. From \Cref{eqn:31} and \Cref{eqn:32},
    \begin{gather}
    \label{eqn:36}
    K(r)= \left(-a\eta+(a^2+r(r+r_2))\right)^2-\Delta f(r)-\mathcal{Q}\ \Delta
    \end{gather}
    \begin{gather}
        \label{eqn:37}
     J(\theta) = \mathcal{Q}-\left(-a  \sin{\theta}+\frac{\eta}{\sin{\theta}}\right)^2-g(\theta)
    \end{gather}
The condition of spherical photon orbits when imposed on the radial geodesic equation,  gives expression for allowed values of constants of motion $\mathcal{Q}$ and $\eta$ for a given spherical photon orbit of radius $r_p$.  For the light ray geodesics of Kerr Sen black hole surrounded by plasma %atisfying separabilitycondition
   \cite{Perlick:2017fio} we obtain the expressions of $\mathcal{Q}$ and $\eta$  as:
   \begin{gather}
       \label{equation of eta}
       \eta(r_p)=-\frac{1}{a\Delta'}\left(2 (a^2-r_p^2)\pm(2r_p+r_2)\Delta\sqrt{1-\frac{f'(r_p) \Delta'}{(2 r_p+r_2)^2}}\right)
       \end{gather}
       \begin{gather}
\label{equation of Q}
           \mathcal{Q}(r_p)=\frac{\Delta(2r_p+r_2)^2}{\Delta'^2}\left(1\pm\sqrt{1-\frac{f'(r_p) \Delta'}{(2 r_p+r_2)^2}}\right)^2-f(r_p)
   \end{gather}An additional condition which needs to be imposed is $J(\theta)\equiv\dot{\theta}^2\geq0$ ,  which gives us
\begin{gather}
\label{eqn:41}
    \mathcal{Q}(r_p)-\left(-a \sin{\theta}+\frac{\eta(r_p)}{\sin{\theta}}\right)^2-g(\theta)\geq0
\end{gather}
or
\begin{gather}
    \label{eqn:42}
    \mathcal{Q}(r_p) a^2 \sin^2{\theta}-(-a\eta(r_p)+a^2\sin^2\theta)^2\geq a^2 \sin^2{\theta} g(\theta)
\end{gather}
When \Cref{equation of eta,equation of Q} are substituted in  \Cref{eqn:41}  or \Cref{eqn:42},   the values of $r_p$ that satisfy the condition give the allowed values of radius of photon orbits (the spherical photon region).  
The stability of spherical photon orbits are inferred by analysis of $K''(r)$ given by
\begin{gather}
    K''(r)=4 \left(a^2-a \eta +r (r+r_2)\right)-\Delta  f''(r)-2 f'(r) \Delta '-f(r) \Delta ''-Q \Delta ''+2
   (2 r+r_2)^2
\end{gather}
For unstable photon orbits $K''(r_p)>0$ needs to be satisfied.

\subsection{Obtaining the shadow outline\label{sec:3.2}}
 In order to obtain the equation of the shadow outline, we follow the procedure as explained in \cite{Perlick:2017fio, Grenzebach:2014fha,Grenzebach:2015oea}.  We consider an observer at a distance $d$ and at an angle of inclination $\theta_i$ . We associate the following ortho-normal tetrads to the observer at $(d, \theta_i)$:
\begin{gather}
\label{eqn:43}
    e^\mu_0=\frac{1}{\sqrt{\rho\Delta}}\left\{r (r+r_2)+a^2, 0, 0, a\right\} \Bigg|_{(d, \theta_i)}
\end{gather}
\begin{gather}
\label{eqn:44}
    e^\mu_1=\frac{1}{\sqrt{\rho}}\left\{0, 0, 1, 0\right\}\Bigg|_{(d, \theta_i)}
\end{gather}
\begin{gather}
\label{eqn:45}
    e^\mu_2=\frac{-1}{\sqrt{\rho}\sin\theta}\left\{a \sin^2\theta, 0, 0, 1\right\}\Bigg|_{(d, \theta_i)}
\end{gather}
\begin{gather}
\label{eqn:46}
    e^\mu_3=-\sqrt{\frac{\Delta}{\rho}}\left\{0, 1, 0, 0\right\}\Bigg|_{(d, \theta_i)}
\end{gather}
Here,  $e^\mu_0$ is the 4-velocity of the observer.   The tangent vector to the light ray geodesic emanating from the observer position $(d,\theta_i)$ is, 
\begin{gather}
\label{eqn:47}
\dot{\Gamma}^\mu(\lambda)=\dot{t}\partial_t+\dot{r}\partial_r+\dot{\theta}\partial_\theta+\dot{\phi}\partial_\phi
\end{gather}
where  overdot implies  derivative with respect to a curve parameter $\lambda$.  At the location of the observer the tangent vector can be expanded in the basis of orthonormal tetrads given by \Cref{eqn:43} -\Cref{eqn:46},
\begin{gather}
\label{eqn:48}
    \dot{\Gamma}^\mu(\lambda)=-\Upsilon e^\mu_0 + \beta (\sin\gamma \cos\delta e^\mu_1+\cos\gamma \cos\delta e^\mu_2+\cos\gamma e^\mu_3)
\end{gather}
Here $\Upsilon$ and $\beta$ are positive factors.  From Hamilton's  equations \Cref{eqn:12}  and \Cref{eqn:13} we have $\dot{x}^\mu=p^\mu$ .  Using the result $g_{\mu\nu}\dot{\Gamma}^\mu\dot{\Gamma}^\nu=-\omega^2_{P}$ ,   we obtain the relation between $\Upsilon\text{ and }\beta$ as:
\begin{gather}
\label{eqn:49}
    \Upsilon^2-\beta^2=\omega^2_{P}(d,\theta_i)
\end{gather}
By projecting $\dot{\Gamma}^\mu$ as given in  \Cref{eqn:47} and \Cref{eqn:48}  along $e^\mu_0$,  and comparing both results   we can  obtain $\Upsilon$.  
\begin{gather}
\label{eqn:50}
    \Upsilon = g_{\mu\nu} \dot{\Gamma}^\mu e^\nu_{0}= -\frac{(r+r_2)r+a^2}{\sqrt{\rho \Delta}}+\frac{a \eta}{\sqrt{\rho \Delta}}%\\
\end{gather}
%\implies\sigma&= -\frac{(r+r_2)r+a^2}{\sqrt{\rho \Delta}}+\frac{a \eta}{\sqrt{\rho \Delta}}\\
%\beta \text{ can be calculate using }\\
%\beta&=\sqrt{\sigma^2-\omega^2_P}
Substituting the expression of $\Upsilon $ in  \Cref{eqn:49} one can obtain the expression of $\beta$. 
\begin{gather}
\label{eqn:51}
    \beta=\sqrt{\frac{(a \eta-(r(r+r_2)+a^2))^2-\Delta (f(r)+g(\theta))}{\rho\Delta}}
\end{gather}
By equating the coefficients of $\partial_r\text{ and }\partial_\phi$ in \Cref{eqn:47} and \Cref{eqn:48} we can obtain the expressions of $\gamma\text{ and }\delta$, respectively.  Equating coefficients of $\partial_r$ we get
\begin{align}\label{eqn:52}
    -\beta \cos\gamma \sqrt{\frac{\Delta}{\rho}}&=\dot{r}%\\
%\implies\cos^2\gamma&=\frac{\rho \dot{r}^2}{\Delta \beta^2}\nonumber\\    
%\implies\sin\gamma&=\sqrt{1-\frac{\rho \dot{r}^2}{\Delta \beta^2}}\nonumber\\
%\text{ using }\rho^2\dot{r}^2=\omega^2_0 K(r)\nonumber\\
%\implies\sin\gamma&=\sqrt{\frac{(\mathcal{Q}-g(\theta))\Delta}{(a  \eta(r_p)-(r(r+r_2)+a^2))^2-\Delta(f(r)+g(\theta_i))}}
\end{align}
Using radial geodesic equation  \Cref{eqn:31} in \Cref{eqn:52} one obtains,
\begin{gather}
\label{eqn:53}
\sin\gamma=\sqrt{\frac{(\mathcal{Q}-g(\theta))\Delta}{(a  \eta(r_p)-(r(r+r_2)+a^2))^2-\Delta(f(r)+g(\theta))}}
\end{gather}
Similarly, for the case of  $\partial_\phi$ we get
\begin{gather}
\label{eqn:54}
    \dot{\phi}=\frac{-a\Upsilon}{\sqrt{\rho\Delta}}-\frac{\beta \sin\gamma\sin\delta}{\sin\theta\sqrt{\rho}}
    \end{gather}
Using \Cref{eqn:33} one obtains,
\begin{gather}
\label{eqn:55}
    \sin\delta=\frac{-\eta+a \sin^2\theta}{\sin\theta\sqrt{\mathcal{Q}-g(\theta)}}\bigg|_{(d, \theta_i)}
\end{gather}
For the observer,  the local coordinates $(\gamma,\delta)$  should be calculated at $(d, \theta_i)$, such that

\begin{gather}
\label{eqn:56}
    \sin\gamma=\sqrt{\frac{(\mathcal{Q}-g(\theta))\Delta}{(a  \eta-(r(r+r_2)+a^2))^2-\Delta(f(r)+g(\theta))}}\Bigg|_{(d, \theta_i)}
\end{gather}
\begin{gather}
\label{eqn:57}
    \sin\delta=\frac{-\eta+a \sin^2\theta}{\sin\theta\sqrt{\mathcal{Q}-g(\theta)}}\bigg|_{(d, \theta_i)}
\end{gather}

The shadow outline is the locus of all points in the observer's sky  which when traced back along the light ray geodesic paths will  reach the spherical photon orbits\cite{Perlick:2017fio}. For a spherical photon radius $r_p$ one needs to calculate the constants of motion $\mathcal{Q}(r_p)$ and $\eta(r_p)$  using \Cref{equation of Q} and \Cref{equation of eta}. Then using \Cref{eqn:56} and \Cref{eqn:57} one calculates the angles $\gamma$ and $\delta$ which  are angular coordinates of the shadow outline in the observer's celestial sphere.
The range of radius of photon orbits which should be used to plot the shadow are obtained by the following condition
\begin{gather}
\label{eqn:58}
    \sin \delta(r_{p_{min/max}}) =\pm 1
\end{gather}
which can be written as
\begin{gather}
    \label{eqn:59}
    \frac{-\eta(r_p)+a \sin^2\theta_i}{\sin\theta_i\sqrt{\mathcal{Q}-g(\theta_i)}}=\pm 1
\end{gather}
The above conditions mean that the shadow outline for the observer at $(d,\theta_i)$ is contributed by the photon region which has turning points at $\theta_i$, i.e. $J(\theta_i)=0.$
 For each value of $r_p$ between  $r_{p_{min}}\text{ and } r_{p_{max}}$ there is a unique $\gamma$ given by \Cref{eqn:56}  but  two values of $\delta$ between   $(-\pi/2,\pi/2)$ and  $(\pi/2,3 \pi/2)$.   In order to obtain the $x\text{ and }y$ coordinates of the shadow we use the stereographic projection of the celestial sphere onto the plane tangent to the celestial sphere at $\gamma=0$ . This gives
\begin{gather}
    \label{cartesian shadow equation}
    x(r_p)=-2 d \tan\left(\frac{\gamma(r_p)}{2}\right)\sin\delta(r_p)
\end{gather}
\begin{gather}
    \label{eqn:61}
    y(r_p)=-2 d \tan\left(\frac{\gamma(r_p)}{2}\right)\cos\delta(r_p)
\end{gather}
the angular coordinate along the $x\text{ and }y$ axes  are given by $X=\frac{x}{D}\text{ and }Y=\frac{y}{D}$\cite{Perlick:2004tq, Grenzebach:2014fha}.  The expressions of angular coordinates are 
\begin{gather}
\label{angular x coordinate of shadow}
    X(r_p)=-2 \tan\left({\frac{\gamma(r_p)}{2}}\right)\sin{\delta(r_p)}
    \end{gather}
    \begin{gather}
    \label{angular y coordinate of shadow}
    Y(r_p)=-2 \tan\left({\frac{\gamma(r_p)}{2}}\right)\cos{\delta(r_p)}
\end{gather}
{In the  equations \Cref{angular x coordinate of shadow},  \Cref{angular y coordinate of shadow},   $X(r_p)$ and $Y(r_p)$ are in the units  of radian and thus can be used directly  to calculate the angular diameter of the shadow outline.  From the equations it can be shown that the shadow outline will be symmetric about the horizontal axis.   The present approach also takes in consideration the effect of finite distance of the observer from the black hole in presence of plasma. }

Thus,  if $r_{p_1}$  represents the radius of the spherical photon orbit corresponding to maximum angular height of the shadow from the horizontal,  i.e,   $Y_{max}=Y(r_{p_1})$,  then \text{the vertical angular diameter} $\Delta\Theta$ \text{of the shadow} can be computed using the formula given below:
\begin{gather}
\label{theoretical angular diameter}
    \Delta\Theta=2 Y_{max}
\end{gather}
In order to find  $r_{p_1}$ we solve the following equation,
\begin{gather}
\label{eqn:60}
    \frac{dY(r_p)}{dr_p}{\Bigg|}_{(r_p=r_{p_1})}=0
\end{gather}
where $r_{p_{min}}<r_{p_1}<r_{p_{max}}$.
In case of a non- rotating black hole we  obtain $r_{p_1}=r_s$,  where $r_s$ can obtained by solving :
\begin{gather}
\label{eqn:66}
    -\frac{1}{\Delta'}\left(2 (a^2-r_p^2)\pm(2r_p+r_2)\Delta\sqrt{1-\frac{f'(r_p) \Delta'}{\Delta'^2}}\right)=0
\end{gather}
The above equation results from the fact that for non-rotating black hole, $a\eta(r_p)=0$ (see \Cref{equation of eta}).

\section{Shadow of Kerr-Sen black hole surrounded by plasma\label{sec:4}}
In this section we study the nature of shadow of Kerr Sen black hole surrounded by a pressureless, non-magnetized plasma. We have already derived the $X\text{ and }Y$  coordinates of the shadow in the previous section. In order to proceed further one needs to specify the plasma profile.
Recall, in \Cref{sec:3.1} we obtained in \Cref{eqn:omegap_form}
\begin{gather}
    \label{eqn:68}
    \omega_{P}(r, \theta)^2=\omega^2_b\left(\frac{f(r)+g(\theta)}{\rho}\right)
\end{gather}
While using \Cref{eqn:22} for the Kerr-Sen black hole, 
\begin{gather}
\label{eqn:69}
    \frac{\omega^2(x)}{\omega^2_0}=\left(1-\frac{2 r}{\rho}\right)^{-1}
\end{gather}
Using the condition \Cref{eqn:18} we get,
\begin{gather}
    \label{eqn:bound}
    \left(1-\frac{2 r}{\rho}\right)^{-1}\geq\alpha\left(\frac{{f}(r)+{g}(\theta)}{\rho}\right)
\end{gather}
where $\alpha=\frac{\omega^2_b}{\omega^2_0}$.
On rearranging the above inequality we obtain
\begin{gather}
\label{eqn:bounds2}
    \alpha\leq\left(1-\frac{2 r}{\rho}\right)^{-1}\left(\frac{f(r)+g(\theta)}{\rho}\right)^{-1}=F(r,\theta)
\end{gather}
Thus, for a given plasma profile surrounding a Kerr Sen black hole of dilaton charge $r_2$ and spin $a$, the maximum possible bound on $\alpha$ is
\begin{gather}
\label{eqn:boundfinal}
    \alpha_{max}=\left(1-\frac{2 r}{\rho}\right)^{-1}\left(\frac{f(r)+g(\theta)}{\rho}\right)^{-1}\Bigg|_{min}
\end{gather}
In our work, we will consider three plasma profiles satisfying the separability condition \Cref{eqn:68} \cite{Perlick:2017fio}:

\begin{gather}
    \label{eqn:p1}
\text{Profile 1:    }\frac{\omega^2_P(r, \theta)}{\omega_0^2}=\alpha_1\left(\frac{\sqrt{ r}}{\rho}\right)\ \text{here    } f(r)=\sqrt{r}\text{ and }g(\theta)=0
\end{gather}

\begin{gather}
    \label{eqn:p2}
\text{Profile 2:      }\frac{\omega^2_P(r, \theta)}{\omega_0^2}=\alpha_2\left(\frac{(1+2 \sin^2\theta)}{\rho}\right)\ \text{here    } f(r)=0\text{ and }g(\theta)=(1+2 \sin^2\theta)
\end{gather}
\begin{gather}
    \label{eqn:p3}
\text{Profile 3:      }\frac{\omega^2_P(r, \theta)}{\omega_0^2}=\alpha_3\left(\frac{r(r+r_2)+a^2\cos^2\theta)}{\rho}\right)\text{ here    } f(r)= r(r+r_2)\text{ and }g(\theta)=a^2 \cos^2\theta
\end{gather}

\subsection{Variation of shadow considering profile 1\label{p1}}

{  The plasma profile in \Cref{eqn:p1} was considered by Shapiro \cite{shapiro1974accretion} when he studied steady-state spherical accretion of interstellar gas/plasma onto Kerr black holes. 
The gas was assumed to be at rest at infinity, but close to the black hole it acquires an angular momentum due to the black
hole's rotation. The accreting gas was assumed to have a polytropic index between $1$ and $5/3$ such that the gas pressure becomes insignificant as the gas approaches the BH. The flow becomes highly supersonic as the gas enters the capture radius $r_c$ and the fluid essentially follows geodesic paths. The velocity
of the plasma in the regime $r_{h_{+}} < R < r_c  $ is comparable to the free-fall velocity $u^r = \sqrt{2GM/R}$, where $r_{h_{+}}$ is the
outer horizon and $r_c = GM/c^2_s$
is the capture radius, $c_s$ being the sound speed \cite{shapiro1974accretion}. The infalling particles are in radial free-fall from rest at infinity where the pressure forces can be neglected.
The fluid particles fall into the BH along a conical surface of constant $\theta_0$, where $\theta_0$ is the polar angle at infinity, such that $u^\theta=\frac{d\theta}{d\tau}\simeq 0$.
This is a reasonable
approximation to model the geometrically thick and optically thin accretion flow generally
prevalent in supermassive black holes accreting from the surrounding ISM, e.g. Sgr A*. 

The accreting fluid obeys conservation of mass and energy momentum tensor.
To arrive at the Profile 1 we use the conservation of mass flux equation,
\begin{align}
\nabla_i(n u^i)=0
\label{20-1}
\end{align}
where, $n$ is the mass density of the accreting fluid measured in a comoving frame and $u^i$ is the 4-velocity of the accreting fluid. For steady-state, axi-symmetric flow Equation \ref{20-1} reduces to,
\begin{align}
\frac{d}{dR}[\sqrt{-g} n u^r]=0
\label{20-2}
\end{align}
where we assumed $u^\theta\simeq 0$. For the Kerr-Sen metric $\sqrt{-g}=\rho\sin\theta$, which when used in Equation \ref{20-2} yields,
\begin{align}
\frac{d}{dR}[\rho n u^r]=0
\end{align}
which upon integration yields,
\begin{align}
4\pi \rho n u^r=\dot{M}
\label{20-3}
\end{align}
where, $\dot{M}$ is the constant mass accretion rate. Considering, $u^r = \sqrt{2GM/R}$ (as discussed above) in Equation \ref{20-3} we get,
\begin{align}
n(r)= \frac{\dot{M}\sqrt{R}}{4\pi\rho\sqrt{2GM}}
\label{20-4}
\end{align}
In this model it was assumed that the accreting fluid comprises of fully ionized hydrogen gas. As the electron mass is negligible compared to the mass of the proton, the plasma density is given by $n=\mathcal{N}m_p$ where $m_p$ is the mass of the proton and $\mathcal{N}$ is the proton number density which is same as the electron number density as the plasma is assumed to be electrically neutral \cite{shapiro1974accretion,Perlick:2015vta}.
Using Equation \ref{20-4} we thus get,
\begin{align}
\mathcal{N}(r)=\frac{\dot{M}\sqrt{R}}{4\pi m_p \rho\sqrt{2GM}}
\label{20-5}
\end{align}
The electron number density $\mathcal{N}$ is also related to the plasma frequency as given in Equation \ref{eqn:9} which yields,
\begin{align}
\omega_P^2&=\frac{e^2 \dot{M}}{m_e m_p \sqrt{2GM}}\frac{\sqrt{R}}{\rho} \nonumber \\
\omega_P^2&=\frac{e^2 \dot{M}c^3}{\sqrt{2} m_e m_p G^2M^2}\frac{\sqrt{r}}{\rho}=\omega_b^2\frac{\sqrt{r}}{\rho}
\label{21}
\end{align}
where, in Equation \ref{21} ${r}=R/r_g$ and $\rho\equiv \rho/r_g^2$. Dividing Equation \ref{21} by $\omega_0^2$ yields plasma profile 1 given in Equation \ref{eqn:p1}, i.e., 
\begin{align}
\frac{\omega_P^2}{\omega_0^2}&=\frac{e^2 \dot{M}c^3}{\sqrt{2} m_e m_p \omega_0^2G^2M^2}\frac{\sqrt{r}}{\rho}=\alpha_1\frac{\sqrt{r}}{\rho}
\label{21-1}
\end{align}
where,
\begin{align}
\alpha_1=\frac{e^2 \dot{M}c^3}{\sqrt{2}\omega_0^2 m_e m_p G^2M^2}
\label{21-2}
\end{align}
}

\iffalse

The plasma is assumed to be at rest at infinity, but close to the black hole it acquires an angular momentum due to the black
hole’s rotation, i.e., at $r >> r_g$ the accreting fluid only has radial velocity. This is a reasonable
approximation to model the geometrically thick and optically thin accretion flow generally
prevalent in supermassive black holes accreting from the surrounding ISM, e.g. Sgr A*. The velocity
of the plasma in the regime $r_{h_{+}} < r < r_c  $ (where the plasma density through which the photon
travels is substantial) is comparable to the free-fall velocity $u^r = \sqrt{2GM/r}$, where $r_{h_{+}}$ is the
outer horizon and $r_c = GM/c^2_s$
is the capture radius, $c_s$ being the sound speed \cite{shapiro1974accretion}. Considering
conservation of mass and energy-momentum tensor, it can be shown that
\begin{gather}
4\pi n u^r \rho = A = constant     
\end{gather}
which in turn implies
\begin{gather}
n = \frac{C_0 \sqrt{r}}{\rho}    
\end{gather}
which is the plasma profile considered in this section. Here, $n$ is the density of the plasma
measured in the comoving frame.
\fi

\begin{figure}[H]
    \centering
     \begin{subfigure}{0.5\textwidth}
        \centering
        \includegraphics[width=\linewidth]{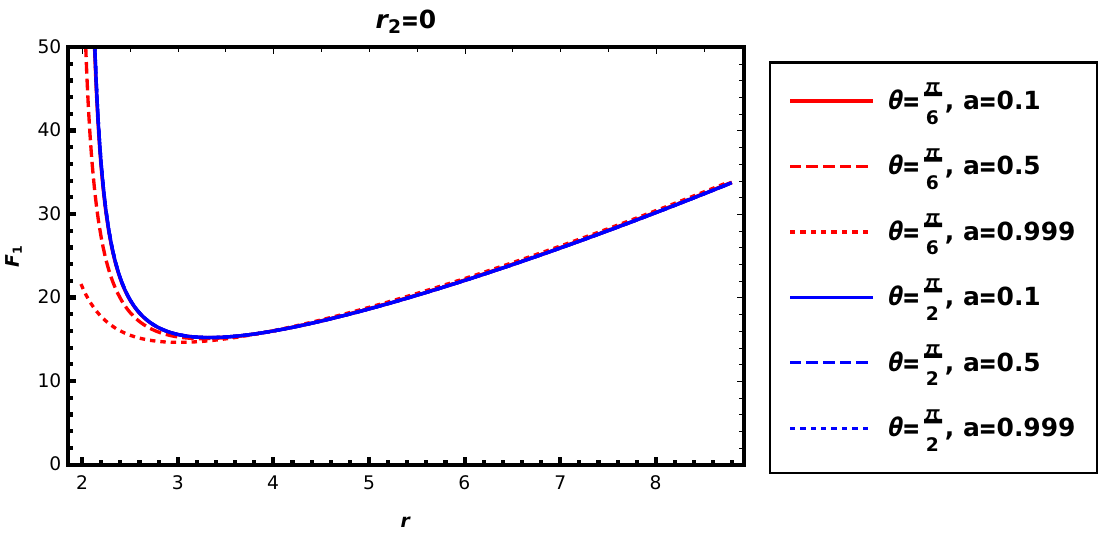}
        \caption{\label{Fr1a}$r_2=0$}
    \end{subfigure}\hspace{0.1cm}
    \begin{subfigure}{0.5\textwidth}
        \centering
        \includegraphics[width=\linewidth]{ 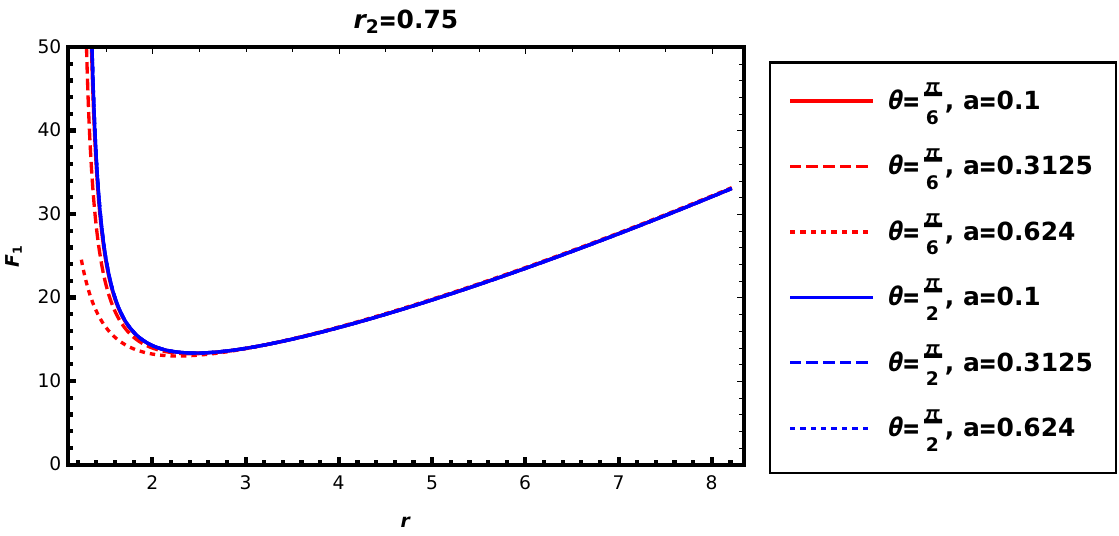}
        \caption{\label{Fr1b}$r_2=0.75$}
    \end{subfigure}\vspace{0.3cm}
    \begin{subfigure}{0.5\textwidth}
        \centering
        \includegraphics[width=\linewidth]{ 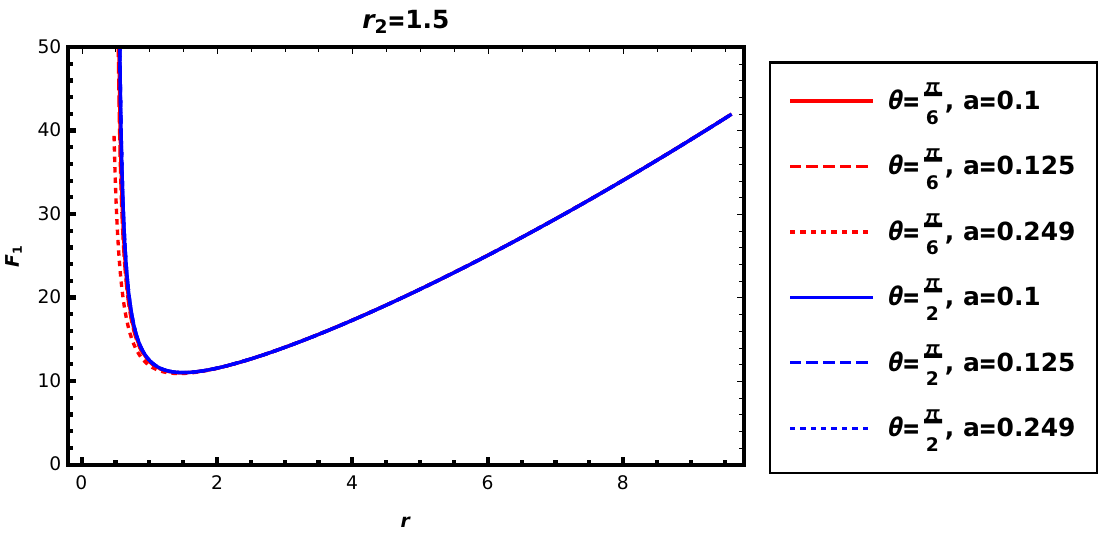}
        \caption{\label{F1c}$r_2=1.5$}
    \end{subfigure}
    \caption{\label{F1}Variation of $F_1(r,\theta)$ with r (in units of $M$) at $a=0.1, 0.5(1-\frac{r_2}{2})\text{ and }
    0.999-\frac{r_2}{2}$  for $\theta=\frac{\pi}{6}\text{ (Red)  and }\frac{\pi}{2}  $ (Blue) for different $r_2$ }
\label{Fig1}
\end{figure}

In profile 1, the plasma parameter $\alpha_1\geq0$ because $\omega^2_P\geq0$. Furthermore, $\alpha_1$ cannot take arbitrary high values because the condition \eqref{eqn:bounds2}  needs to be satisfied for light propagation in plasma. The maximum value of $\alpha_1$ is determined by condition \eqref{eqn:boundfinal}. We denote $F(r,\theta)$ by $F_1(r,\theta)$ for profile 1. In \Cref{F1} we plot the variation of $F_1(r,\theta)$ with $r$ at $\theta=\frac{\pi}{6}\text{ (Red) and }\frac{\pi}{2}\text{ (Blue)}$. \Cref{Fr1a,Fr1b,F1c} represent plots for $r_2=0,0.75\text{ and }1.5$, respectively. Furthermore, for each $\theta$ and $r_2$ we make plots for spins $a=0.1\text{(Solid)}, 0.5(1-\frac{r_2}{2})\text{ (Dashed) and }(0.999-\frac{r_2}{2}))\text{ (Dotted)}$.  

From \Cref{Fr1a,Fr1b,F1c} we make the following observations:
\begin{itemize}
    \item   For each choice of $a$  and  $\theta$ for a given $r_2$ in  \Cref{Fr1a,Fr1b,F1c} there exists a minimum.  
    \item $F_1(r,\theta)$ is sensitive to $a$ at lower values of $\theta$, and close to the black hole.
    \item $F_1(r,\theta)$ for a given $r_2$ and $a$ is nearly insensitive to the variation in $\theta$ as we go away from the horizon.
    \item The minimum of $F_1(r,\theta)$ (which corresponds to $\alpha_{1max}$) in \Cref{Fr1a,Fr1b,F1c}, decreases with an increase in $r_2$.
    \item  The minimum value of $F_1(r,\theta)$ can be used to estimate the maximum value of $\alpha_1$ for a given $r_2$, that is, for $r_2=0, 0.75\text{ and }1.5$ the corresponding $\alpha_{1max}\approx 16, 14\text{ and }10$, respectively. 
\end{itemize}
We now study the effect of  the dilaton charge, the spin of black hole, and the plasma parameter $\alpha_1$  on the shadow at different inclination angles. We describe the necessary details about our study below :
\begin{itemize}
\item \Cref{1,2,3} plots the  variation of shadow with $\alpha_1$ for different values of $r_2\text{ and }a$ assuming $\theta_i=15^\circ ,45^\circ\text{ and } 90^\circ$, respectively.
    \item All the plots are done assuming a black hole of mass $M=6.2\times10^{9} M_{\odot}$ and seen by an observer at a distance $D\approx5.6\times10^{10} M$ (or $16.8$ Mpc), which correspond to mass and distance measurements of M87* \cite{EventHorizonTelescope:2019dse}, respectively.

    \item Each column in \Cref{1,2,3} is associated with a fixed $r_2$. Column 1 corresponds to $r_2=0$, column 2 corresponds to $r_2=0.75$ and column 3 corresponds to $r_2=1.5$.
    \item Each column contains three plots for the chosen $r_2$ with spins $a=0$, half maximal spin $a=0.5\left(1-\frac{r_2}{2}\right)$ and near maximal spin $a=(0.999-\frac{r_2}{2})$  as we go vertically downwards.
    \item  In each sub figure the individual plots show   shadow outlines for $\alpha_1=0$ (Magenta curve),  $\alpha_1=5$ (Blue curve) and $\alpha_1=10$ (Red curve) for a given  $r_2$, $a$ and $\theta_i$. The $X$ and $Y$ angular coordinates are in units of $\mu as$, while the spin $a$ and the dilaton charge $r_2$ are in  geometrised units.    
\end{itemize}

From \Cref{1} we draw the following conclusions:
\begin{itemize}
    \item The contraction of the shadow size due to the dilaton charge  $r_2$ can be seen from the   \Cref{r1a,r1b,r1c}  for all values of $\alpha_1$, i.e,  $\alpha_1=0$, $\alpha_1=5$ and $\alpha_1=10$, which was also reported in\cite{Sahoo:2023czj}. This contraction due to  $r_2$ can also be seen in case of non-zero $a$ and $\alpha_1=0,5\text{ and }10$ as we go horizontally from \Cref{r1d,r1e,r1f} and then from \Cref{r1g,r1e,r1f}. Thus we observe  contraction of shadow size due to $r_2$  both in absence and presence of plasma and irrespective of the spin $a$.    
    \item  \Cref{r1a}  shows purely the effect of $\alpha_1$ on the shadow size since $r_2=0\text{ and }a=0$ at $\theta=15^\circ$.  We observe that as $\alpha_1$ increases, the size of the shadow decreases. This effect of $\alpha_1$ is quite generic irrespective of the choice of  $r_2\text{ and }a$. We also observe  from \Cref{r1b,r1c,r1d,r1e,r1f,r1g,r1h,r1i} that with non-zero dilaton charge $r_2$ the shadow size decreases more strongly with increase in $\alpha_1$.
    
   \item   The effect of spin $a$ on shadow outline can be seen for given $(r_2,\alpha_1,\theta_i)$ if we go vertically  downwards along each column in \Cref{1}. In each column $r_2$ and $\theta_i$ are fixed. Comparing same colored shadow outline plots (which means same $\alpha_1$), we observe that as spin $a$ increases there is change of geometric center of the shadow. 
    
    \item  Note that all shadow outlines in \Cref{1}  are nearly circular. This is because of the small angle inclination $\theta_i=15^\circ$.

\end{itemize}

 We next consider shadow plots in \Cref{2}. From analysis of    \Cref{2} we draw the following conclusions:
\begin{itemize}
    \item The increase in the dilaton charge $r_2$ causes a contraction in the shadow size (\Cref{2a,2b,2c}), as was observed in the case of  $\theta_i=15^\circ$(see \Cref{1}). Interestingly, even though $\theta_i$ has changed from $15^\circ$  to $45^\circ$ , the degree of contraction in vertical angular width of shadow due to $r_2$ are nearly the same. %This is because although $\theta_i$ is different for \Cref{1} and  \Cref{2}, the vertical angular widths of the shadow   in both are nearly same. 

\begin{figure}[H]
    \centering
    \begin{subfigure}{0.333\textwidth}
        \centering
        \includegraphics[width=\linewidth]{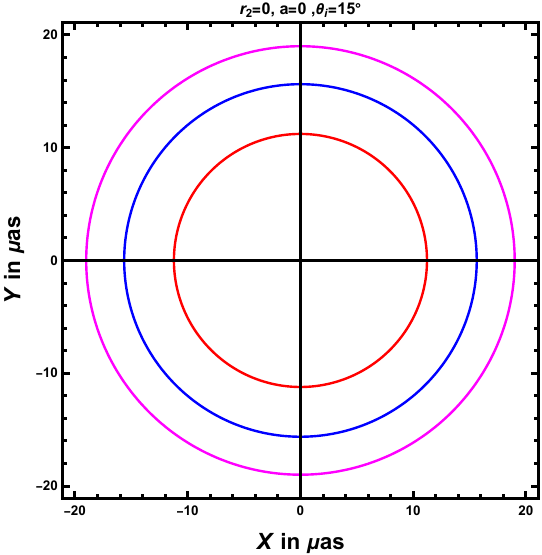}
        \caption{$r_2=0, \ a=0$}
  \label{r1a}  
    
    \end{subfigure}\hfill
    \begin{subfigure}{0.333\textwidth}
        \centering
        \includegraphics[width=\linewidth]{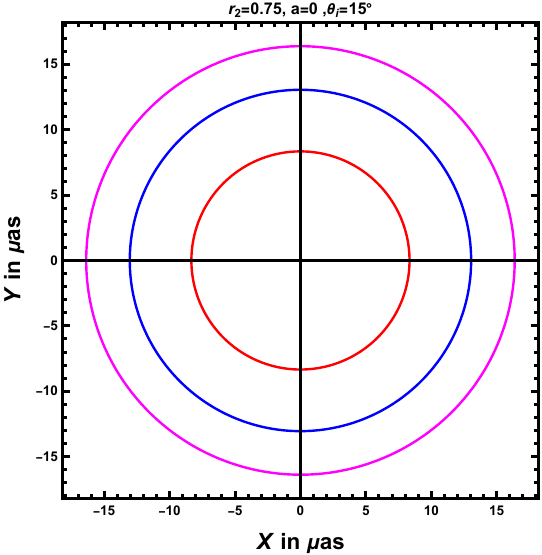}
        \caption{$r_2=0. 75, \ a=0$}
    
    \label{r1b}  
    \end{subfigure}\hfill
    \begin{subfigure}{0.333\textwidth}
        \centering
        \includegraphics[width=\linewidth]{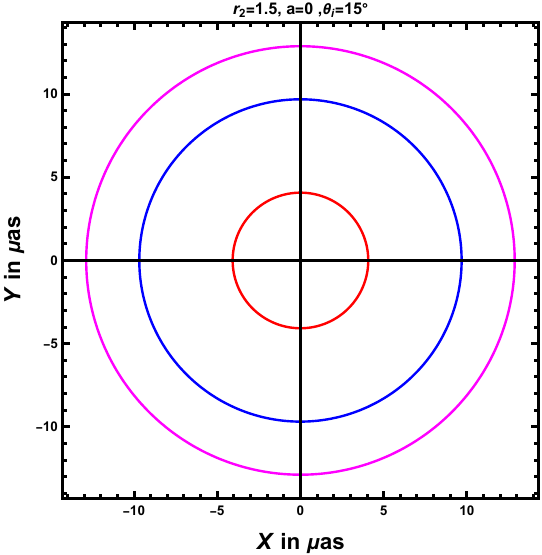}
        \caption{$r_2=1.5, \ a=0$}
    \label{r1c}  
    
    \end{subfigure}

    \vspace{0.5cm}

    \begin{subfigure}{0.333\textwidth}
        \centering
        \includegraphics[width=\linewidth]{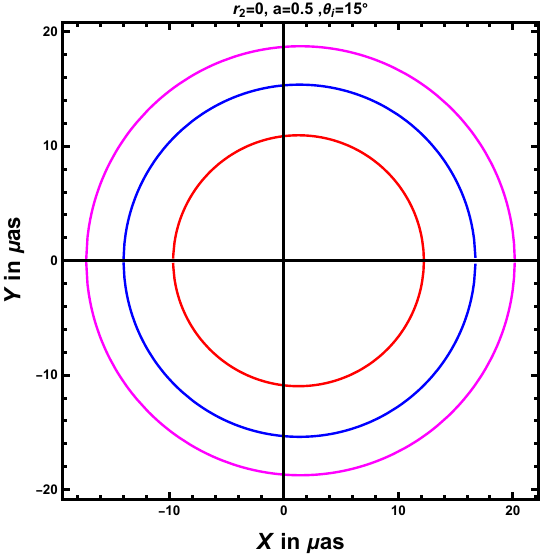}
        \caption{$r_2=0, \ a=0.5$}
    \label{r1d}  
    
    \end{subfigure}\hfill
    \begin{subfigure}{0.333\textwidth}
        \centering
        \includegraphics[width=\linewidth]{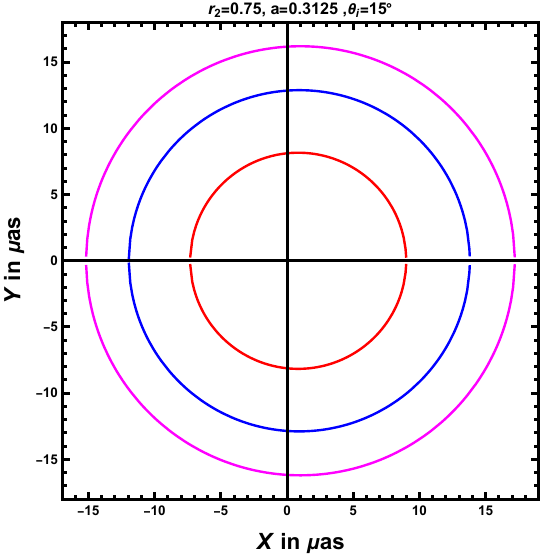}
        \caption{$r_2=0.75, \ a=0.3125$}
    \label{r1e}  
    
    \end{subfigure}\hfill
    \begin{subfigure}{0.333\textwidth}
        \centering
        \includegraphics[width=\linewidth]{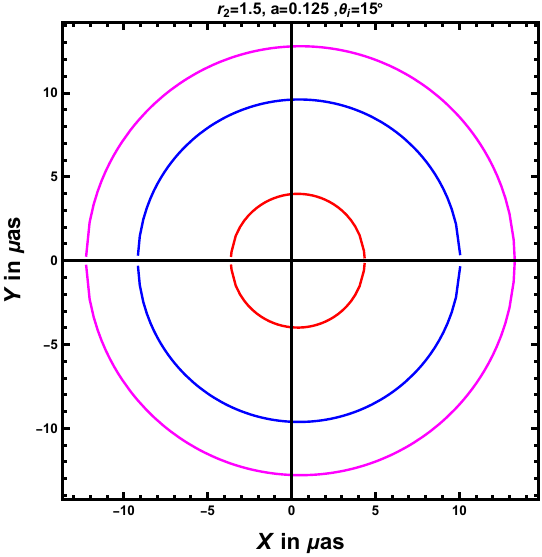}
        \caption{$r_2=1.5, \ a=0.125$}
    \label{r1f}  
    
    \end{subfigure}

    \vspace{0.5cm}

    \begin{subfigure}{0.333\textwidth}
        \centering
        \includegraphics[width=\linewidth]{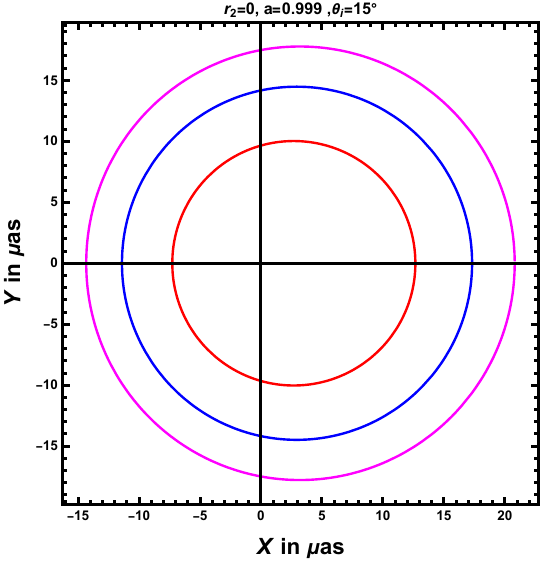}
        \caption{$r_2=0, \ a=0.999$}
    \label{r1g}  
    \end{subfigure}\hfill
    \begin{subfigure}{0.333\textwidth}
        \centering
        \includegraphics[width=\linewidth]{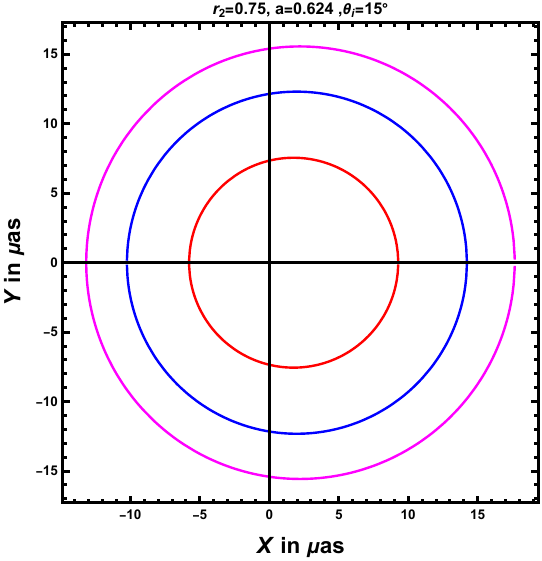}
        \caption{$r_2=0.75, \ a=0.624$}
    \label{r1h}  
    \end{subfigure}\hfill
    \begin{subfigure}{0.333\textwidth}
        \centering
        \includegraphics[width=\linewidth]{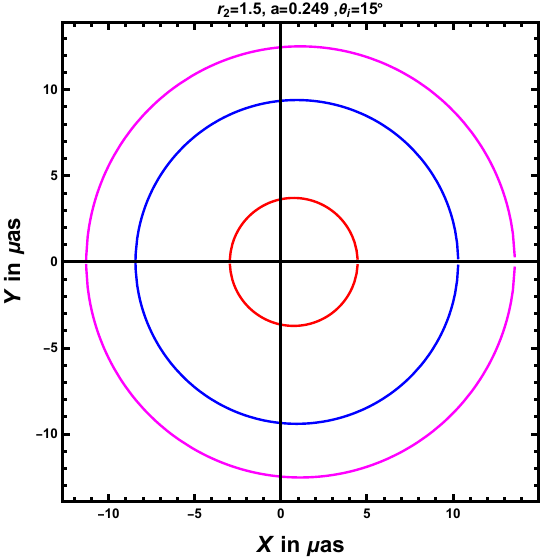}
        \caption{$r_2=1.5, \ a=0.249$}
 \label{r1i}  
    \end{subfigure}
    \vspace{0.1cm}
 \begin{subfigure}{1\textwidth}
        \centering
        \includegraphics[width=0.5\linewidth]{ 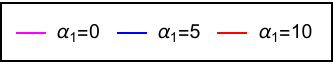}
          
    \end{subfigure}
    \caption{\label{1}Variation of  shadow of Kerr Sen black hole in presence of plasma profile 1 at inclination angle $\theta_i=15^\circ$.}
\end{figure}

\begin{figure}[H]
    \centering
    \begin{subfigure}{0.333\textwidth}
        \centering
        \includegraphics[width=\linewidth]{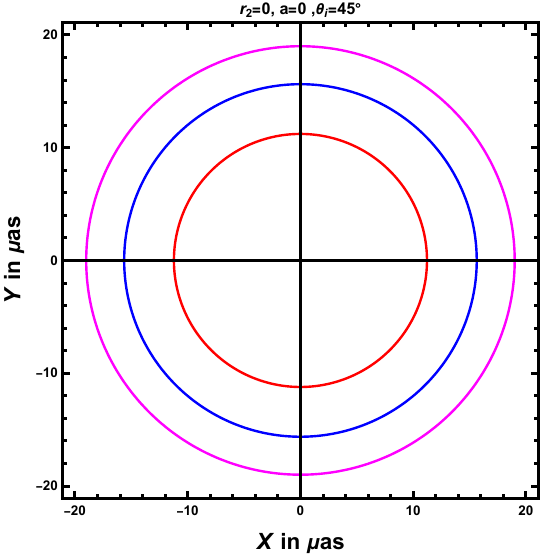}
        \caption{$r_2=0, \ a=0$}
  \label{2a}  
    
    \end{subfigure}\hfill
    \begin{subfigure}{0.333\textwidth}
        \centering
        \includegraphics[width=\linewidth]{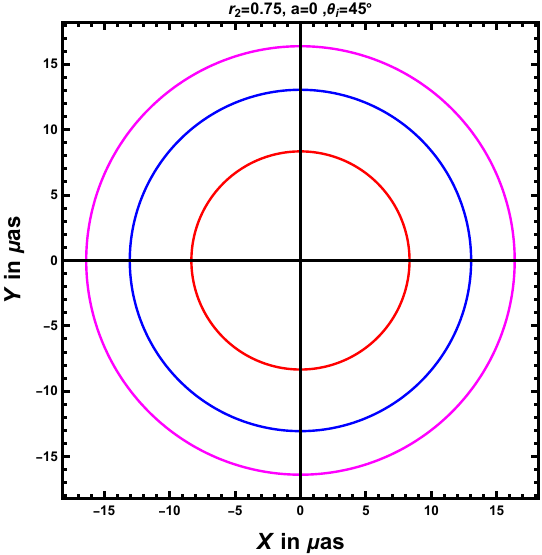}
        \caption{$r_2=0. 75, \ a=0$}
    
    \label{2b}  
    \end{subfigure}\hfill
    \begin{subfigure}{0.333\textwidth}
        \centering
        \includegraphics[width=\linewidth]{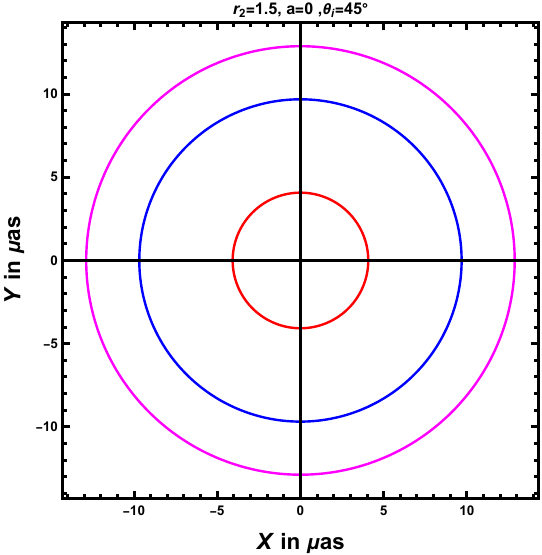}
        \caption{$r_2=1.5, \ a=0$}
    \label{2c}  
    
    \end{subfigure}

    \vspace{0.5cm}

    \begin{subfigure}{0.333\textwidth}
        \centering
        \includegraphics[width=\linewidth]{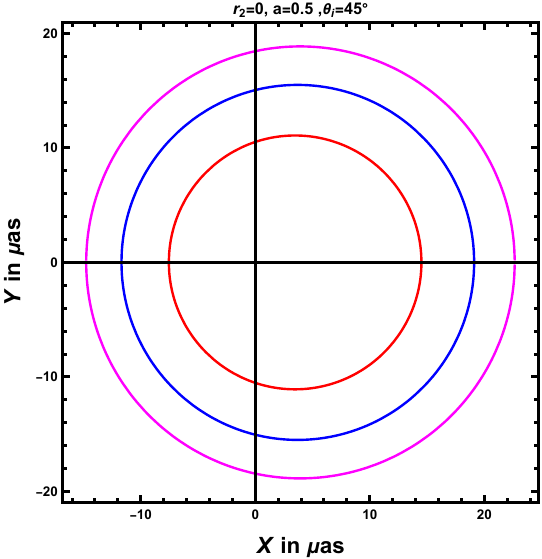}
        \caption{$r_2=0, \ a=0.5$}
    \label{2d}  
    
    \end{subfigure}\hfill
    \begin{subfigure}{0.333\textwidth}
        \centering
        \includegraphics[width=\linewidth]{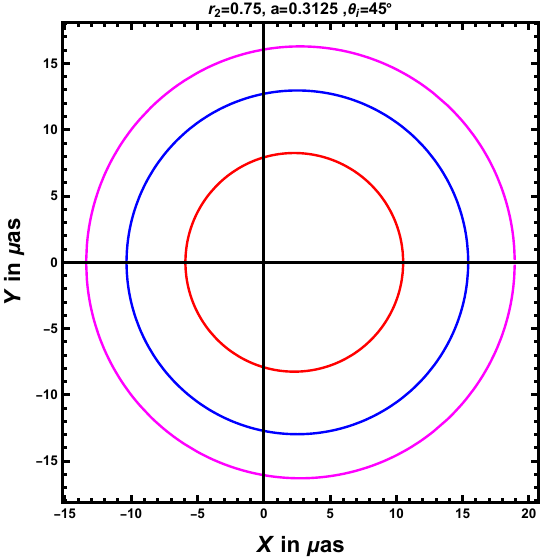}
        \caption{$r_2=0.75, \ a=0.3125$}
    \label{2e}  
    
    \end{subfigure}\hfill
    \begin{subfigure}{0.333\textwidth}
        \centering
        \includegraphics[width=\linewidth]{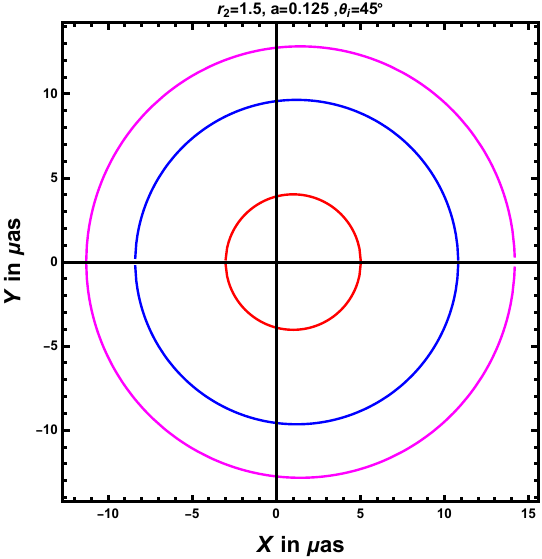}
        \caption{$r_2=1.5, \ a=0.125$}
    \label{2f}  
    
    \end{subfigure}

    \vspace{0.5cm}

    \begin{subfigure}{0.333\textwidth}
        \centering
        \includegraphics[width=\linewidth]{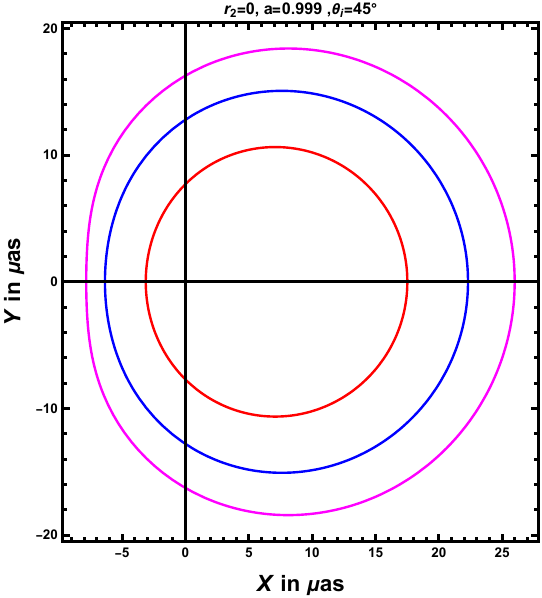}
        \caption{$r_2=0, \ a=0.999$}
    \label{2g}  
    \end{subfigure}\hfill
    \begin{subfigure}{0.333\textwidth}
        \centering
        \includegraphics[width=\linewidth]{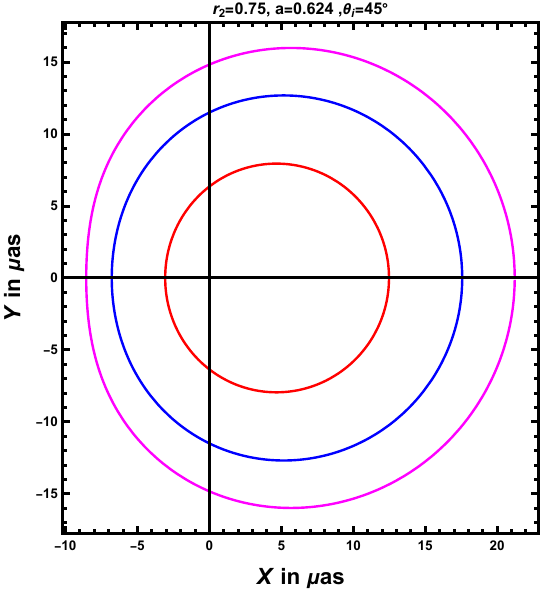}
        \caption{$r_2=0.75, \ a=0.624$}
    \label{2h}  
    \end{subfigure}\hfill
    \begin{subfigure}{0.333\textwidth}
        \centering
        \includegraphics[width=\linewidth]{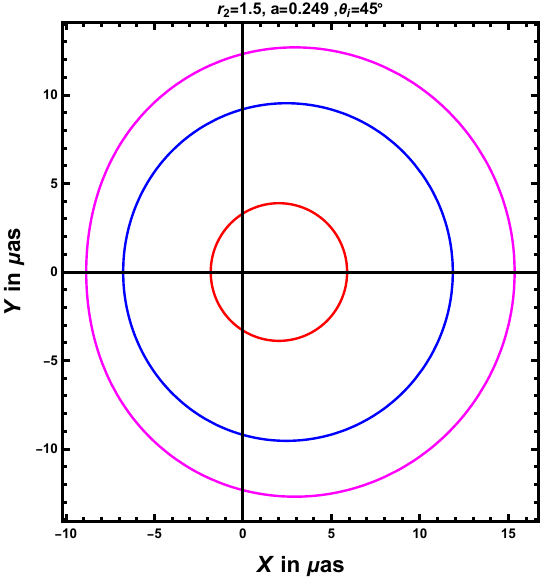}
 
        \caption{$r_2=1.5, \ a=0.249$}
    \label{2i}
    \end{subfigure}
    \vspace{0.1cm}
 \begin{subfigure}{1\textwidth}
        \centering
        \includegraphics[width=0.5\linewidth]{ legend1.pdf}
          
    \end{subfigure}
    
    \caption{\label{2}Variation of  shadow of Kerr Sen black hole in presence of plasma profile 1 at inclination angle $\theta_i=45^\circ$.}
\end{figure}

\begin{figure}[H]
    \centering
    \begin{subfigure}{0.333\textwidth}
        \centering
        \includegraphics[width=\linewidth]{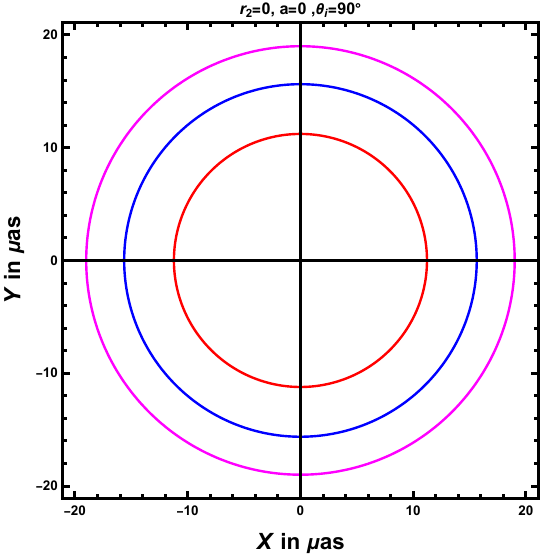}
        \caption{$r_2=0, \ a=0$}
  \label{3a}  
    
    \end{subfigure}\hfill
    \begin{subfigure}{0.333\textwidth}
        \centering
        \includegraphics[width=\linewidth]{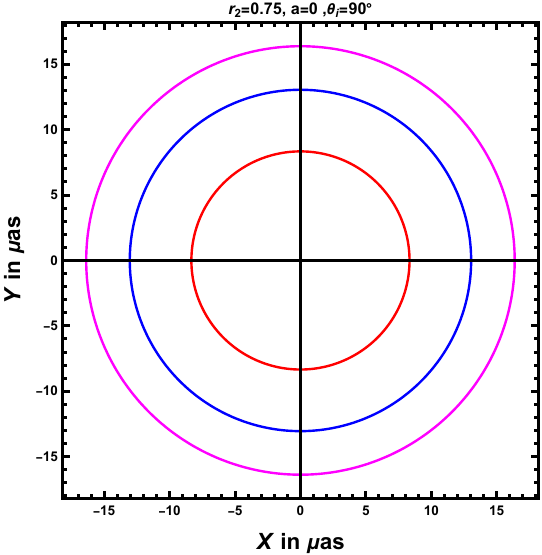}
        \caption{$r_2=0. 75, \ a=0$}
    
    \label{3b}  
    \end{subfigure}\hfill
    \begin{subfigure}{0.333\textwidth}
        \centering
        \includegraphics[width=\linewidth]{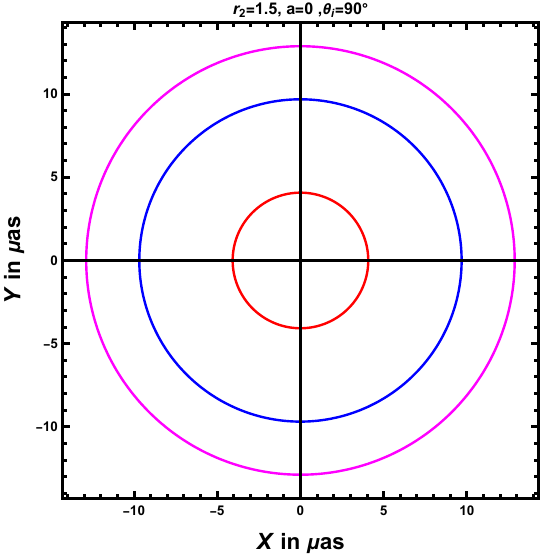}
        \caption{$r_2=1.5, \ a=0$}
    \label{3c}  
    
    \end{subfigure}

    \vspace{0.5cm}

    \begin{subfigure}{0.333\textwidth}
        \centering
        \includegraphics[width=\linewidth]{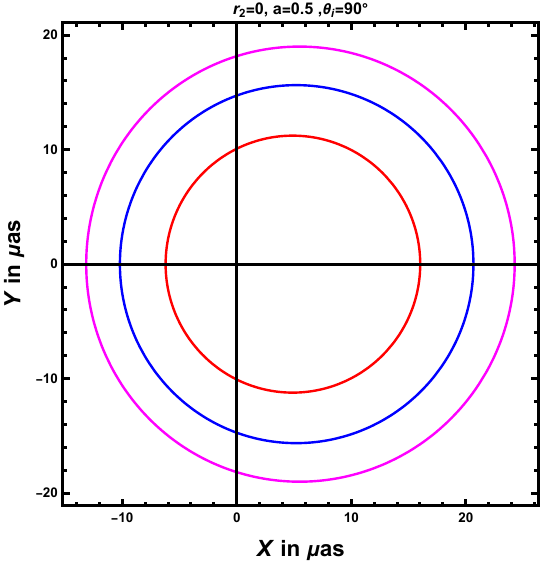}
        \caption{$r_2=0, \ a=0.5$}
    \label{3d}  
    
    \end{subfigure}\hfill
    \begin{subfigure}{0.333\textwidth}
        \centering
        \includegraphics[width=\linewidth]{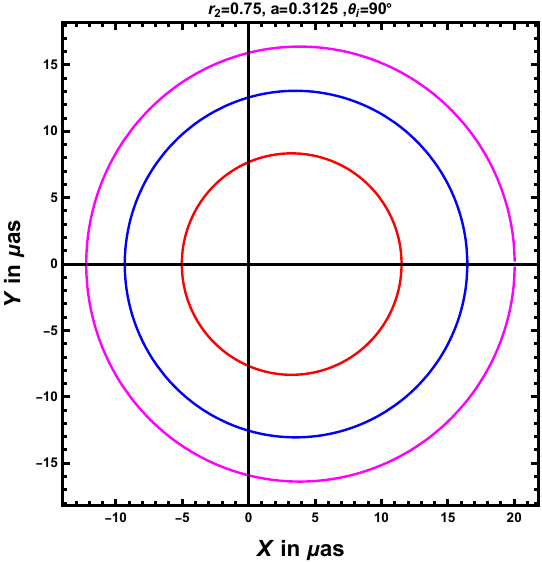}
        \caption{$r_2=0.75, \ a=0.3125$}
    \label{3e}  
    
    \end{subfigure}\hfill
    \begin{subfigure}{0.333\textwidth}
        \centering
        \includegraphics[width=\linewidth]{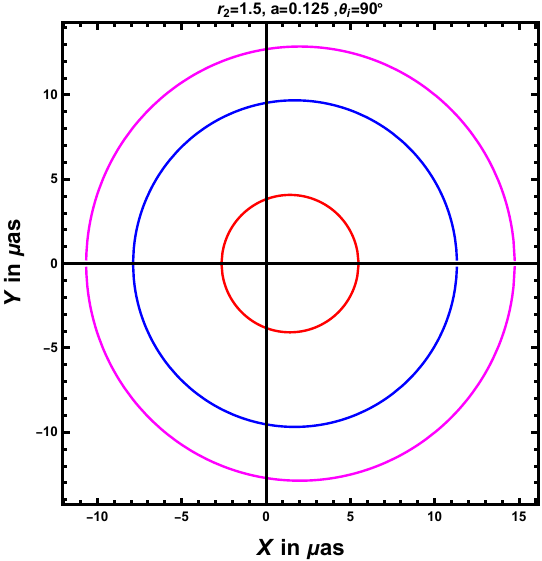}
        \caption{$r_2=1.5, \ a=0.125$}
    \label{3f}  
    
    \end{subfigure}

    \vspace{0.5cm}

    \begin{subfigure}{0.333\textwidth}
        \centering
        \includegraphics[width=\linewidth]{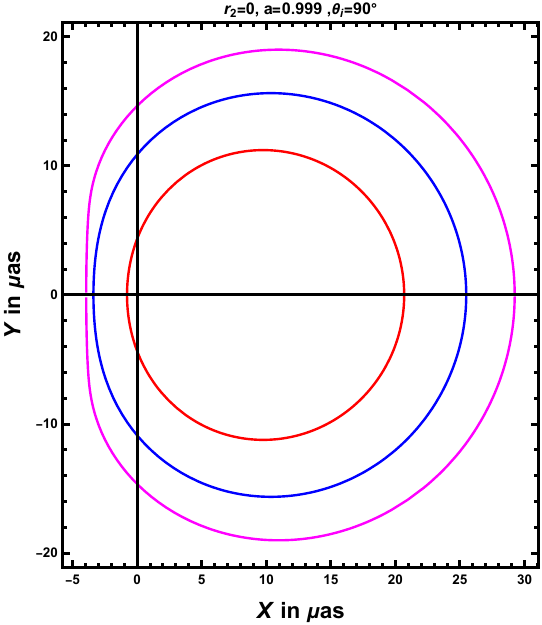}
        \caption{$r_2=0, \ a=0.999$}
    \label{3g}  
    \end{subfigure}\hfill
    \begin{subfigure}{0.333\textwidth}
        \centering
        \includegraphics[width=\linewidth]{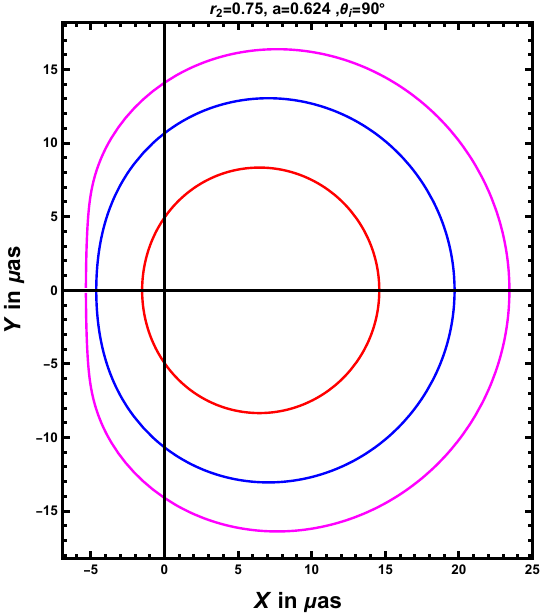}
        \caption{$r_2=0.75, \ a=0.624$}
    \label{3h}  
    \end{subfigure}\hfill
    \begin{subfigure}{0.333\textwidth}
        \centering
        \includegraphics[width=\linewidth]{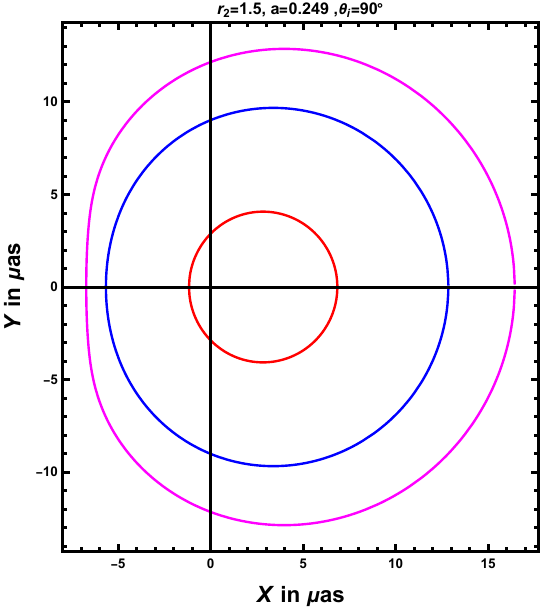}
        \caption{$r_2=1.5, \ a=0.249$}
 \label{3i}
    \end{subfigure}
   
  \vspace{0.1cm}
 \begin{subfigure}{1\textwidth}
        \centering
        \includegraphics[width=0.5\linewidth]{ legend1.pdf}
          
    \end{subfigure}
    
    \caption{\label{3}Variation of  shadow of Kerr Sen black hole in presence of plasma profile 1 at inclination angle $\theta_i=90^\circ$.}
\end{figure}

\item  From analysis of the \Cref{2a,2b,2c,2d,2e,2f,2g,2h,2i} we observe an increase in the plasma parameter $\alpha_1$ decreases the shadow size. This effect was also observed in \Cref{1}.

    \item     In addition to the contracting effect of the plasma parameter $\alpha_1$ on the shadow size, we also  observe in all the sub \Cref{2a,2b,2c,2d,2e,2f,2g,2h,2i} that, for sufficiently high value of $\alpha_1$ (observe red curve in each sub figure) the shadow becomes circular irrespective of the choice of $r_2$ and $a$.  
    \item   We  observe  that with increase in the spin $a$  of the black hole, the shape of the shadow increasingly deviates from circular shape in addition to the shift in geometric center of the shadow. However, the effect deviation from circularity decreases with increase in $\alpha_1$ and nearly negligible for sufficiently high values of $\alpha_1$. 
\end{itemize}

\Cref{3} shows the shadow plots for an observer in the equatorial plane.
\begin{itemize}
    \item  The contraction of shadow size due to the dilaton charge $r_2$ and the plasma parameter $\alpha_1$ as was observed in \Cref{1,2} is also observed in \Cref{3}.  
    \item The deviation from circularity of the shadow  and the shift in the geometric center with increase in the spin $a$ of the black hole  can be observed as we go vertically downwards along each column in \Cref{3} . However, comparing the deviation in circularity of the shadow observed in  \Cref{2,3,1} , we note that the shadow becomes increasingly non-circular when $a$ and $\theta_i$ are simultaneously increased.  
    \item Similar to \Cref{2} we note that for sufficiently high value of plasma parameter $\alpha_1$, the shadow becomes nearly circular irrespective of the value of $r_2$, $a$ and $\theta_i$.
\end{itemize}

\subsection{Variation of shadow in case of profile 2\label{p2}}
In this section we consider profile 2 given in \Cref{eqn:p2} which  was discussed in\cite{Perlick:2017fio}. The density associated with $\omega^2_P$ in profile 2 qualitatively  represents black hole accretion tori in the asymptotic limit\cite{Mosallanezhad:2013gre,Rees:1982pe,Komissarov:2006nz}. In profile 2 $g(\theta)\geq0$ which is consistent with the requirement $\omega^2_{P}\geq0$.   The maximum value of  the plasma parameter $\alpha_2$ is determined from the  \Cref{eqn:boundfinal}. For profile 2 we denote $F(r,\theta)$
by $F_2(r,\theta)$.  We plot variation of $F_2(r,\theta)$ with $r$ for $r_2=0,0.75\text{ and }1.5$.
As $\alpha_2$ can not exceed the minimum of $F_2(r,\theta)$ from condition \eqref{eqn:bounds2} thus the  \Cref{F2a,F2b,F2c} can be used to infer the bound on $\alpha_2$.
We make the following observations from \Cref{F2}:
\begin{itemize}
    \item Similar to the case of profile 1 we find that for a given $r_2$ and each case of $a$  and $\theta$  in \Cref{F2a,F2b,F2c} there exists a minimum.

    \item The minimum of $F_2(r,\theta)$ decreases with increase in $\theta$  (refer \Cref{F2a,F2b,F2c}).    
    \item    For a given $r_2\text{ and }a$ we observe that, $F_2(r,\theta)$ (refer \Cref{F2a,F2b,F2c}) is more sensitive to $\theta$  compared to $F_1(r,\theta)$ ( refer \Cref{Fr1a,Fr1b,F1c}).
    \item    $F_2(r,\theta)$ is sensitive to  $a$  as we go near the black hole when $\theta$ is small.
    \item The sensitivity of $F_2(r,\theta)$ with respect to $a$ decreases with increase in  $r_2$ ( refer  \Cref{F2a,F2b,F2c}).
    \item  The minimum value of $F_2(r,\pi/2)$ gives an estimate of the maximum value of plasma parameter, that is, $\alpha_{2max}$.

\begin{figure}[H]
    \centering
     \begin{subfigure}{0.5\textwidth}
        \centering
        \includegraphics[width=\linewidth]{ 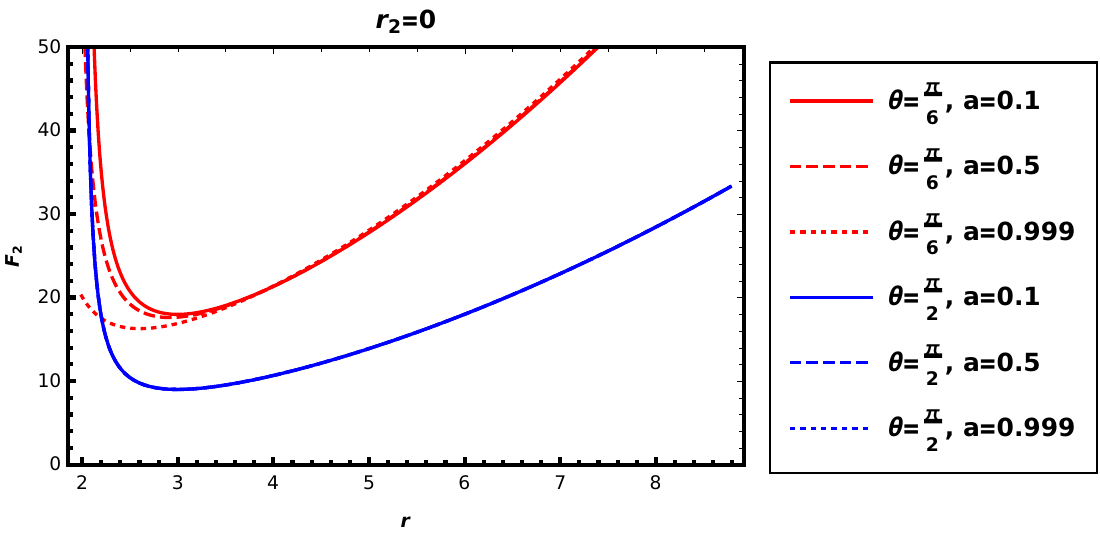}
        \caption{\label{F2a}$r_2=0$}
    \end{subfigure}\hspace{0.1cm}
    \begin{subfigure}{0.5\textwidth}
        \centering
        \includegraphics[width=\linewidth]{ 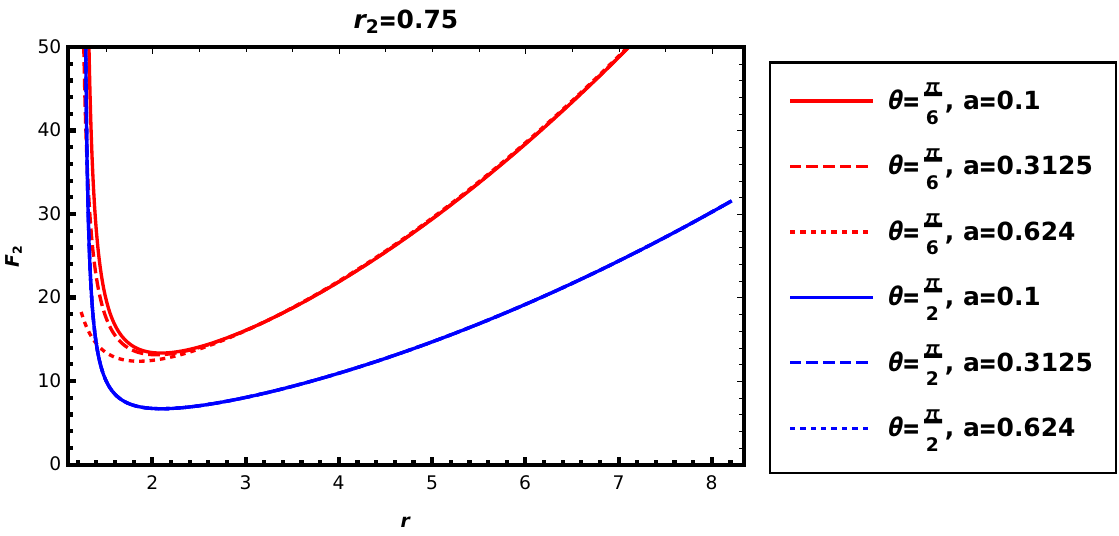}
        \caption{\label{F2b}$r_2=0.75$}
    \end{subfigure}\vspace{0.3cm}
    \begin{subfigure}{0.5\textwidth}
        \centering
        \includegraphics[width=\linewidth]{ 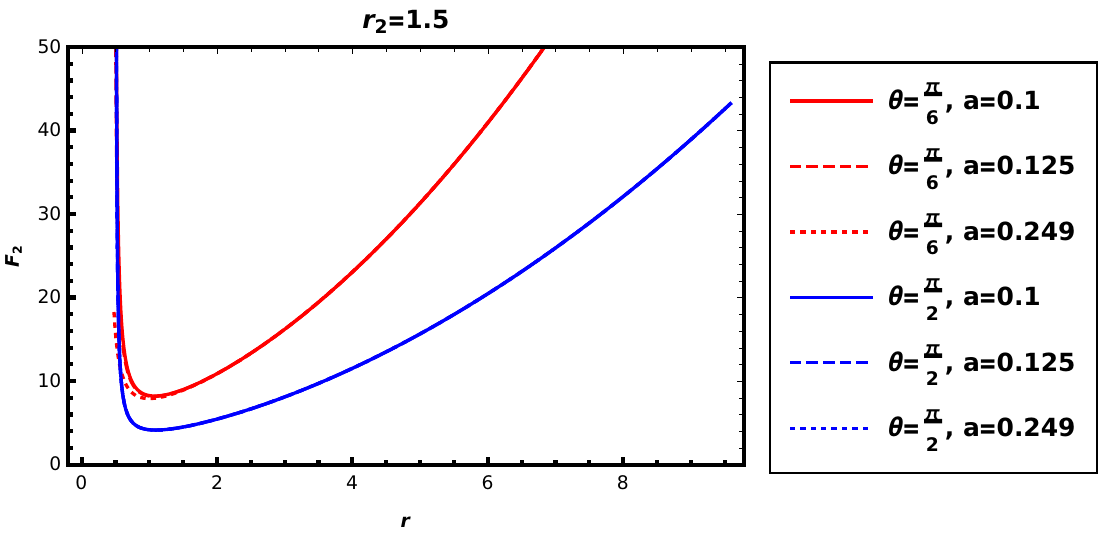}
        \caption{\label{F2c}$r_2=1.5$}
    \end{subfigure}
    \caption{\label{F2}Variation of $F_2(r,\theta)$ with r (in units of $M$)at $a=0.1, 0.5(1-\frac{r_2}{2})\text{ and }
    0.999-\frac{r_2}{2}$  for $\theta=\frac{\pi}{6}\text{ (Red) and }\frac{\pi}{2}  $ (Blue) for different $r_2$ }
\label{Fig5}
\end{figure}

    \item From \Cref{F2a,F2b,F2c} we find for $r_2=0,0.75\text{ and }1.5$ the corresponding $\alpha_{2max}\approx9, 7\text{ and }4.12$, respectively.
   % \item If one observes  a black hole with a dense plasma environment then such environments are less likely to host Kerr Sen black hole with a large dilaton charge.   
\end{itemize} 
Furthermore, the photon region for profile 2 can be determined from \Cref{eqn:41}. Using $g(\theta)$ from  \Cref{eqn:p2} in \Cref{eqn:41}  we obtain
 
 \begin{gather}
     \label{p2region}
     \frac{ \mathcal{Q}(r_p) a^2 \sin^2{\theta}-(-a\eta(r_p)+a^2\sin^2\theta)^2}{a^2 \sin^2{\theta}}\geq  \alpha_2 (1+2 \sin^2\theta)
 \end{gather}
In the above equation $\mathcal{Q}(r_p)\text{ and }\eta(r_p)$ are given by \Cref{equation of Q,equation of eta}  with $f(r)=0$. The region in which the inequality \eqref{p2region} is satisfied is the photon region for profile 2.

\begin{figure}[H]
    \centering
    \begin{subfigure}{0.333\textwidth}
        \centering
        \includegraphics[width=\linewidth]{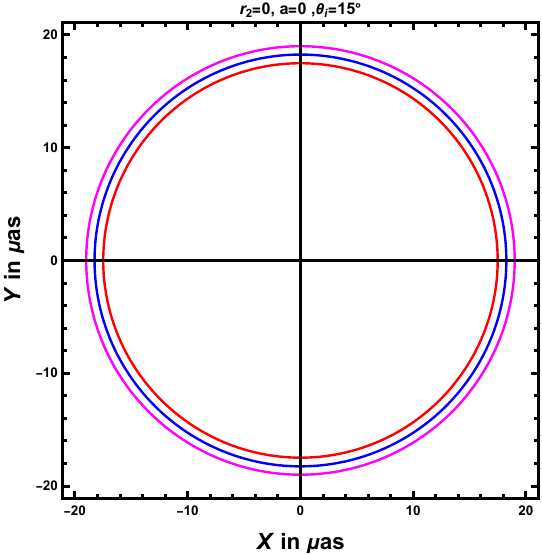}
        \caption{\label{t1a}$r_2=0, \ a=0  $}

    \end{subfigure}\hfill
    \begin{subfigure}{0.333\textwidth}
        \centering
        \includegraphics[width=\linewidth]{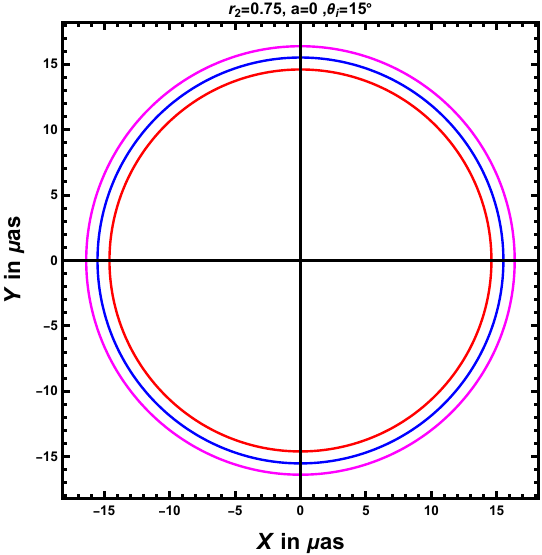}
\caption{$r_2=0.75, \ a=0   $\label{t1b}}

    \end{subfigure}\hfill
    \begin{subfigure}{0.333\textwidth}
        \centering
        \includegraphics[width=\linewidth]{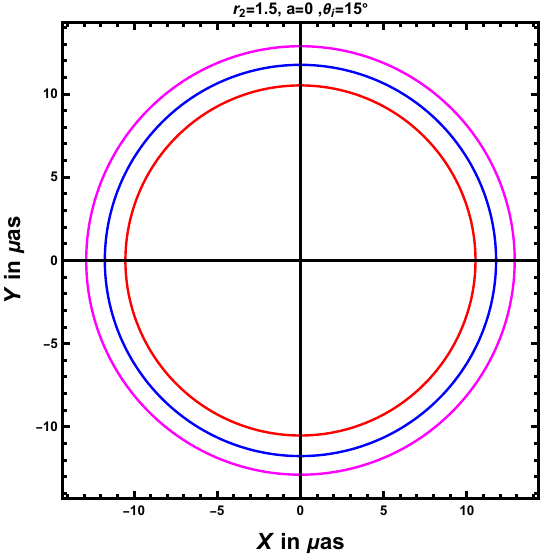}
        \caption{$r_2=1.5, \ a=0   $\label{t1c}}

    \end{subfigure}

    \vspace{0.5cm}

    \begin{subfigure}{0.333\textwidth}
        \centering
        \includegraphics[width=\linewidth]{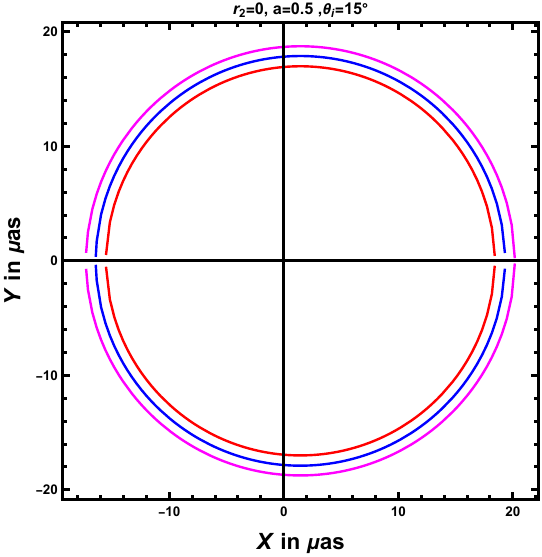}
        \caption{$r_2=0, \ a=0.5   $\label{t1d}}

    \end{subfigure}\hfill
    \begin{subfigure}{0.333\textwidth}
        \centering
        \includegraphics[width=\linewidth]{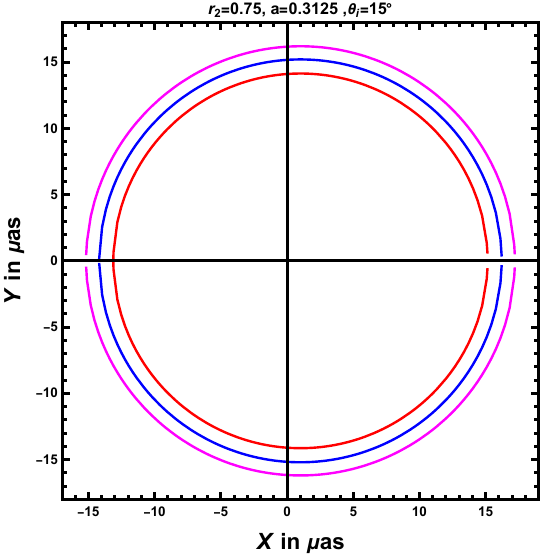}
\caption{\label{t1e}$r_2=0.75, \ a=0.3125   $}
  
    \end{subfigure}\hfill
    \begin{subfigure}{0.333\textwidth}
        \centering
        \includegraphics[width=\linewidth]{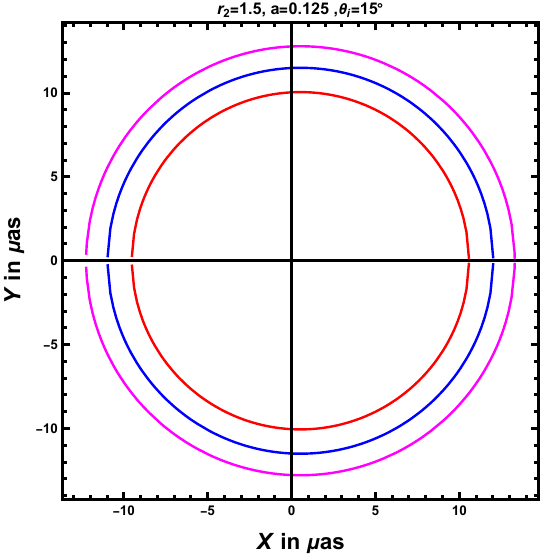}
        \caption{\label{t1f}$r_2=1.5, \ a=0.125   $}
    \end{subfigure}

    \vspace{0.5cm}

    \begin{subfigure}{0.333\textwidth}
        \centering
        \includegraphics[width=\linewidth]{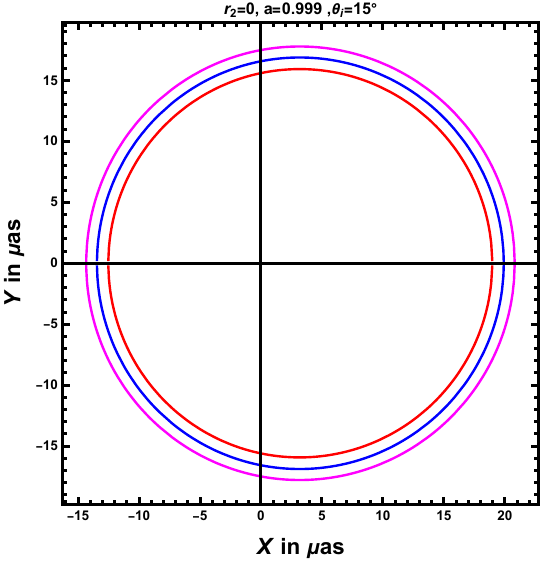}
        \caption{$r_2=0, \ a=0.999   $\label{t1g}}
    \end{subfigure}\hfill
    \begin{subfigure}{0.333\textwidth}
        \centering
        \includegraphics[width=\linewidth]{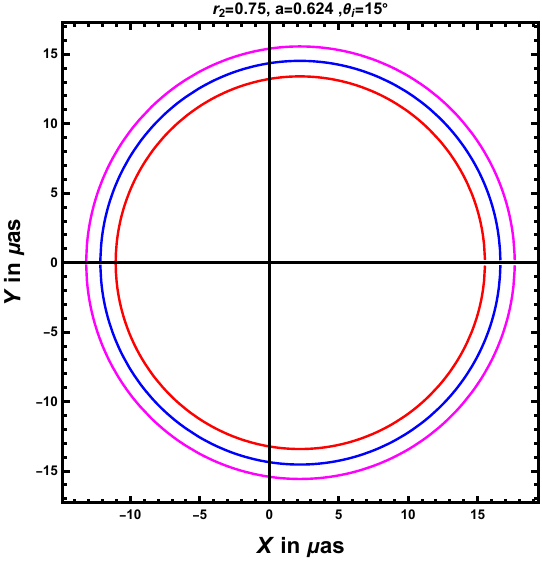}
        \caption{$r_2=0.75, \ a=0.62499   $\label{t1h}}
    \end{subfigure}\hfill
    \begin{subfigure}{0.333\textwidth}
        \centering
        \includegraphics[width=\linewidth]{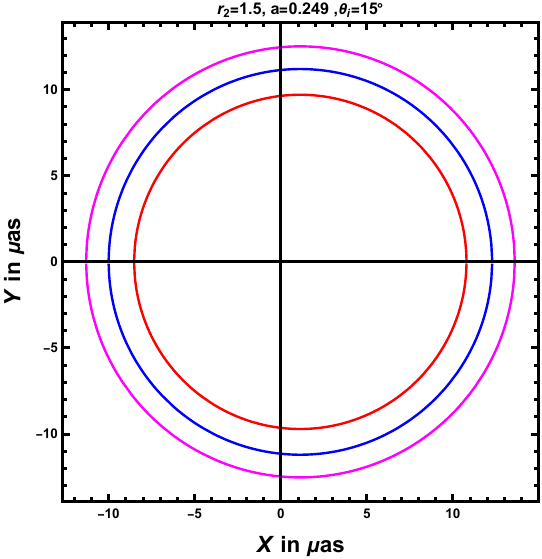}
        \caption{$r_2=1.5, \ a=0.24   $\label{t1i}}
    \end{subfigure}
     \vspace{0.1cm}
 \begin{subfigure}{1\textwidth}
        \centering
        \includegraphics[width=0.5\linewidth]{ 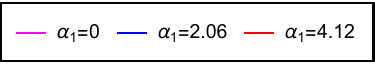}
          
    \end{subfigure}

    \caption{\label{t1}Variation of  shadow of Kerr Sen black hole in presence of plasma profile 2 at inclination angle $\theta_i=15^\circ$.}
\end{figure}

\begin{figure}[H]
    \centering
    \begin{subfigure}{0.333\textwidth}
        \centering
        \includegraphics[width=\linewidth]{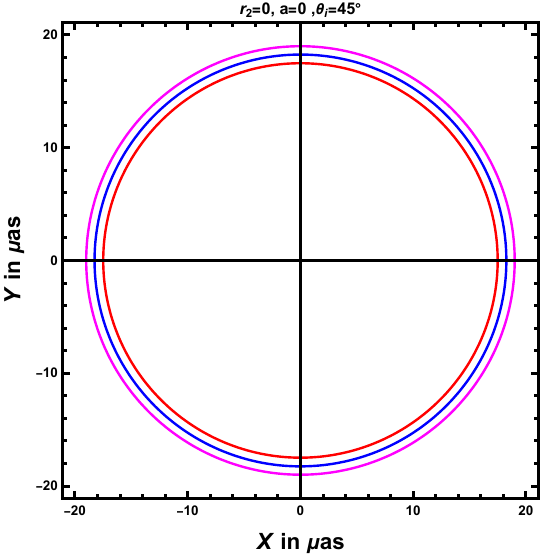}
        \caption{$r_2=0, \ a=0   $\label{t2a}}
    \end{subfigure}\hfill
    \begin{subfigure}{0.333\textwidth}
        \centering
        \includegraphics[width=\linewidth]{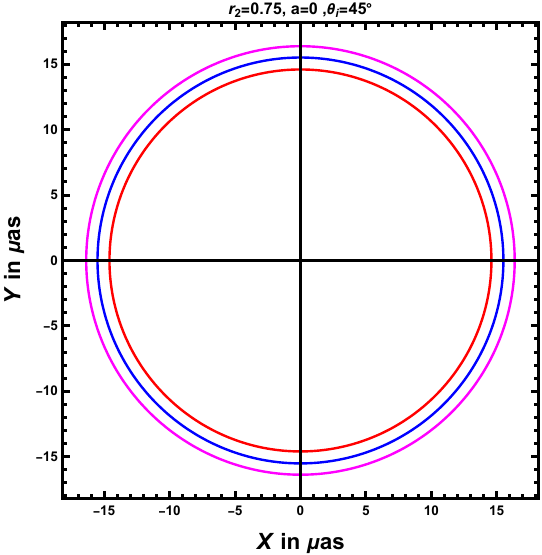}
        \caption{$r_2=0.75, \ a=0   $\label{t2b}}
    \end{subfigure}\hfill
    \begin{subfigure}{0.333\textwidth}
        \centering
        \includegraphics[width=\linewidth]{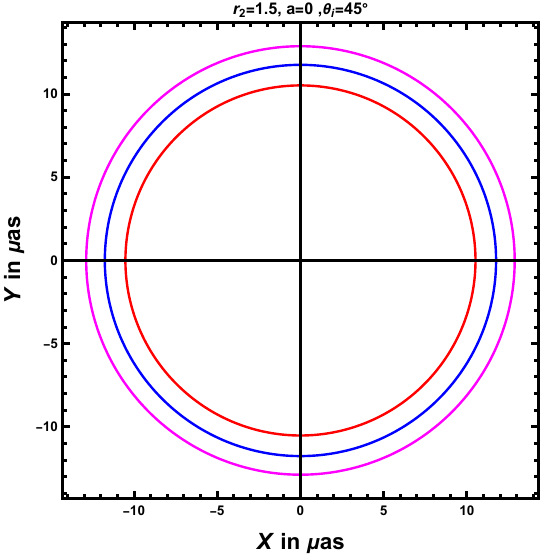}
        \caption{\label{t2c}$r_2=1.5, \ a=0   $}
    \end{subfigure}

    \vspace{0.5cm}

    \begin{subfigure}{0.333\textwidth}
        \centering
        \includegraphics[width=\linewidth]{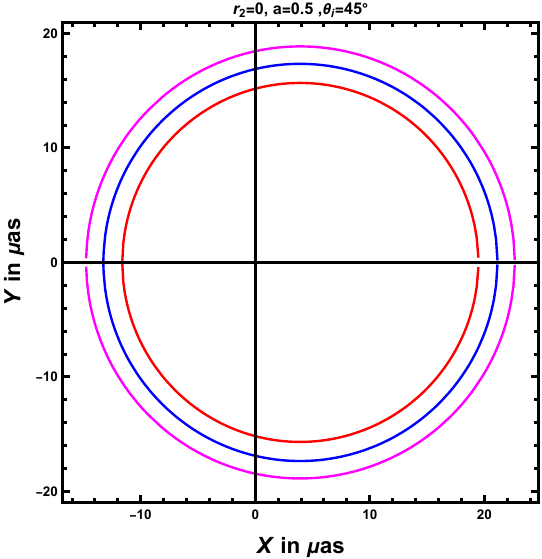}
        \caption{\label{t2d}$r_2=0, \ a=0.5   $}
    \end{subfigure}\hfill
    \begin{subfigure}{0.333\textwidth}
        \centering
        \includegraphics[width=\linewidth]{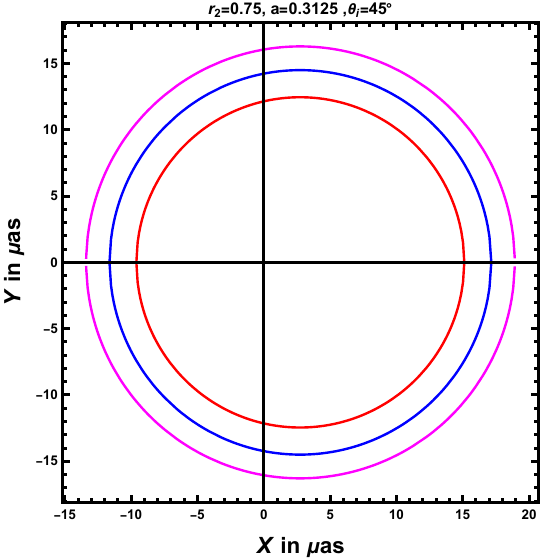}
        \caption{\label{t2e}$r_2=0.75, \ a=0.3125   $}
    \end{subfigure}\hfill
    \begin{subfigure}{0.333\textwidth}
        \centering
        \includegraphics[width=\linewidth]{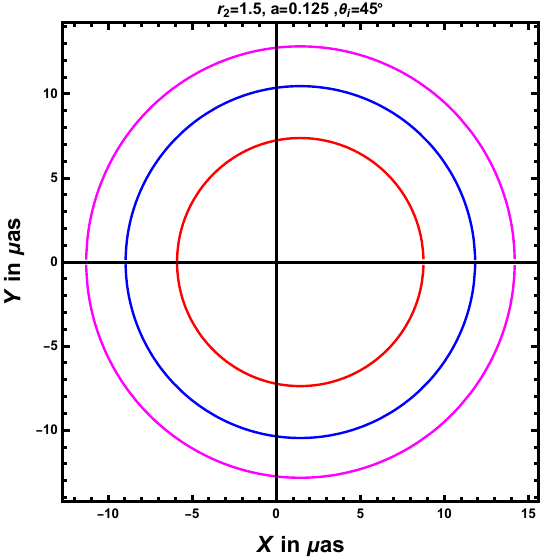}
        \caption{\label{t2f}$r_2=1.5, \ a=0.125   $}
    \end{subfigure}

    \vspace{0.5cm}

    \begin{subfigure}{0.333\textwidth}
        \centering
        \includegraphics[width=\linewidth]{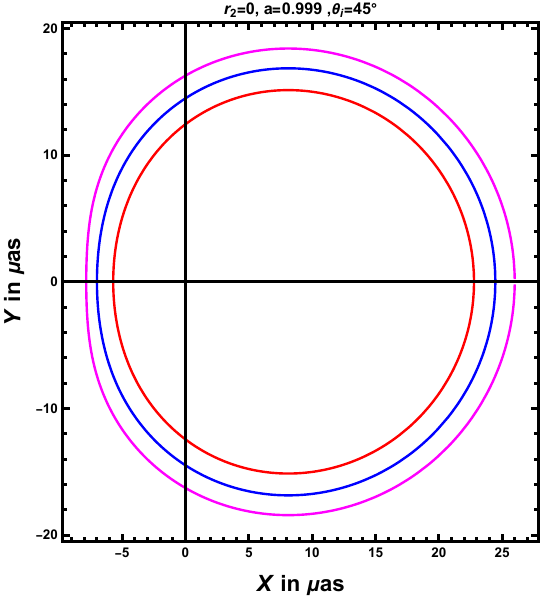}
        \caption{\label{t2g}$r_2=0, \ a=0.999   $}
    \end{subfigure}\hfill
    \begin{subfigure}{0.333\textwidth}
        \centering
        \includegraphics[width=\linewidth]{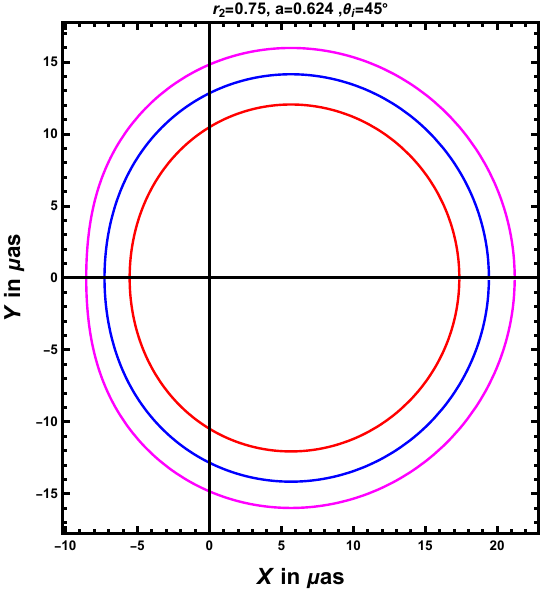}
        \caption{\label{t2h}$r_2=0.75, \ a=0.62499   $}
    \end{subfigure}\hfill
    \begin{subfigure}{0.333\textwidth}
        \centering
        \includegraphics[width=\linewidth]{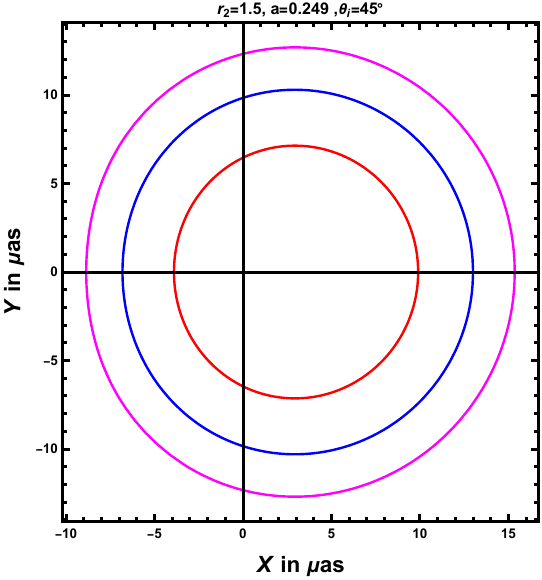}
        \caption{\label{t2i}$r_2=1.5, \ a=0.249   $}
    \end{subfigure}
 \vspace{0.1cm}
 \begin{subfigure}{1\textwidth}
        \centering
        \includegraphics[width=0.5\linewidth]{ legend2.pdf}
          
    \end{subfigure}
    \caption{\label{t2}Variation of  shadow of Kerr Sen black hole in presence of plasma profile 2 at inclination angle $\theta_i=45^\circ$.}
\end{figure}

\begin{figure}[H]
    \centering
    \begin{subfigure}{0.333\textwidth}
        \centering
        \includegraphics[width=\linewidth]{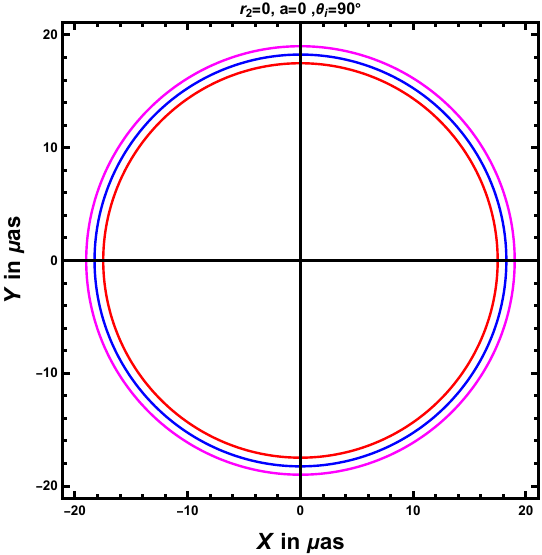}
        \caption{\label{t3a}$r_2=0, \ a=0   $}
    \end{subfigure}\hfill
    \begin{subfigure}{0.333\textwidth}
        \centering
        \includegraphics[width=\linewidth]{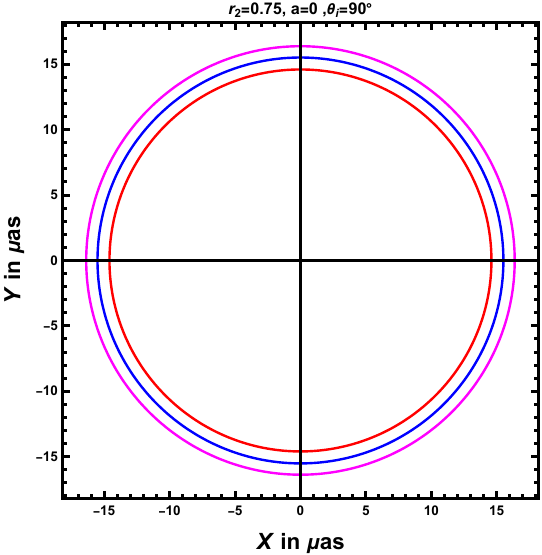}
        \caption{$r_2=0.75, \ a=0   $\label{t3b}}
    \end{subfigure}\hfill
    \begin{subfigure}{0.333\textwidth}
        \centering
        \includegraphics[width=\linewidth]{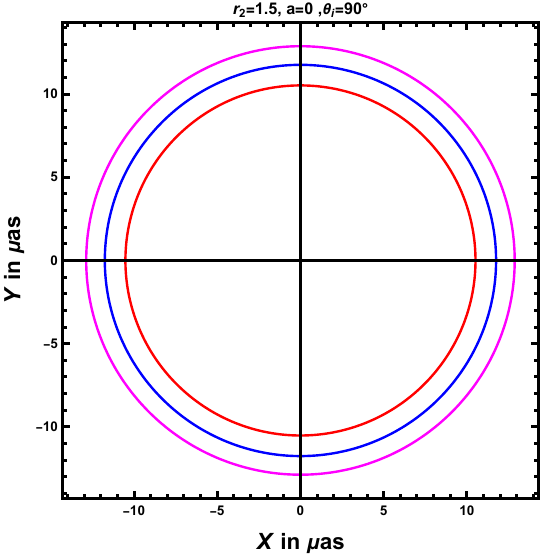}
        \caption{$r_2=1.5, \ a=0   $\label{t3c}}
    \end{subfigure}

    \vspace{0.5cm}

    \begin{subfigure}{0.333\textwidth}
        \centering
        \includegraphics[width=\linewidth]{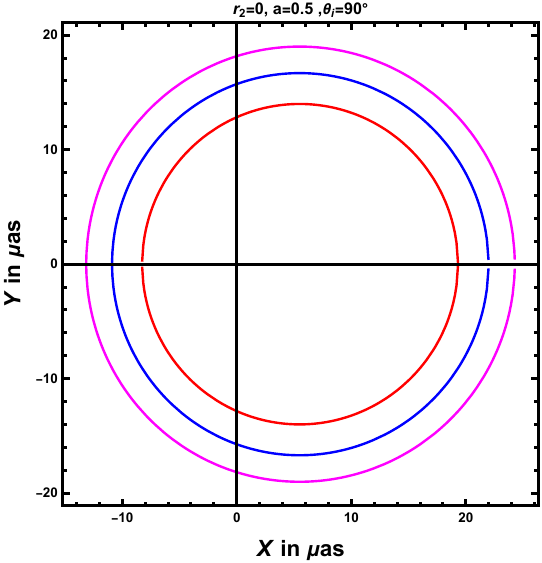}
        \caption{$r_2=0, \ a=0.5   $\label{t3d}}
    \end{subfigure}\hfill
    \begin{subfigure}{0.333\textwidth}
        \centering
        \includegraphics[width=\linewidth]{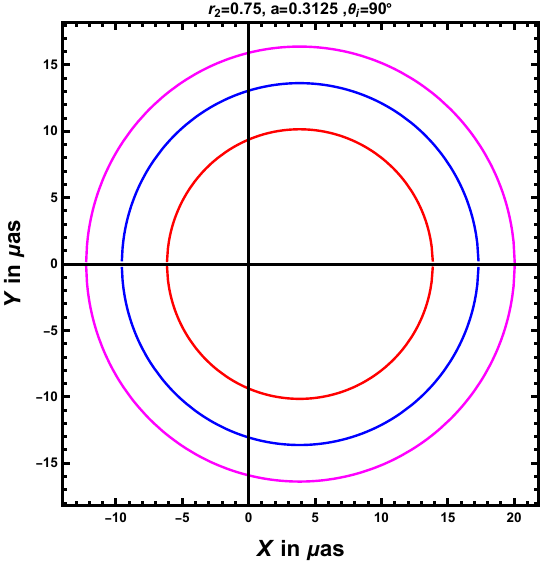}
        \caption{$r_2=0.75, \ a=0.3125   $\label{t3e}}
    \end{subfigure}\hfill
    \begin{subfigure}{0.333\textwidth}
        \centering
        \includegraphics[width=\linewidth]{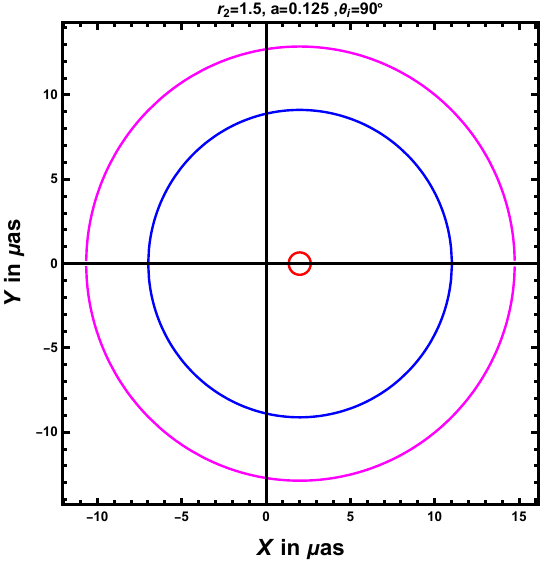}
        \caption{$r_2=1.5, \ a=0.125   $\label{t3f}}
    \end{subfigure}
    \vspace{0.5cm}

    \begin{subfigure}{0.333\textwidth}
        \centering
        \includegraphics[width=\linewidth]{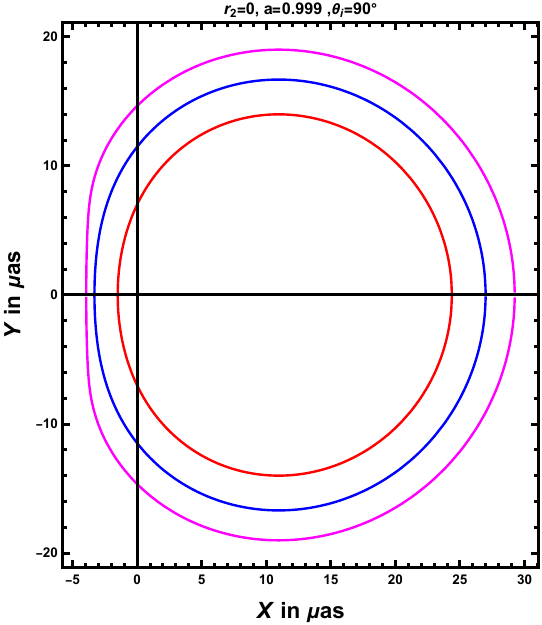}
        \caption{$r_2=0, \ a=0.999   $\label{t3g}}
    \end{subfigure}\hfill
    \begin{subfigure}{0.333\textwidth}
        \centering
        \includegraphics[width=\linewidth]{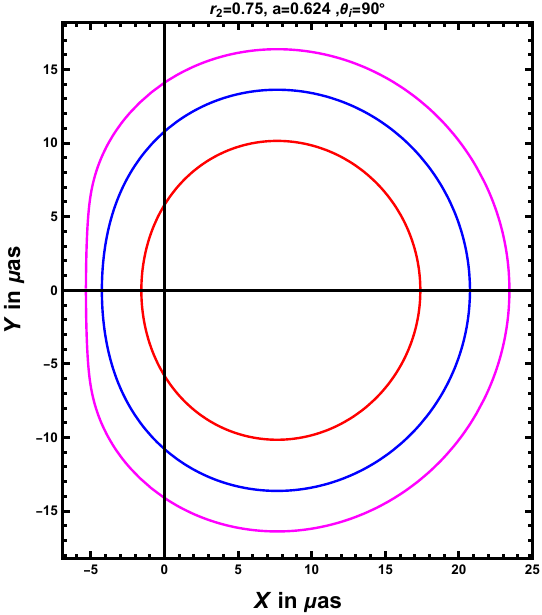}
        \caption{\label{t3h}$r_2=0.75, \ a=0.62499   $}
    \end{subfigure}\hfill
    \begin{subfigure}{0.333\textwidth}
        \centering
        \includegraphics[width=\linewidth]{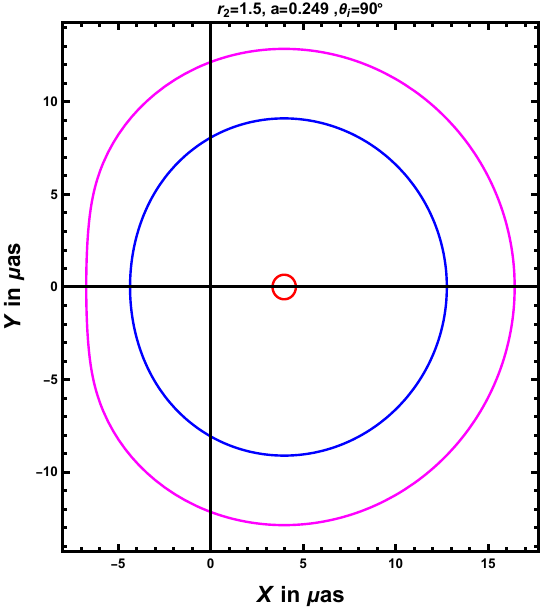}
        \caption{\label{t3i}$r_2=1.5, \ a=0.24   $}
    \end{subfigure}
     \vspace{0.1cm}
 \begin{subfigure}{1\textwidth}
        \centering
        \includegraphics[width=0.5\linewidth]{ legend2.pdf}
          
    \end{subfigure}
    \caption{\label{t3}Variation of  shadow of Kerr Sen black hole in presence of plasma profile 2 at inclination angle $\theta_i=90^\circ$.}
\end{figure}

We now proceed to discuss the variation of shadow of Kerr Sen black hole considering plasma profile 2 given by  \Cref{eqn:p2} .  \Cref{t1,t2,t3} show the variation of shadow with $r_2,a \text{ and }\alpha_2$ for $\theta_i=15^\circ, 45^\circ\text{ and }90^\circ$, respectively. The arrangement  of sub-figures in \Cref{t1,t2,t3} is same as in \Cref{1,2,3} (which is described in \cref{p1}). In each case, we can study the variation of  shadow with dilaton $r_2$ by going from left to right  along each row. By going vertically downwards along each column one can study effect of $a$ on the shadow. By comparing shadow plots for $\alpha_2=0$ (Magenta curve), $\alpha_2=2.06$ (Blue curve) and $\alpha_2=4.12$ (Red curve) in each subfigure we can study the effect of $\alpha_2$ on the shadow for given $r_2,a\text{ and }\theta_i$. 

We first discuss the variation of shadow as shown in \Cref{t1}:
\begin{itemize}
    \item The effect of dilaton charge  $r_2$ on the shadow can be observed from \Cref{t1a,t1b,t1c}. The increase in $r_2$ decreases the shadow size  as discussed in  \cref{p1}. The increase in  contraction due to increase of $r_2$ is observed irrespective of the choice of $a\text{ and }\alpha_2$. This was also observed in \Cref{1},\textbf{ }however, the contraction of shadow due to $r_2$ in case plasma profile 2 (\Cref{t1}) is weaker than what was observed in case of profile 1 (\Cref{1})  at $\theta_i=15^\circ$. 
    \item      \Cref{t1a}  shows the variation of shadow purely due to plasma parameter  $\alpha_2$. We observe that, with increase in $\alpha_2$ there is decrease in shadow size as was observed in case of profile 1 (\Cref{r1a}). Furthermore, the decrease in shadow size with increase in $\alpha_2$ is observed  irrespective of our choice of $r_2\text{ and }a$ (\Cref{t1a,t1b,t1c,t1d,t1e,t1f,t1g,t1h,t1i}). 
    \item By moving vertically downwards along each column in \Cref{t1} we observe that, as we increase the spin $a$ for a given $(r_2,\alpha_2,\theta_i)$ the geometric center of shadow shifts away from the geometric centre in case $a=0$.
\end{itemize}

We next discuss the variation of shadow as shown in \Cref{t2} for $\theta_i=45^\circ$. We observe the following:
\begin{itemize}
    \item The previously observed effect  of decrease in  shadow size (refer \Cref{t1a,t1b,t1c,t1d,t1e,t1f,t1g,t1h,t1i}) with an increase in $r_2$  and also with $\alpha_2$  can be observed in (refer \Cref{t2a,t2b,t2c,t2d,t2e,t2f,t2g,t2h,t2i}).   

    \item Moving vertically downwards along each column in \Cref{t2} we observe that, as we increase the spin $a$ for a given $(r_2,\alpha_2,\theta_i)$ the geometric centre of shadow shifts away from the geometric centre in case of $a=0$ and also there is deviation in circularity which increase with increase in $a$. But deviation in circularity in the shadow was not observed in  \Cref{t1g,t1h,t1i} where $a$ was also increased. Thus, increase in both $\theta_i$ and $a$  leads to the distortion in the shape of the shadow.  
   
\item  With increase in  $\alpha_2$ the deviation in circularity due to the spin $a$  and $\theta_i$ decreases, but the shift in the geometric center remains (refer \Cref{t2g,t2h,t2i}). Thus, presence of plasma makes the shadow more circular irrespective of large $a$ and $\theta_i$. 
\end{itemize}


The variation of shadow as observed by an equatorial observer is shown in  \Cref{t3}. 
We observe the following from \Cref{t3}:
\begin{itemize}
\item The generic effect of decrease in  shadow size with an increase in $r_2$  and also with $\alpha_2$  can be observed in \Cref{t3}.  
\item From  comparison of  \Cref{t3c,t3f,t3i}  with  \Cref{t2c,t2f,t2i} we observe  that,  for higher $r_2$ the effect of contraction of shadow size due to increase in $\alpha_2$ is more intense, for \Cref{t3} (i.e, at higher inclination angle).

\item The combined effect of high $\theta_i$ and  $a$  which causes increase in the deviation in circularity of shadow can be seen   as we go vertically downwards  along  each column in  \Cref{t3}.

    \item  As was observed in \Cref{t2} with increase in  $\alpha_2$ the deviation in circularity due to the spin $a$ decreases, but the shift in the geometric centre persists (refer \Cref{t3g,t3h,t3i}).
\end{itemize}

\subsection{Variation of shadow outline in case of homogeneous plasma \label{p3}}
For completeness, we now consider  the case of homogeneous plasma (\Cref{eqn:p3})  in which the plasma density is constant. \Cref{h1},  \Cref{h2} and \Cref{h3} plots the variation  of shadow outline with $\alpha_3$ for angle of observation $\theta_i=15^\circ, 45^\circ\text{ and }90^\circ$ respectively.  
For a homogeneous plasma, \Cref{eqn:boundfinal} becomes
\begin{gather}
    \label{homogeneous bound equation}
    \alpha_{3max}=\left(1-\frac{2\ r}{\rho}\right)^{-1}\Bigg|_{min}
\end{gather}
The RHS of the above equation is non negative for $\rho>2r$ (which is true outside the horizon). Furthermore, as $\rho$ increases, the RHS decreases asymptotically approaching 1. Thus, for light rays to propagate to an observer at any finite distance outside the horizon in a homogeneous plasma, $\alpha_{3max}\approx1$. 
In homogeneous plasma both unstable and stable photon orbits may exist which has been studied in \cite{Bisnovatyi-Kogan:2010flt,Kulsrud:1991jt}. The unstable photon orbits are one of the main reasons behind shadow formation as light rays when perturbed in these orbits fall into the black hole. Light rays in stable photon orbits on the other hand, may continue to stay on the orbit after being perturbed and thus do not contribute in the formation of shadow. 

The arrangement style of subfigures in \Cref{h1,h2,h3} is same as that in \Cref{1,2,3} and \Cref{t1,t2,t3} (refer \cref{p1,p2} for details). In contrary to previous profiles of non homogeneous plasma,  we observe that in case of homogeneous plasma  the  shadow size expands with increase in $\alpha_3$.        

From \Cref{h1,h2,h3} we can observe the following interesting features:
\begin{itemize}
    \item From all subfigures in \Cref{h1}, \Cref{h2,h3}, it is clear that with increase in $\alpha_3$  the size of the shadow increases which is opposite to the case of profile 1 (refer \cref{p1}) and profile 2 (refer \cref{p2}).   
    \item   Unlike profile 1 and profile 2, we observe from \Cref{h2,h3} that, even when we  increase $\alpha_3$ the shadow continues to be non-circular when $a$ and $\theta_i$ are large. 
    
\begin{figure}[H]
    \centering
    \begin{subfigure}{0.333\textwidth}
        \centering
        \includegraphics[width=\linewidth]{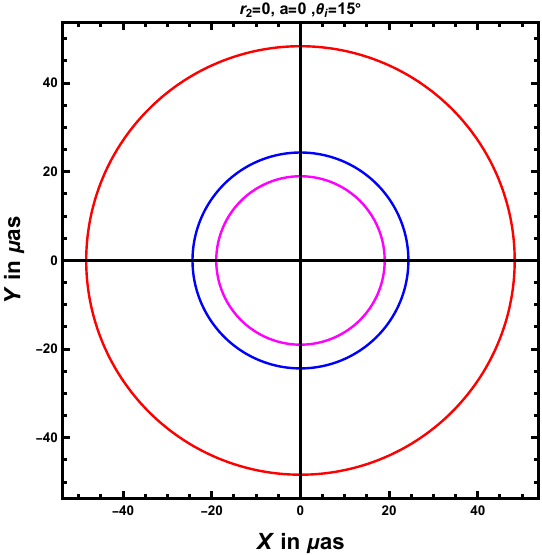}
        \caption{\label{h1a}$r_2=0, \ a=0   $}
    \end{subfigure}\hfill
    \begin{subfigure}{0.333\textwidth}
        \centering
        \includegraphics[width=\linewidth]{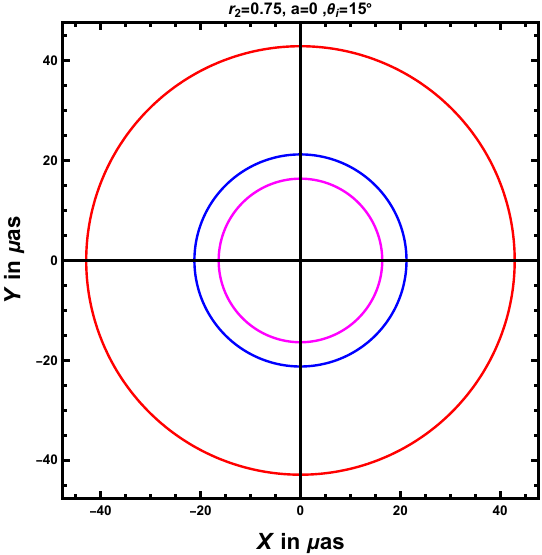}
        \caption{$r_2=0.75, \ a=0   $\label{h1b}}
    \end{subfigure}\hfill
    \begin{subfigure}{0.333\textwidth}
        \centering
        \includegraphics[width=\linewidth]{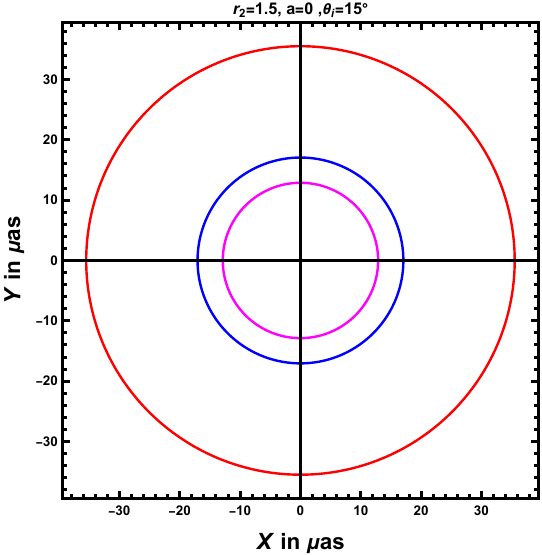}
        \caption{$r_2=1.5, \ a=0   $\label{h1c}}
    \end{subfigure}

    \vspace{0.5cm}

    \begin{subfigure}{0.333\textwidth}
        \centering
        \includegraphics[width=\linewidth]{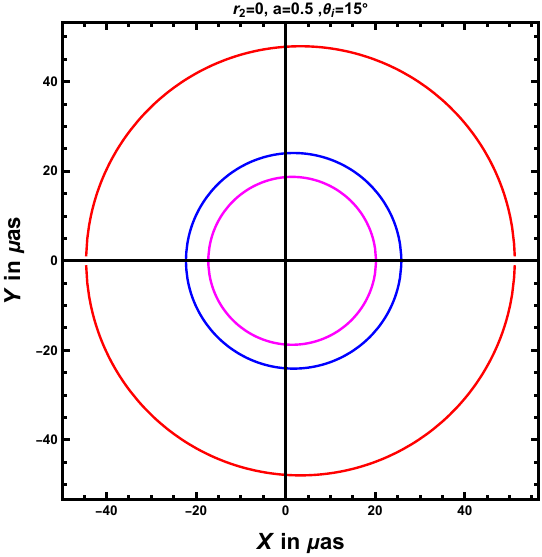}
        \caption{$r_2=0, \ a=0.5   $\label{h1d}}
    \end{subfigure}\hfill
    \begin{subfigure}{0.333\textwidth}
        \centering
        \includegraphics[width=\linewidth]{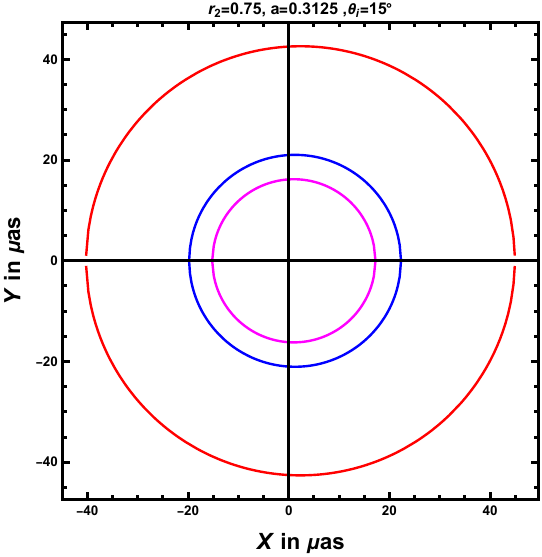}
        \caption{$r_2=0.75, \ a=0.3125   $\label{h1e}}
    \end{subfigure}\hfill
    \begin{subfigure}{0.333\textwidth}
        \centering
        \includegraphics[width=\linewidth]{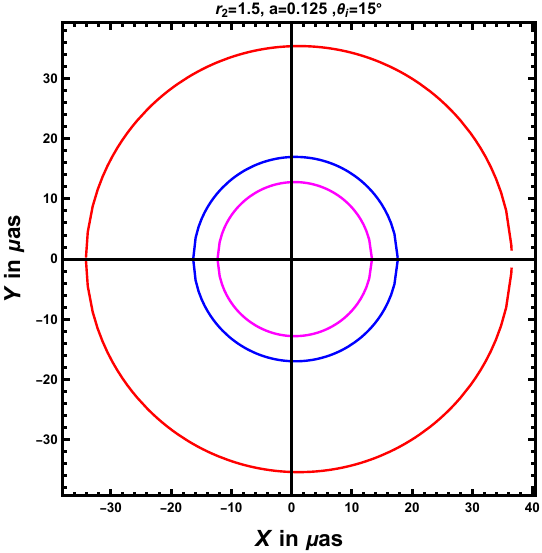}
        \caption{$r_2=1.5, \ a=0.125   $\label{h1f}}
    \end{subfigure}

    \vspace{0.5cm}

    \begin{subfigure}{0.333\textwidth}
        \centering
        \includegraphics[width=\linewidth]{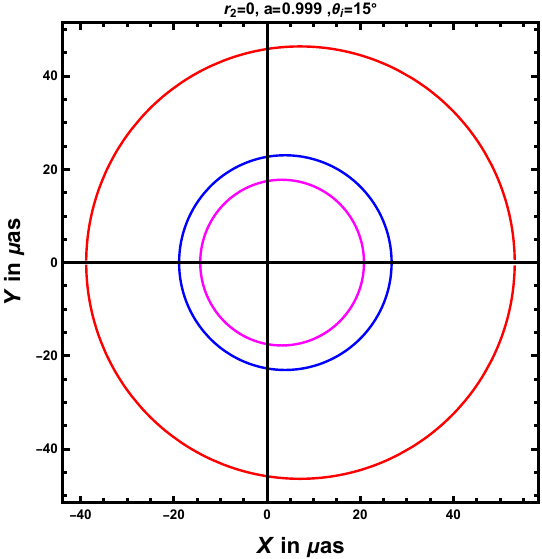}
        \caption{$r_2=0, \ a=0.999   $\label{h1g}}
    \end{subfigure}\hfill
    \begin{subfigure}{0.333\textwidth}
        \centering
        \includegraphics[width=\linewidth]{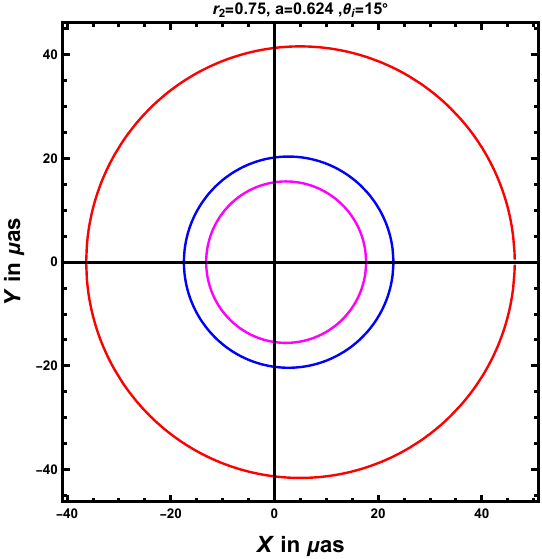}
        \caption{$r_2=0.75, \ a=0.624   $\label{h1h}}
    \end{subfigure}\hfill
    \begin{subfigure}{0.333\textwidth}
        \centering
        \includegraphics[width=\linewidth]{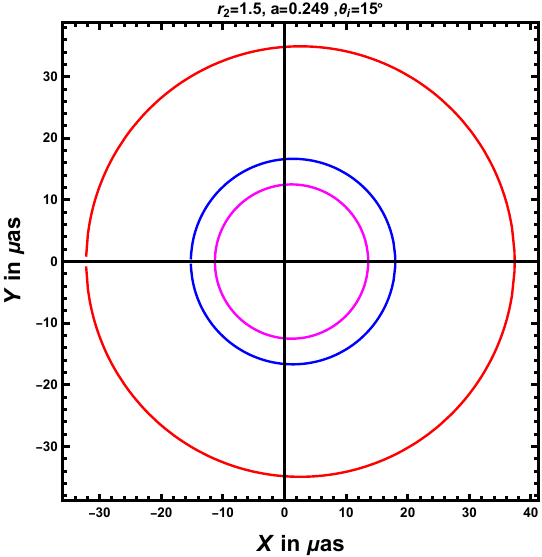}
        \caption{$r_2=1.5,  \ a=0.249   $\label{h1i}}
    \end{subfigure}
\vspace{0.1cm}
 \begin{subfigure}{1\textwidth}
        \centering
        \includegraphics[width=0.5\linewidth]{ 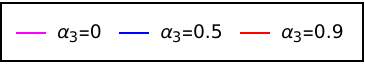}
          
    \end{subfigure}
    \caption{\label{h1}Variation of  shadow of Kerr Sen black hole in presence of homogeneous plasma at $\theta_i=15^\circ$}
\end{figure}

\begin{figure}[H]
    \centering
    \begin{subfigure}{0.333\textwidth}
        \centering
        \includegraphics[width=\linewidth]{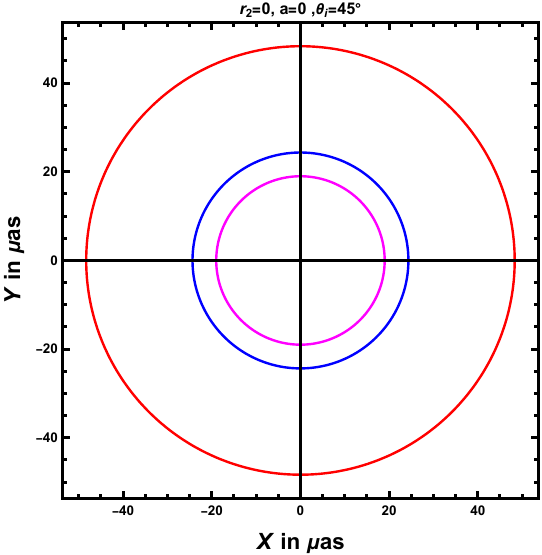}
        \caption{$r_2=0, \ a=0   $\label{h2a}}
    \end{subfigure}\hfill
    \begin{subfigure}{0.333\textwidth}
        \centering
        \includegraphics[width=\linewidth]{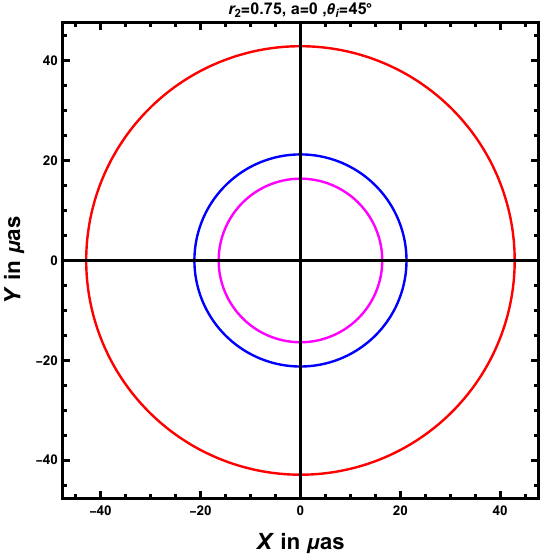}
        \caption{$r_2=0.75, \ a=0   $\label{h2b}}
    \end{subfigure}\hfill
    \begin{subfigure}{0.333\textwidth}
        \centering
        \includegraphics[width=\linewidth]{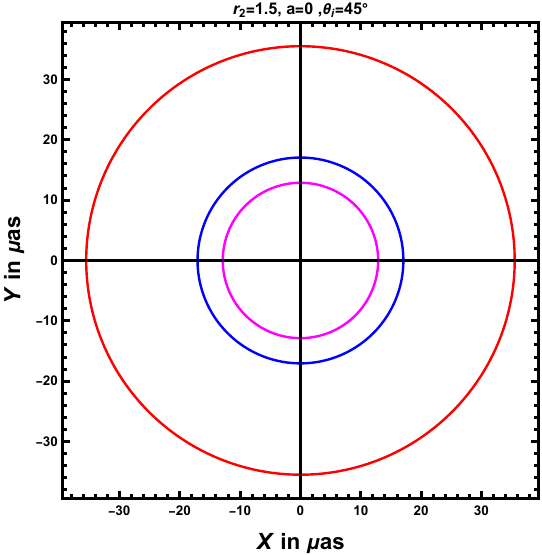}
        \caption{$r_2=1.5, \ a=0   $\label{h2c}}
    \end{subfigure}

    \vspace{0.5cm}

    \begin{subfigure}{0.333\textwidth}
        \centering
        \includegraphics[width=\linewidth]{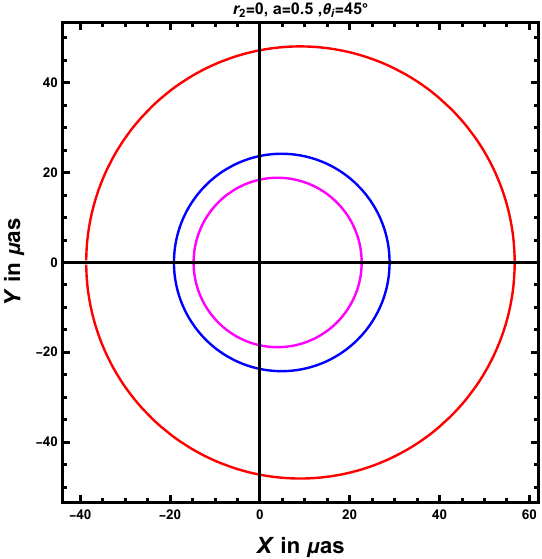}
        \caption{$r_2=1.5, \ a=0.5   $\label{h2d}}
    \end{subfigure}\hfill
    \begin{subfigure}{0.333\textwidth}
        \centering
        \includegraphics[width=\linewidth]{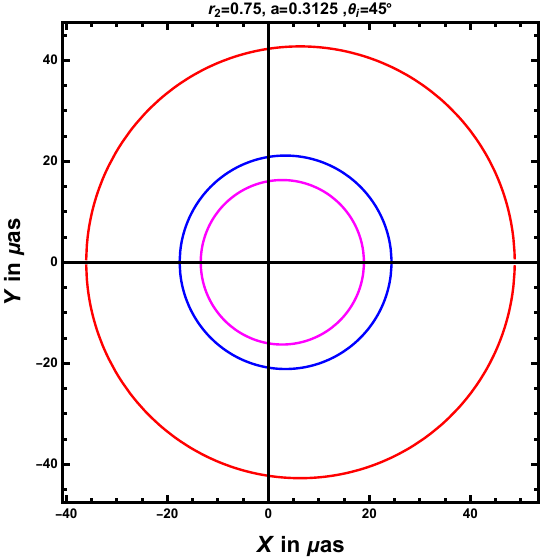}
        \caption{$r_2=0.75, \ a=0.3125   $\label{h2e}}
    \end{subfigure}\hfill
    \begin{subfigure}{0.333\textwidth}
        \centering
        \includegraphics[width=\linewidth]{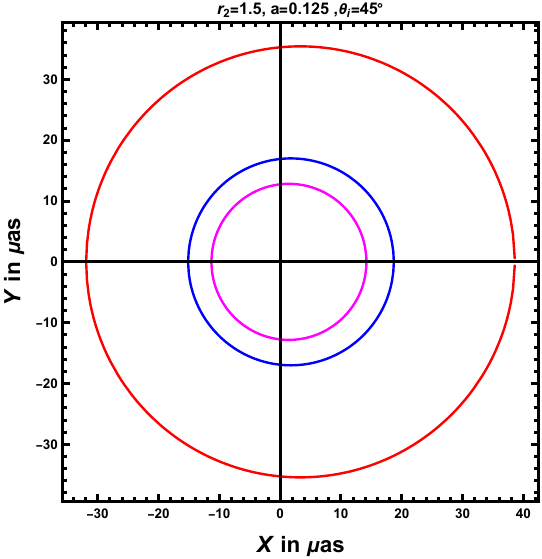}
        \caption{$r_2=1.5, \ a=0.125   $\label{h2f}}
    \end{subfigure}

    \vspace{0.5cm}

    \begin{subfigure}{0.333\textwidth}
        \centering
        \includegraphics[width=\linewidth]{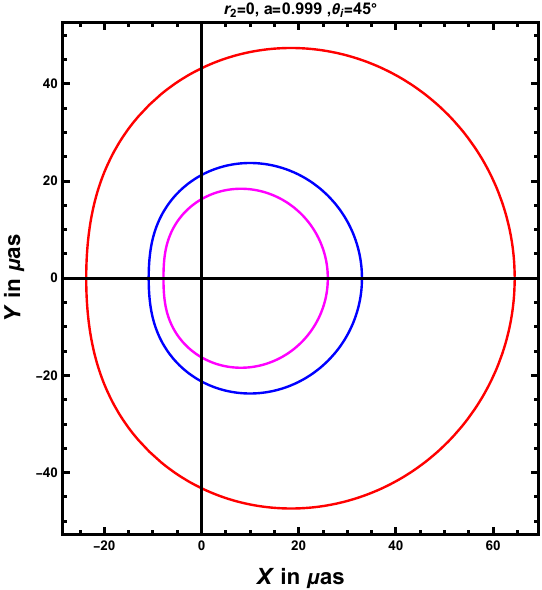}
        \caption{$r_2=0, \ a=0.999   $\label{h2g}}
    \end{subfigure}\hfill
    \begin{subfigure}{0.333\textwidth}
        \centering
        \includegraphics[width=\linewidth]{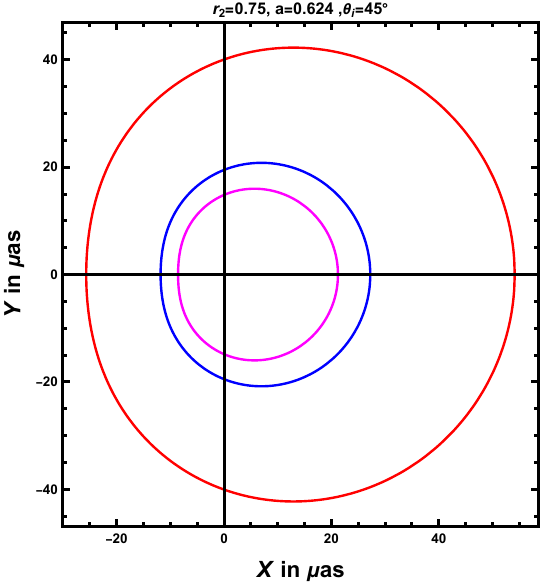}
        \caption{$r_2=0, \ a=0.999   $\label{h2h}}
    \end{subfigure}\hfill
    \begin{subfigure}{0.333\textwidth}
        \centering
        \includegraphics[width=\linewidth]{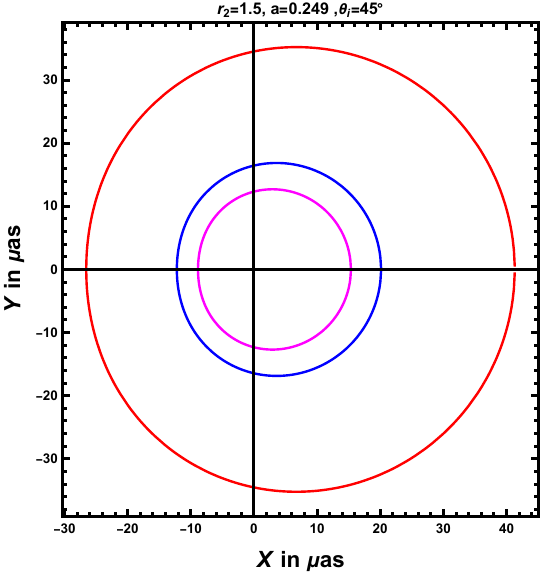}
        \caption{$r_2=1.5,  \ a=0.249   $}
    \end{subfigure}
\vspace{0.1cm}
 \begin{subfigure}{1\textwidth}
        \centering
        \includegraphics[width=0.5\linewidth]{ legend3.pdf}
          
    \end{subfigure}
 
    \caption{\label{h2}Variation of  shadow of Kerr Sen black hole in presence of homogeneous plasma at $\theta_i=45^\circ$}
\end{figure}

\begin{figure}[H]
    \centering
    \begin{subfigure}{0.333\textwidth}
        \centering
        \includegraphics[width=\linewidth]{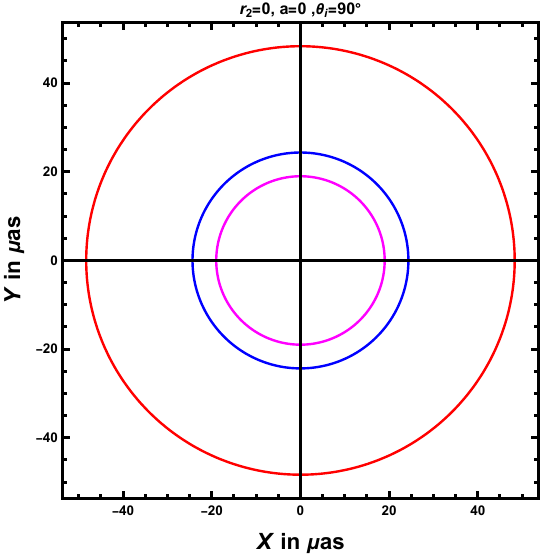}
        \caption{$r_2=0, \ a=0   $\label{h3a}}
    \end{subfigure}\hfill
    \begin{subfigure}{0.333\textwidth}
        \centering
        \includegraphics[width=\linewidth]{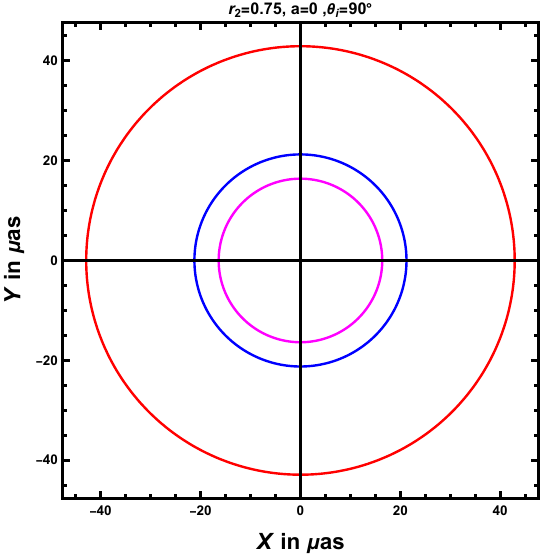}
        \caption{$r_2=0.75, \ a=0   $\label{h3b}}
    \end{subfigure}\hfill
    \begin{subfigure}{0.333\textwidth}
        \centering
        \includegraphics[width=\linewidth]{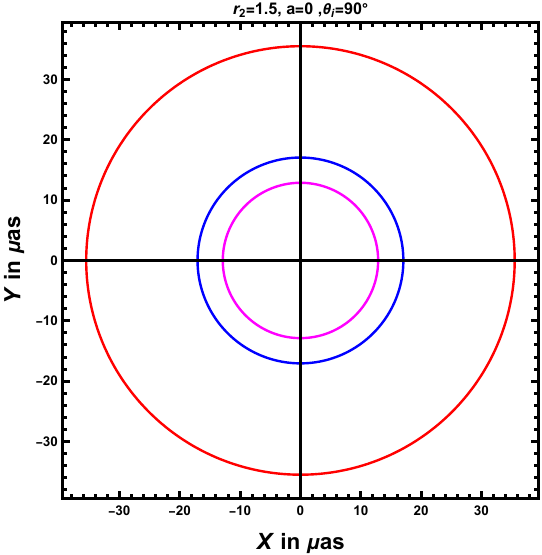}
        \caption{$r_2=1.5, \ a=0   $\label{h3c}}
    \end{subfigure}

    \vspace{0.5cm}

    \begin{subfigure}{0.333\textwidth}
        \centering
        \includegraphics[width=\linewidth]{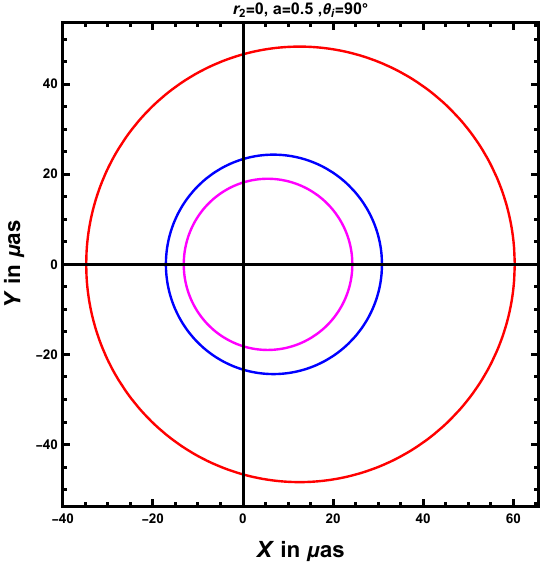}
        \caption{$r_2=0, \ a=0.5   $\label{h3d}}
    \end{subfigure}\hfill
    \begin{subfigure}{0.333\textwidth}
        \centering
        \includegraphics[width=\linewidth]{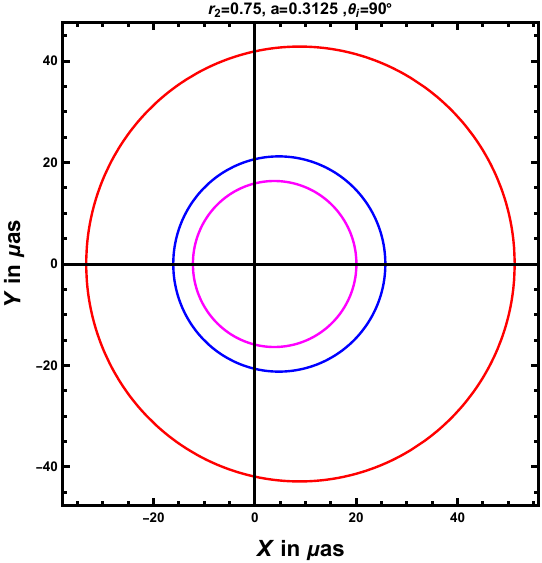}
        \caption{$r_2=0.75, \ a=0.3125   $\label{h3e}}
    \end{subfigure}\hfill
    \begin{subfigure}{0.333\textwidth}
        \centering
        \includegraphics[width=\linewidth]{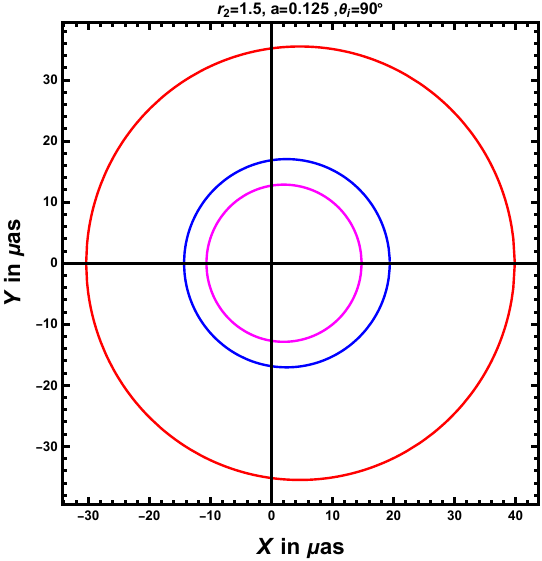}
        \caption{$r_2=1.5, \ a=0.125   $\label{h3f}}
    \end{subfigure}

    \vspace{0.5cm}

    \begin{subfigure}{0.333\textwidth}
        \centering
        \includegraphics[width=\linewidth]{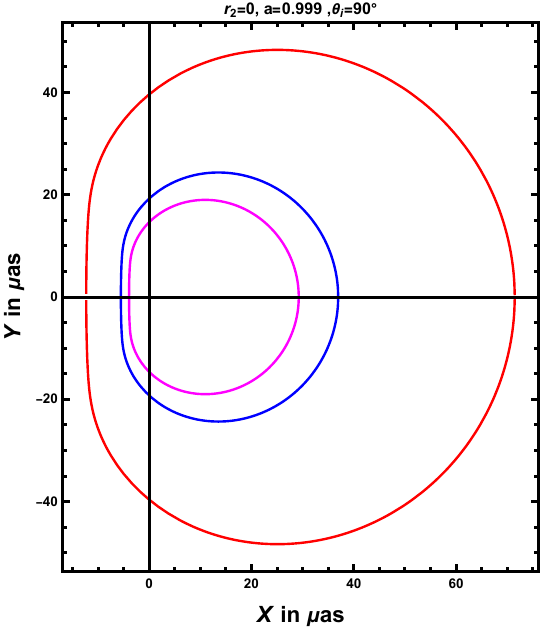}
        \caption{$r_2=0, \ a=0.999   $\label{h3g}}
    \end{subfigure}\hfill
    \begin{subfigure}{0.333\textwidth}
        \centering
        \includegraphics[width=\linewidth]{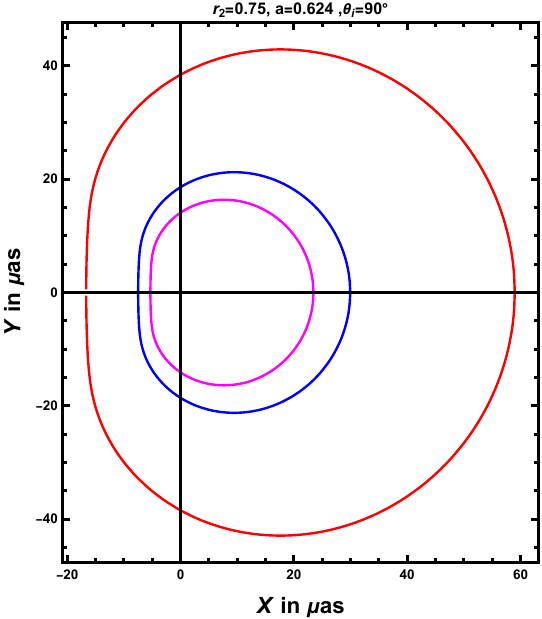}
        \caption{$r_2=0, \ a=0.999   $\label{h3h}}
    \end{subfigure}\hfill
    \begin{subfigure}{0.333\textwidth}
        \centering
        \includegraphics[width=\linewidth]{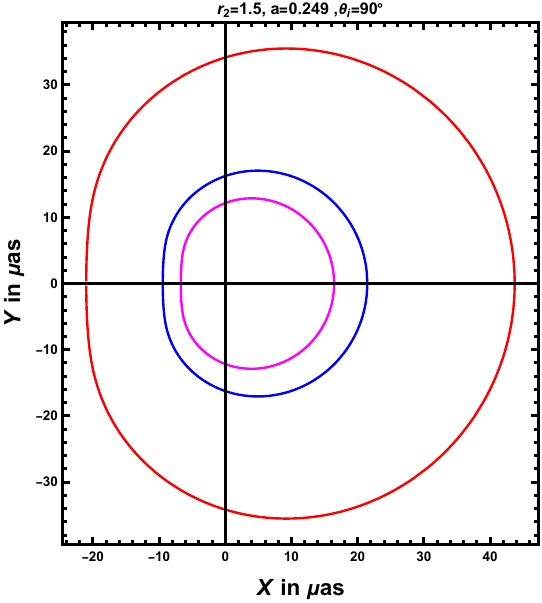}
        \caption{$r_2=1.5,  \ a=0.249   $\label{h3i}}
    \end{subfigure}
\vspace{0.1cm}
 \begin{subfigure}{1\textwidth}
        \centering
        \includegraphics[width=0.5\linewidth]{ legend3.pdf}
          
    \end{subfigure}
 
    \caption{\label{h3}Variation of  shadow of Kerr Sen black hole in presence of homogeneous plasma at $\theta_i=90^\circ$}
\end{figure}

\item The decreasing of shadow size due to increase in $r_2$  irrespective of presence or absence of plasma  can be observed in \Cref{h1a,h1b,h1c}, \Cref{h2a,h2b,h2c} and \Cref{h3a,h3b,h3c}. Thus, effect of $r_2$ on the shadow is generic irrespective of the three plasma environments considered in here. 
\end{itemize}

\section{EHT observations for shadows of M87* and Sgr A*\label{eht}}
In this section we first describe the observations related to shadows of M87* and Sgr A* as reported by the EHT collaboration. We also discuss briefly about the mass $M$, distance $D$ and angle of inclination angle $\theta_i$ measurements of M87* and Sgr A* as reported from previous observations.

For M87*,  the EHT collaboration reported  the image and   estimated the vertical angular diameter of  the bright ring $\Delta\Theta=(42\pm 3)\mu as$ \cite{EventHorizonTelescope:2019dse}. They also reported in \cite{EventHorizonTelescope:2019dse,EventHorizonTelescope:2019ggy} a maximum offset of $10\%$ between the ring diameter and the shadow (dark patch) diameter. If we take into account the maximum offset, the shadow angular diameter of M87* is $\Delta\Theta=(37.8\pm 3)\mu as$. Along with $\Delta\Theta$, the EHT collaboration has also reported bounds on the deviation in circularity $\Delta C\lesssim 10\%$ \cite{EventHorizonTelescope:2019dse} and the axis ratio $\Delta A\lesssim4/3$\cite{EventHorizonTelescope:2019dse} . Assuming M87* as a Kerr black hole, the EHT estimated the mass of  M87* to be $M=(6.5\pm 0.7)\times 10^9  M_{\odot}$\cite{EventHorizonTelescope:2019dse,EventHorizonTelescope:2019ggy}. The mass of M87* has also been previously measured,  $M=6.2^{+1.1}_{-0.6}\times10^9 M_{\odot}$ by analyzing  stellar brightness and dispersion in stellar velocity\cite{EventHorizonTelescope:2019dse,Gebhardt:2009cr,Gebhardt:2000fk}. From gas dynamics studies of M87* the mass was estimated to be $M=3.5^{+0.9}_{-0.3}\times 10^9 M_{\odot}$\cite{EventHorizonTelescope:2019dse,Walsh:2013uua} .  The   distance $D$ and angle of inclination $\theta_i$ for M87* , as reported from previous measurements are $D=(16.8\pm 0.8)$Mpc \cite{Blakeslee:2009tc,Bird:2010rd} and  $\theta_i=(17\pm 2)^\circ$\cite{Tamburini:2019vrf}, respectively.

The EHT collaboration has also  reported observations for image of Sgr A*. The  angular diameter estimate of the primary ring of Sgr A* as reported by the EHT is $\Delta\Theta=(51.8\pm 2.3)\mu as$ \cite{EventHorizonTelescope:2022wkp,EventHorizonTelescope:2022apq,EventHorizonTelescope:2022exc,EventHorizonTelescope:2022urf,EventHorizonTelescope:2022wok,EventHorizonTelescope:2022xqj} and the angular diameter of the shadow is estimated to be $\Delta\Theta=(48.7\pm 7)\mu as$\cite{EventHorizonTelescope:2022wkp}.  The mass and distance measurements of Sgr A* have been reported by the Keck and the GRAVITY collaboration.  Keeping  the  redshift parameter free, the Keck collaboration estimated the mass and distance of Sgr A* to be $M=(3.975\pm0.058\pm0.026)\times 10^6 M_{\odot}$ \cite{Do:2019txf} and $D=(7959\pm59\pm 32)$pc\cite{Do:2019txf}, respectively. When redshift parameter was set to unity the Keck collaboration reported  mass $M=(3.951\pm0.047)\times10^6 M_{\odot}$\cite{Do:2019txf} and distance $D=(7935\pm50)$pc\cite{Do:2019txf} for Sgr A*, respectively. Another set of mass and distance measurements for Sgr A* was reported by the GRAVITY collaboration  from astrometry observations. The GRAVITY collaboration reported the mass and distance of Sgr A* to be $M=(4.261\pm0.012)\times 10^6 M_{\odot}$\cite{GRAVITY:2020gka,GRAVITY:2021xju} and $=(8246.7\pm9.3)$pc\cite{GRAVITY:2020gka,GRAVITY:2021xju}, respectively. When effects of optical aberrations were taken into account, GRAVITY collaboration estimates the  mass and distance of Sgr A* to be $M=(4.297\pm0.012\pm0.040)\times 10^6 M_{\odot}$\cite{GRAVITY:2020gka,GRAVITY:2021xju} and $D=(8277\pm9\pm33)$pc\cite{GRAVITY:2020gka,GRAVITY:2021xju}. The angle of inclination for Sgr A* $\theta_i=46^\circ$  has been estimated in \cite{abuter2019geometric}.     
{  It is important to note that the masses and distances of M87* and Sgr A* were determined assuming these objects to be Kerr BHs. We now discuss that even if they are considered to be Kerr-Sen BHs, it is reasonable to assume the previously estimated masses and distances in determining the theoretical angular diameter of their shadow.  

When one determines the mass of Sgr A* from the radial velocity profile of S0-2 star \cite{Do:2019txf,GRAVITY:2020gka}, one essentially considers the radial geodesic equation of a massive test particle moving in the spacetime associated with Sgr A*. The radial geodesic equation of S0-2 star is expanded in post-Newtonian orders and one considers the leading order term, i.e, terms upto 1PN order \cite{Do:2019txf,GRAVITY:2020gka}, which does not contain any effect of the BH's spin, even if Sgr A* is described by a Kerr BH. The effect of the BH's spin becomes important when one considers higher post-Newtonian order terms \cite{Will:2007pp,Will:2016pgm}, which however are not significant at distances $\gtrsim {\rm 120 A.U\approx 2800 ~R_g}$ (where 120 A.U. corresponds to the periastron position of S0-2 star from Sgr A*). In such a scenario the trajectory of S0-2 star is mainly governed by the mass of the central object and hence the Schwarzschild metric is sufficient to explain its trajectory. Considering the Schwarzschild spacetime, the radial geodesic equation of S0-2 star upto 1-PN order is given by \cite{Do:2019txf,GRAVITY:2020gka},
\begin{align}
\frac{d^2\vec{R}}{dt^2}=-\frac{GM}{R^3}\vec{R} + \frac{GM}{c^2R^3}\Big(4\frac{GM}{R}-v^2\Big)\vec{R} + 4\frac{GM(\vec{R}.\vec{V})}{c^2R^3}\vec{V}
\label{8}
\end{align}
This equation is mainly used by the Keck Team and the GRAVITY collaboration to determine the mass and distance of Sgr A*. 

It is worthwhile to mention that the GRAVITY team has also provided an estimate of the precession of the pericentre ($f_{SP}$) of S0-2 star, by considering a modified version of Equation (\ref{8}) \cite{GRAVITY:2020gka},
\begin{align}
\frac{d^2\vec{R}}{dt^2}=-\frac{GM}{R^3}\vec{R} + f_{SP}\frac{GM}{c^2R^3}\Big[\Big(4\frac{GM}{R}-v^2\Big)\vec{R} + 4(\vec{R}.\vec{V})\vec{V}\Big]
\label{8-1}
\end{align}
where $f_{SP}$ encodes the change in periastron precession due to deviations from the Schwarzschild scenario, such that, the net effect of the precession is given by,
\begin{align}
\Delta \phi_{{\rm per~orbit}}=f_{SP}\frac{6\pi GM}{c^2a(1-e^2)}
\label{8-2}
\end{align}
It is clear from Equation (\ref{8-2}) that $f_{SP}=1$ corresponds to GR while $f_{SP}=0$ is associated with the Newtonian framework. A non-zero value of $f_{SP}$ apart from unity represents a deviation from the standard Schwarzschild scenario. Interestingly, the GRAVITY collaboration reported $f_{SP}=1.10\pm 0.19$ in close agreement with GR, but may also include signatures of alternatives to GR. 

In \cite{Do:2019txf}, the Keck team provided an estimate of the relativistic redshift experienced by S0-2 while passing through the gravitational potential of Sgr A*. They used spectroscopic and astrometric measurements of S0-2 from 1995–2017 along with the data from March to September 2018 which includes its periastron passage. They fitted the radial velocity profile of S0-2 containing a red-shift parameter $\Upsilon$ (which marks deviation from GR) with the aforesaid data and reported that $\Upsilon=0.88\pm 0.17$. Note that, if the trajectory of S0-2 is purely governed by the Schwarzschild metric in GR then $\Upsilon=1$, while $\Upsilon=0$ in Newtonian gravity \cite{Do:2019txf}. They determined the mass and distance of Sgr A* from the orbital data of S0-2 star, by keeping the redshift parameter free as well as setting by it to unity and reported that both the estimates exhibit only mild variation \cite{Do:2019txf}. We have used both sets of mass and distance estimates of Sgr A* provided by the Keck team to derive the theoretical angular diameter of the shadow of Sgr A*. We have also used the mass and distance estimates of Sgr A* provided by the GRAVITY collaboration in determining the theoretical angular diameter of its shadow. Thus, while deriving the theoretical angular diameter of Sgr A*, we have accounted for the variations in the mass and distance estimates of Sgr A*, in case there are deviations from GR.

The above discussion further elucidates that if Sgr A* is described by the Kerr-Sen metric, at distances $\gtrsim {\rm 120 A.U}$ this would essentially reduce to its non-rotating version, i.e., a dilaton BH described by,
\begin{align}
ds^2=-\Big[1-\frac{2M}{(r+r_2)}\Big]dt^2 +\frac{dr^2}{\Big[1-\frac{2M}{(r+r_2)}\Big]}+r(r+r_2) (d\theta^2 +\sin^2\theta d\phi^2)
\label{9}
\end{align}
In the large $r$ limit, Equation (\ref{9}) reduces to the Schwarzschild metric (since $0\leq r_2\leq 2$),
\begin{align}
ds^2&=-\Big[1-\frac{2M}{r}\Big(1-\frac{r_2}{r}\Big)\Big] dt^2 +\frac{dr^2}{\Big[1-\frac{2M}{r} \Big(1-\frac{r_2}{r}\big)\Big]} +r^2 \Big(1+\frac{r_2}{r}\Big)(d\theta^2 +\sin^2\theta d\phi^2) \nonumber \\
&\approx -\Big[1-\frac{2M}{r}\Big]dt^2 +\frac{dr^2}{\Big[1-\frac{2M}{r}\Big]}+r^2 (d\theta^2 +\sin^2\theta d\phi^2) ~~~~~{ \rm{(when ~r_2<<r)}}
\label{11} 
\end{align}
which implies that at distances $r\gtrsim {\rm 120 A.U}$ it will be difficult to distinguish the non-rotating dilaton BH from the Schwarzschild BH. Thus, the forms of Equation (\ref{8}) or Equation (\ref{8-1}) remain unchanged in the large $r$ limit and hence whether we assume Sgr A* to be a dilaton BH or a Schwarzschild BH, the estimated mass is not expected to change.

Such an analysis has been done in \cite{Fernandez:2023kro} where the authors attempt to constrain the dilaton charge of Sgr A* from the trajectory of S0-2 star. Using the publicly available astrometric data for S0-2, they derive an upper bound, $r_2\leq 24M$, which reduces to $r_2\leq 2.8M$, when they additionally use the data related to 
the relativistic orbital precession for S0-2 \cite{GRAVITY:2018ofz,GRAVITY:2020gka}. But, both the bounds are greater than the theoretical bound $0\leq r_2 \leq 2M$ and hence their analysis reveals that the available data associated with the trajectory of S0-2 star cannot distinguish between the Schwarzschild and the dilaton BH. The distance of Sgr A* estimated in this work is $\rm 8.23\pm 0.23$ kpc, in agreement with previous estimates \cite{Do:2019txf,GRAVITY:2020gka}.

In another work \cite{DellaMonica:2021fdr}, the authors attempted to test the nature of the compact object in Sgr A* by assuming that Sgr A* is governed by the black bounce (Simpson-Visser) metric which may represent a black hole or a wormhole (WH) depending on the value of the parameter $\alpha$ which marks the deviation from the Schwarzschild scenario. Using an MCMC algorithm, the geodesic equations for a massive test particle in the aforesaid background are compared with the publicly available data for the orbital motion of the S0-2 star to establish constrains on the parameter $\alpha$. Their analysis reveals that the present data cannot establish whether Sgr A* is a Schwarzschild BH/a Simpson-Visser BH/a  Simpson-Visser WH. The distance of Sgr A* predicted from their analysis is in agreement with earlier estimates \cite{Do:2019txf,GRAVITY:2020gka}. 

In \cite{Bambhaniya:2022xbz}, Sgr A* is assumed to be governed by the Janis-Newman-Winicour (JNW) spacetime which exhibits deviations from the Schwarzschild metric due to the presence of a scalar charge $q$. Using the orbital data of S0-2 star the authors established constrains on the mass of Sgr A* in presence of the scalar charge. Their analysis reveals that even in the presence of a non-zero $q$ the mass of Sgr A* comes out to be $\sim 4.63\times 10^6 M_\odot$, which is similar to the mass estimates provided by the Keck Team and the GRAVITY collaboration \cite{Do:2019txf,GRAVITY:2020gka}.
Thus, even if one undertakes a re-evaluation of the observations related to the trajectory of S0-2 star assuming that Sgr A* is a non-Kerr BH, the results of mass estimates of Sgr A* is not expected to change significantly. 

Also, the presence of plasma around M87* and Sgr A* is not expected to change the mass and distance estimates of these BHs as 
the plasma density near these BHs is very low (which will be discussed later). Moreover, the mass and distances
are estimated from the motion of stars surrounding these objects where the relevant distances are $\gtrsim$ few thousands of gravitational radii, where the plasma density becomes much lower compared to the density near the source. For Sgr A* the closest the S0-2 star can get $\sim 2800 R_g$ \cite{Do:2019txf} while for M87* the distance probed is $\sim$ 17 pc to 170 kpc \cite{Gebhardt:2011yw}(which implies distance $\gtrsim 5000 R_g$ assuming M87* to possess a mass $M\sim 6.2\times 10^9 M_\odot$).
At such distances the effect of the plasma becomes inconspicuous and the spacetime metric plays a dominant role in determining the stellar trajectory.
Hence, it seems reasonable to use the previously determined masses and distances of M87* and Sgr A*  to estimate the theoretical angular diameter of the shadow of these objects both in the absence and presence of plasma, which has also been done in previous works \cite{Vagnozzi:2022moj,KumarWalia:2024omf,Shaikh:2021yux,Shaikh:2022ivr,Pal:2023wqg,Islam:2024sph,Islam:2022wck,Kumar:2018ple,Afrin:2021imp,Raza:2024zkp}.}

In order to obtain the constrains on $r_2$ and $\alpha_i$ (where $i=1,2,3$ for profiles 1,2 and 3, respectively),  we follow the procedure described below:
\begin{enumerate}
    
    \item We choose a plasma profile from the profiles given in \Cref{eqn:p1,eqn:p2,eqn:p3} and calculate the maximum possible value of $\alpha_i$ using \Cref{eqn:boundfinal} (refer \cref{p1,p2,p3} for details).     
    \item     Fixing $\alpha_i$ , we choose a value of $0\leq r_2\leq2$ and vary the spin in the allowed range $0\leq a\leq1-\frac{r_2}{2}$.   For each combination of $\alpha_i, r_2\text{ and }a$ we calculate the theoretical vertical angular diameter $\Delta\Theta_{th}$ {for M87* and Sgr A* } using \Cref{theoretical angular diameter}.    
    \item In order to calculate the $\Delta\Theta_{th}$ we use previously determined mass $M$, distance $D$ and angle of inclination $\theta_i$ as discussed above. In particular, we use the central values of $M,D$ and $\theta_i$.

    \item We compare the observed angular diameter $\Delta\Theta_{obs}$  with the theoretical one $\Delta\Theta_{th}$ and calculate the $\chi^2$  given by,
\begin{gather}
    \label{chi-square}
    \chi^2=\left(\frac{\Delta\Theta_{obs}-\Delta\Theta_{th}(\alpha_i,r_2,a_{min})}{\sigma}\right)^2
\end{gather}
      In the above equation $\Delta\Theta_{obs}$ and $\sigma$ for M87* is $37.8\mu as$ and $3\mu as$, respectively. For Sgr A*,  $\Delta\Theta_{obs}=48.7\mu as$ and $\sigma=7\mu as$.

\item  The $\chi^2$ in \Cref{chi-square} is determined by first computing $\chi^2$ for allowed values of spin in the range $0\leq a\leq 1-\frac{r_2}{2}$  for a given value of  $r_2\text{ and }\alpha_i$, and then finding the spin  $a_{min}$ corresponding to the  lowest value of $\chi^2$. Thus, $a_{min}$  in \Cref{chi-square}  corresponds to spin for which $\chi^2$ is minimum for a given $\alpha_i$ and $r_2$\cite{1976ApAvni}. 
    \item We follow steps 2-5 for all the allowed values of $\alpha_i\text{ and }r_2$.

    \item We draw contour plots of $\chi^2\leq1$ to obtain the observationally favored parameter space of $\alpha_i $ and $r_2$. 
    \item We follow steps 2-7 for another plasma profile.
 \end{enumerate}

\subsection{Constraining the dilaton charge and plasma environment from M87* shadow\label{Constraining dilaton charge and plasma environment from M87* shadow} \label{Sec5-1}}
For the gas dynamics mass measurement of M87* ($M=3.5\times10^9 M_{\odot}$),   $\Delta\Theta_{th}$ evaluated with all combinations   of $r_2$ and $\alpha_1$ in the allowed range give $\chi^2>1$ and thus, no constraints are obtained on $r_2$ and $\alpha_1$. This implies that using this mass one cannot reproduce the observed angular diameter of M87*, which possibly indicates that this mass measurement needs to be revisited.  This result is also consistent with our previous finding in \cite{Sahoo:2023czj} . The constraints on $\alpha_1$ and $r_2$ obtained using the EHT mass measurement of M87* (refer \Cref{contour plots for M87* 6.5 considering profile 1})is for the sake of comparison and completeness and should not be considered to constrain $r_2$. This is because the mass estimate was obtained assuming  M87* is a  Kerr black hole.
We now use the methodology described previously to obtain constraints on the plasma parameter $\alpha_1$ and the dilaton charge $r_2$ from the EHT observations of M87*. 
\Cref{contour plots for M87* 6.2 considering profile 1,contour plots for M87* 6.5 considering profile 1}  show contours of $\chi^2$ for M87* assuming different values of $\alpha_1$ and $r_2$ for $M=6.2\times 10^9 M_{\odot}$  and $M=6.5\times10^9 M_{\odot}$ respectively, considering plasma profile 1. 

\begin{itemize}
    \item In \Cref{contour plots for M87* 6.2 considering profile 1,contour plots for M87* 6.5 considering profile 1},  we observe that, there is an upper bound on $\alpha_1$ and $r_2$ corresponding to each $\chi^2$ contour. More importantly, in  the parameter space of $(\alpha_1,r_2)$ corresponding to the white region ( $\chi^2>1$) $\Delta\Theta_{th}$ lies beyond the $1-\sigma$ interval of $\Delta\Theta_{obs}$,  and therefore values of  $r_2$ and $\alpha_1$ corresponding to the white region are excluded outside $1-\sigma$. 
    
    \item   In \Cref{contour plots for M87* 6.2 considering profile 1},  for region $\chi^2\leq1$ (which corresponds to $\Delta\Theta_{th}\approx34.8\mu as$ in the present case)  we observe $\alpha_1\leq2.5$ and $r_2\leq0.48$.  For  the region $\chi^2\leq0.5$ (which corresponds to $\Delta\Theta_{th}\approx35.68\mu as$),  we find $\alpha_1\leq1.8$ and $r_2\leq0.34$. The the region $\chi^2\leq0.1$ (which corresponds to $\Delta\Theta_{th}\approx36.85\mu as$)  restricts $\alpha_1\leq0.9$ and $r_2\leq0.16$ .  The constraints on $\alpha_1$ and $r_2$ obtained from the region in blue ($\chi^2\leq0.1$) are the ones  most observationally favored. From \Cref{contour plots for M87* 6.2 considering profile 1} we rule out $r_2>0.48$ and $\alpha_1>2.5$ outside $1-\sigma$ . We note that, the bounds on $r_2$ obtained in our previous work\cite{Sahoo:2023czj} are consistent with the present bounds.

\begin{figure}[H]
\begin{subfigure}{0.5\textwidth}
        \centering
        \includegraphics[width=\linewidth]{ 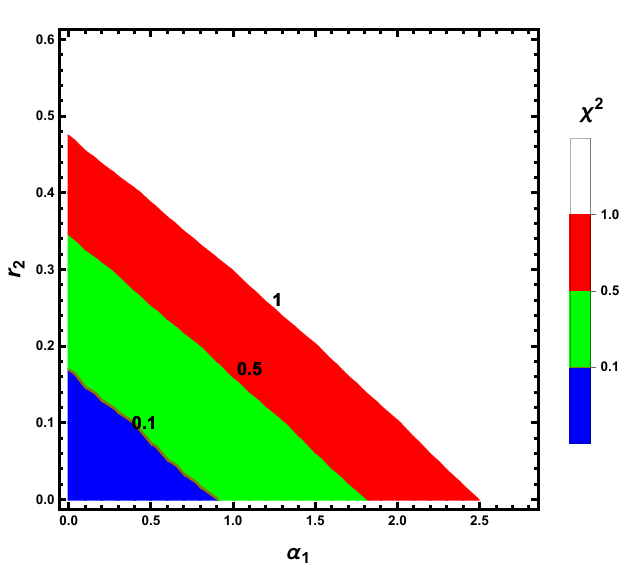}
        \caption{\label{contour plots for M87* 6.2 considering profile 1}$M=6.2\times10^9 M_{\odot}$}
    \end{subfigure}\hfill
    \begin{subfigure}{0.5\textwidth}
        \centering
        \includegraphics[width=\linewidth]{ 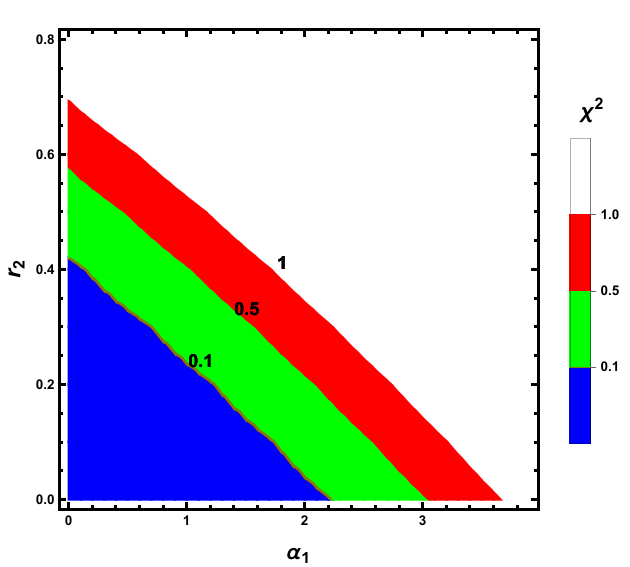}
        \caption{\label{contour plots for M87* 6.5 considering profile 1}$M=6.5\times10^9 M_{\odot}$}
    \end{subfigure}
    \caption{\label{contour plots for M87* considering profile 1}The figures represent  contour plots of $\chi^2$ (calculated using \Cref{chi-square} corresponding to $a_{min}$) for various choices of $\alpha_1$ and $r_2$,  associated with  plasma profile 1 for M87*. In order to evaluate the $\chi^2$ we use $D=16.8$Mpc, $\theta_i=17^\circ$ and black hole mass (a) $M=6.2\times 10^9 M_{\circ}$ and (b) $M=6.5\times 10^9 M_{\odot}$ (EHT estimate).  In the figures,  the red color region in the $\alpha_1- r_2$ plane corresponds to  $0.5\leq\chi^2\leq1$, the region with $0.1\leq\chi^2\leq0.5$ is colored green and the blue region represents $0\leq\chi^2\leq0.1$. }
    \label{Fig12-1}
\end{figure}

    \item For stellar dynamics mass  measurement $\alpha_1=0.1, r_2=0$ and $a=0.2$ gives the lowest $\chi^2$. Also, for $\chi^2\leq1$    the allowed spin range is $0.1\leq a\leq0.3$. For completeness, when mass estimate by the EHT collaboration was considered, $\alpha_1=1.3, r_2=0$ and $a=0.4$ gave lowest $\chi^2$ and the allowed spin range for $\chi^2\leq1$ is $0.1\leq a\leq0.9$. 

   \item For each $\chi^2$ contour in \Cref{contour plots for M87* 6.2 considering profile 1,contour plots for M87* 6.5 considering profile 1} we  observe that, as $\alpha_1$ is increased in the allowed range, the upper bound on $r_2$ corresponding to the $\alpha_1$,  keeps decreasing and vice-versa. This is an artifact of the decreasing effect produced  due to $\alpha_1$ and $r_2$ on the shadow size (refer \Cref{1,2,3}).

    \item We further observe that there is only one contour for each $\chi^2$ in contrast with \Cref{homogeneous_3.5,homogeneous_6.2,homogeneous_6.5} for homogeneous plasma (for example, contours of $\chi^2=0.1$ appears twice).  This is because in the case of plasma profile 1 both $\alpha_1$ and $r_2$ have a contracting effect on the shadow size, whereas homogeneous plasma has expanding effect on the shadow . Therefore the upper $1-\sigma$ value of $\Delta\Theta_{obs}$ is not obtained for any $(\alpha_1, r_2)$ using mass measurements in \Cref{contour plots for M87* 6.2 considering profile 1,contour plots for M87* 6.5 considering profile 1} .

\end{itemize}
 
\begin{figure}[H]
\begin{subfigure}{0.5\textwidth}
        \centering
        \includegraphics[width=\linewidth]{ 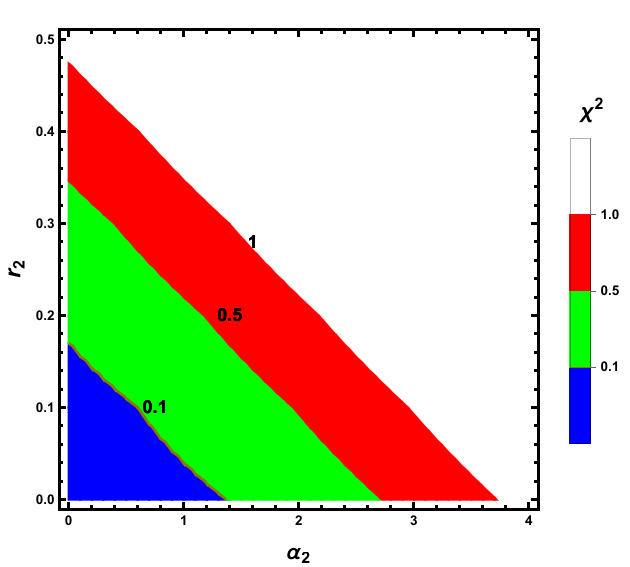}
        \caption{\label{angular_6.2}$M=6.2\times10^9 M_{\odot}$}
    \end{subfigure}\hfill
    \begin{subfigure}{0.5\textwidth}
        \centering
        \includegraphics[width=\linewidth]{ 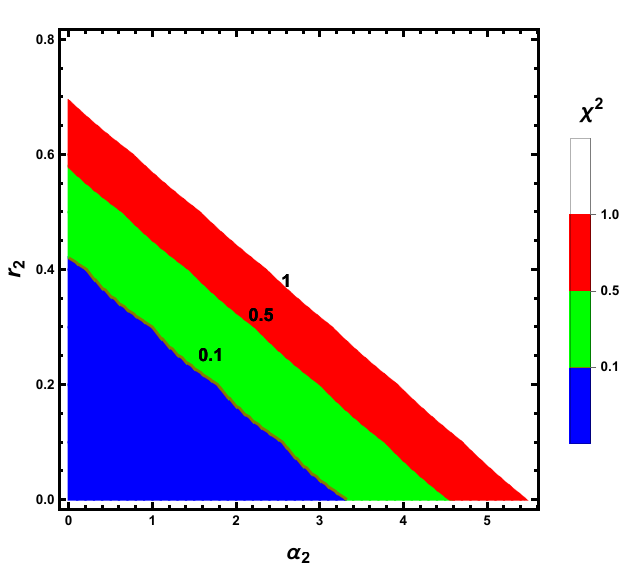}
        \caption{\label{angular_6.5}$M=6.5\times10^9 M_{\odot}$}
    \end{subfigure}
    \caption{\label{contour plots for M87* considering profile 2}The figures represent  contour plots of $\chi^2$ (calculated using \Cref{chi-square} corresponding to $a_{min}$) for various choices of $\alpha_2$ and $r_2$,  corresponding to plasma profile 2 for M87*. In order to evaluate the $\chi^2$ we use $D=16.8$Mpc, $\theta_i=17^\circ$ and black hole mass (a) $M=6.2\times 10^9 M_{\circ}$ and (b) $M=6.5\times 10^9 M_{\odot}$.  In the figures,  the red color region in $\alpha_1- r_2$ plane corresponds to  $0.5\leq\chi^2\leq1$, the region with $0.1\leq\chi^2\leq0.5$ is colored green and the blue region represents $0\leq\chi^2\leq0.1$. }
\label{Fig13-1}
\end{figure}

\begin{figure}[H]
\begin{subfigure}{0.3333\textwidth}
        \centering
        \includegraphics[width=\linewidth]{ 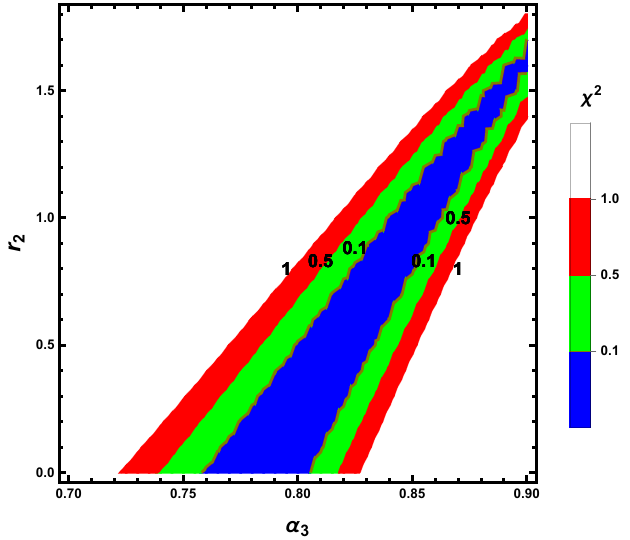}
         \caption{\label{homogeneous_3.5}$M=3.5\times10^9 M_{\odot}$}
    \end{subfigure}\hfill
    \begin{subfigure}{0.3333\textwidth}
        \centering
        \includegraphics[width=\linewidth]{ 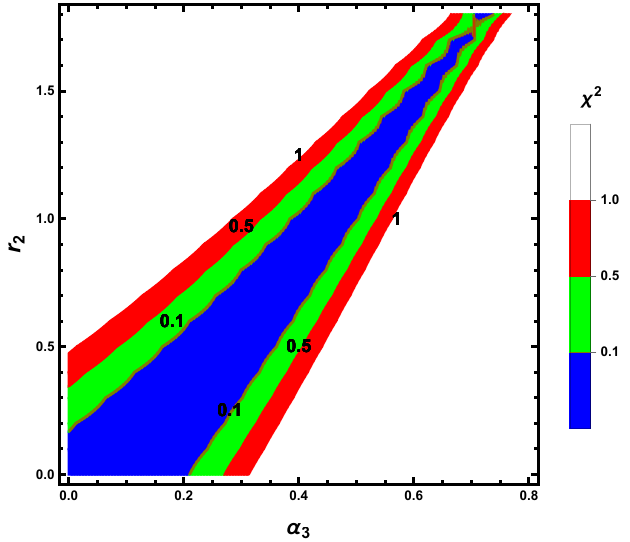}
            \caption{\label{homogeneous_6.2}$M=6.2\times10^9 M_{\odot}$}
    \end{subfigure}\hfill
    \begin{subfigure}{0.3333\textwidth}
        \centering
        \includegraphics[width=\linewidth]{ 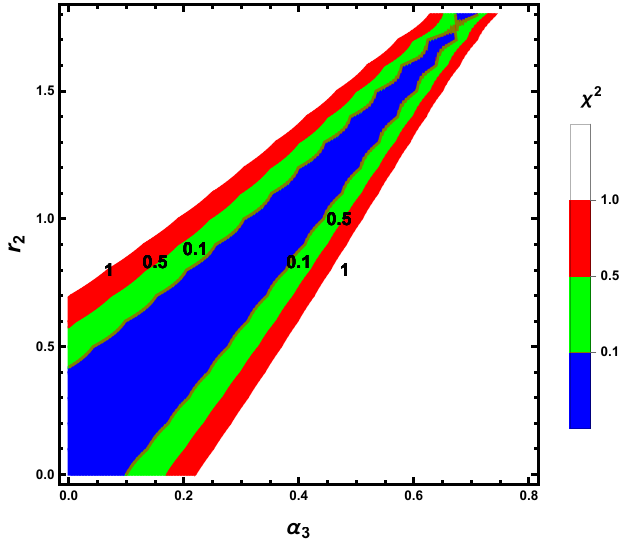}
            \caption{\label{homogeneous_6.5}$M=6.5\times10^9 M_{\odot}$}
    \end{subfigure}
    \caption{\label{contour plots for M87* considering profile 3}The figures represent  contour plots of $\chi^2$ (calculated using \Cref{chi-square} corresponding to $a_{min}$) for (a) $M=3.5\times10^9 M_{\odot}$,(b) $M=6.2\times10^9 M_{\odot}$ and (c) $M=6.5\times10^9 M_{\odot}$  for M87* surrounded by homogeneous plasma. In the figures, the red color region of $\alpha_1- r_2$ plane corresponds to $0.5\leq\chi^2\leq1$, the region with $0.1\leq\chi^2\leq0.5$ is colored green and the blue region represents $0\leq\chi^2\leq0.1$. For the figures the distance $D=16.8$Mpc and angle of inclination $\theta_i=17^\circ$.}
\label{Fig14}
\end{figure}

Next we discuss the constraints on the dilaton charge $r_2$ and $\alpha_2$ for plasma profile 2 using the EHT observations of M87*.  In \Cref{angular_6.2,angular_6.5} ,  the $\chi^2$- contour plots for M87* considering plasma profile 2 are given. For reasons already discussed in case of profile 1 previously, we will consider constraints obtained from \Cref{angular_6.2} only. In case of profile 2 also we obtain $\chi^2>1$ for all choices of $\alpha_2$ and $r_2$ when gas dynamics mass measurement of M87* is considered, and thus we get no constraints. Hence, the inability to reproduce the observed shadow angular diameter of  M87* with this mass ($M=3.5\times10^9 M_{\odot}$) which was highlighted in our previous work\cite{Sahoo:2023czj} is consistent with  plasma profile 2 as well. Furthermore, the mass of M87* estimated from gas dynamics measurement is clearly not in agreement with the EHT constraint $M=6.5\times10^9 M_{\odot}$. The constraints on the dilaton parameter and the plasma parameter are very similar to that of profile 1.

     In \Cref{angular_6.2}, from the region $\chi^2\leq1$ (which in case of profile 2 will represent the lower $1-\sigma$ value of $\Delta\Theta_{obs}$ for M87*) we find $\alpha_2\leq3.8$ and $r_2\leq0.48$. However, the upper bound of $\alpha_2$ is more than that of $\alpha_1$ in \Cref{contour plots for M87* 6.2 considering profile 1}.  For the region $\chi^2\leq0.5$  in \Cref{angular_6.2} we obtain the upper bounds on $\alpha_2\leq2.8$ and $r_2\leq0.34$ . From the region $\chi^2\leq0.1$  (which is most observationally favored) we obtain $\alpha_2\leq1.4$ and $r_2\leq0.16$. We again observe that the upper bounds on $\alpha_2$ are more than the upper bounds of $\alpha_1$.  This is because, profile 1 has a stronger contraction effect on the shadow size compared to profile 2  in case of M87* (see   \Cref{p1,p2}).  This result is sensitive to the mass and distance of the black hole and hence may not be generic (\Cref{angular x coordinate of shadow,angular y coordinate of shadow,theoretical angular diameter}). 
     For profile 2, when mass measurement from stellar dynamics is considered  $\alpha_2=0.2, r_2=0$  and $a=0.1$ gives the lowest 
     $\chi^2$ and the allowed range of spin for $\chi^2\leq1$ is $0\leq a\leq0.3$. When mass estimated by the EHT collaboration is used $\alpha_2=2.2, r_2=0$ and $a=0.2$ gives the lowest $\chi^2$ and the allowed range of spin within $\chi^2\leq1$ in this case, $0\leq a\leq0.9$.    

From our results obtained from \Cref{contour plots for M87* considering profile 1,contour plots for M87* considering profile 2}, we can say that the presence of plasma affects the constraints on the dilaton charge, because of the dispersive effect of the plasma on the shadow. Secondly, in both cases (plasma profiles 1 and 2), the highest bound on $\alpha_1$ and $\alpha_2$ obtained from the EHT observations is far less than the theoretical upper bounds on $\alpha_1\text{ and }\alpha_2$ (\Cref{eqn:bound,eqn:boundfinal} and \Cref{F1,F2}). This means that high plasma densities are not favored by the EHT observations of M87* when profiles 1 and 2 are considered. Also, for the highest value of plasma parameters $\alpha_1$ and $\alpha_2$ , we get $r_2\approx0$ (for all $\chi^2$ contours). Thus, higher density plasma environments described by profile 1 and 2 favor Kerr black hole compared to Kerr Sen black hole scenario.   Furthermore, the persistence of M87* shadow \cite{EventHorizonTelescope:2024dhe}  indicates that the effect of the metric on the shadow diameter is predominant compared to the accretion environment. Our analysis  indicates that, the spin estimate for M87* based on  stellar dynamics  mass measurements is  $0\lesssim a\leq0.3$ and from  the EHT mass estimate is $0\leq a\leq0.9 $. The constraints on spin are more or less same in case of profile 1 and profile 2. The spin of M87* has been estimated previously  based on jet power in \cite{Nemmen:2019idv} to be $|a|\geq0.4$ (in prograde case) and $|a|\geq0.5$ (in retrograde case) and  Tamburini et al.\cite{Tamburini:2019vrf} reported the spin to be $a=0.9\pm0.05$ with approximately 95\% confidence level. These previous estimates  are higher than spin constraints obtained by us using stellar dynamics mass estimate of M87*. It must be noted that, constraining the spin of a  black hole  from its shadow is not easy as the angular diameter of the shadow is  not very sensitive to the spin particularly when the inclination angle is low (which is $17^\circ$ for M87*)  (refer \Cref{p1,p2}) . Also in the Kerr black hole case the maximum variation of angular diameter of the shadow is 4\% for the allowed spin range\cite{Nemmen:2019idv,Takahashi:2004xh,Johannsen:2010ru}.    Our results based on profile 1 and profile 2 indicate that the EHT data for M87* favors the Kerr black hole scenario although Kerr Sen black hole with dilaton charge $r_2\leq0.48$ is allowed within $1-\sigma$.

We now consider the  homogeneous plasma  case, where  we have reported contour plots considering  gas dynamics, stellar dynamics and the EHT mass measurement in \Cref{homogeneous_3.5,homogeneous_6.2,homogeneous_6.5} respectively. We  highlight  some distinct features of the contour plots in case of homogeneous plasma.

\begin{itemize}
    \item We observe that each contour plot in \Cref{homogeneous_3.5,homogeneous_6.2,homogeneous_6.5} has two contours corresponding to each $\chi^2=0.1,0.5\text{ and }1$. This happens because the homogeneous plasma parameter $\alpha_3$ has an expanding effect on the shadow size which can dominate over contracting effect produced due to $r_2$ (refer \Cref{h1,h2,h3}). This allows one to reproduce the $\chi^2$ corresponding to the upper $1-\sigma$ and the lower $1-\sigma$ bounds related to $\Delta\Theta_{obs}$.
    \item Secondly, here we observe from \Cref{homogeneous_3.5}, unlike the case of profile 1 and  profile 2, when homogeneous plasma is considered, even considering the gas dynamics mass measurement for M87* one can produce the the central as well as $1-\sigma$ values of  $\Delta\Theta_{obs}=(37.8\pm3)\mu as$. This is possible because, although a smaller mass decreases the shadow (one of the causes of smaller shadow size \cite{Sahoo:2023czj}),  the expansive effect of $\alpha_3$ allows one to   reproduce $\Delta\Theta_{obs}$ .
% This is because while smaller mass $M$ is very less compared to stellar dynamics mass estimate,  $\alpha_3$ has a very strong increasing effect on the shadow which allows reproduction of the observed data using $M=3.5\times10^9 M_{\odot}$.  
    \item    When $\alpha_3\leq0.72$, we fail to reproduce $\Delta\Theta_{obs}$ within $1-\sigma$ . For $\alpha_3>0.83$ a non-zero $r_2$ is required to  explain the observed shadow.
    \item   We now consider $M=6.2\times10^9 M_{\odot}$ to obtain $\Delta\Theta_{th}$. Since, mass has an increasing effect on the shadow
we require smaller $\alpha_3$ to reproduce $\Delta\Theta_{obs}$  (compare \Cref{homogeneous_3.5,homogeneous_6.2}). For a given $\alpha_3$, this mass allows us to encompass larger $r_2$ (\Cref{homogeneous_3.5,homogeneous_6.2}), since $M\text{ and  }\alpha_3$ increase the shadow, and $r_2$ decreases the shadow. Same observation holds good for \Cref{homogeneous_6.5} with $M=6.5\times10^9 M_{\odot}$. 
\end{itemize}

\subsection{Constraining the dilaton charge and plasma environment from Sgr A* shadow\label{Constraining dilaton charge and plasma environment from Sgr A* shadow} \label{Sec5-2}}
We now use the methodology to obtain constraints on the dilaton charge $r_2$ and $\alpha_1$ (profile 1) for Sgr A* using the EHT observations of its shadow. The EHT collaboration reported the angular diameter of the shadow to be $\Delta\Theta_{obs}=(48.7\pm7)\mu as$. 
\Cref{profile1_3.951,profile1_3.975,profile1_4.261,profile1_4.297} show the contours of $\chi^2$ for Sgr A* considering plasma profile 1 for previously reported mass and distance measurements. \Cref{profile1_3.951,profile1_3.975} are plotted using Keck team measurements\cite{Do:2019txf} $M=3.951\times10^6 M_{\odot}\text{, } D=7935$pc and $M=3.975\times10^6 M_{\odot}\text{, } D=7959$pc, respectively. \Cref{profile1_4.261,profile1_4.297} are plotted using GRAVITY collaboration measurements\cite{GRAVITY:2020gka,GRAVITY:2021xju} $M=4.261\times10^6 M_{\odot}\text{, } D=8246.7$pc and $M=4.297\times10^6 M_{\odot}\text{,
} D=8277$pc, respectively.

\begin{itemize}

    \item The contour plots in \Cref{contour plots for Sgr A* considering profile 1} and \Cref{contour plots for M87* considering profile 1} are similar in nature.   For  each $\chi^2$ contour,  as $\alpha_1$ increases in the allowed range, the upper bound of $r_2$ decreases and vice-versa.
    \item In \Cref{profile1_3.951,profile1_3.975}, $\alpha_1\leq5.2$ and $r_2\lesssim1$ is allowed within $1-\sigma$  (which corresponds to $\Delta\Theta_{th}\gtrsim41.7\mu as$). We obtain the bounds, $\alpha_1\leq4.2$ and $r_2\leq0.8$, when $\chi^2\leq0.5$ (which corresponds to $\Delta\Theta_{th}\gtrsim43.75 \mu as$). For the most observationally favored  region $\chi^2\leq0.1$ (corresponding to $\Delta\Theta_{th}\gtrsim46.49\mu as$), $\alpha_1\leq2.8$ and $r_2\leq0.5$. Furthermore, $\alpha_1>5.2$ and $r_2>1$ are ruled outside of $1-\sigma$. 
   \item For the Keck team mass and distance measurement $M=3.951\times10^6M_{\odot},D=7935$pc$,\alpha_1=0.3, r_2=0.2 $ and $a=0.45$ gives lowest $\chi^2$ and for the mass and distance measurement $M=3.975\times10^6 M_{\odot}, D=7959$pc, $\alpha_1=1.2,r_2=0 $ and $a=0.7$ gives lowest $\chi^2$.  However, for both measurements, the allowed range of spin within $\chi^2\leq1$ is $0\lesssim   a\leq1$.  
    
\begin{figure}[H]
\begin{subfigure}{0.5\textwidth}
        \centering
        \includegraphics[width=\linewidth]{ 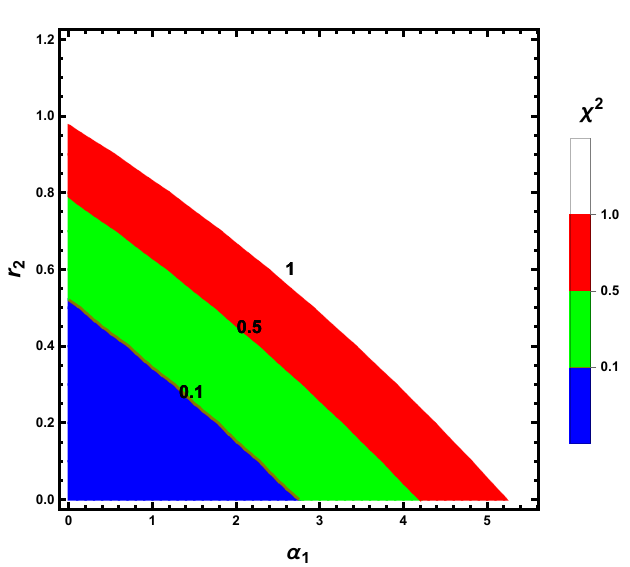}
        \caption{\label{profile1_3.951}$M=3. 951\times10^6 M_{\odot}$ and $D=7935$pc}
    \end{subfigure}\hfill
    \begin{subfigure}{0.5\textwidth}
        \centering
        \includegraphics[width=\linewidth]{ 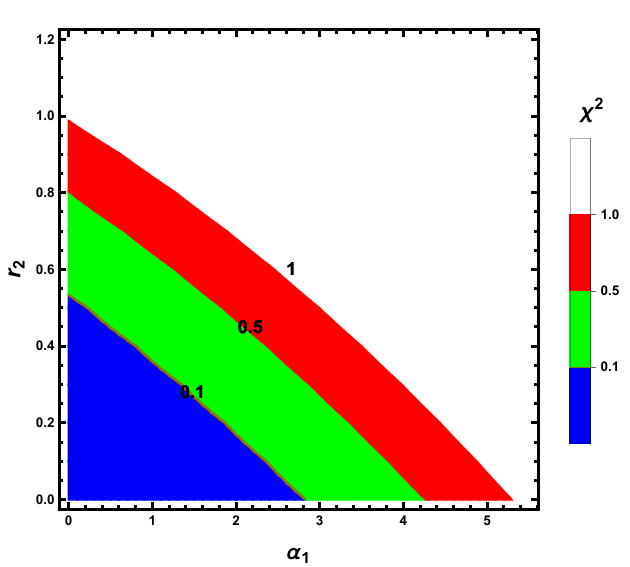}
        \caption{\label{profile1_3.975}$M=3. 975\times10^6 M_{\odot}$ and $D=7959$pc}
     \end{subfigure}\hfill
    \begin{subfigure}{0.5\textwidth}
        \centering
        \includegraphics[width=\linewidth]{ 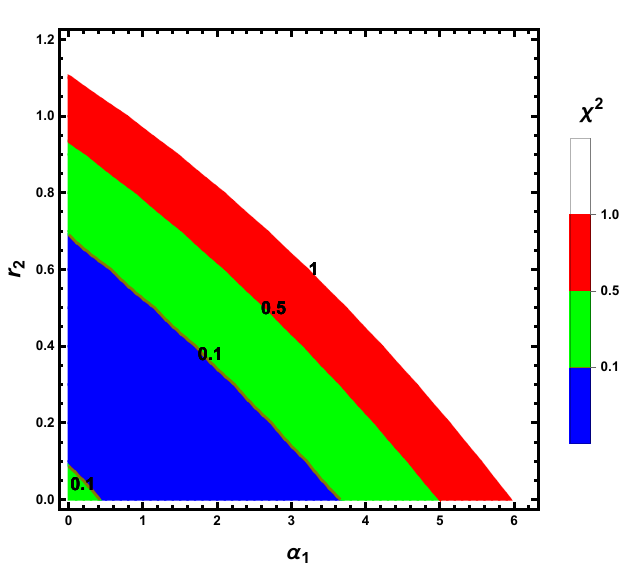}
        \caption{\label{profile1_4.261}$M=4. 261\times10^6 M_{\odot}$ and $D=8246. 7$pc}
    \end{subfigure}\hfill
    \begin{subfigure}{0.5\textwidth}
        \centering
        \includegraphics[width=\linewidth]{ 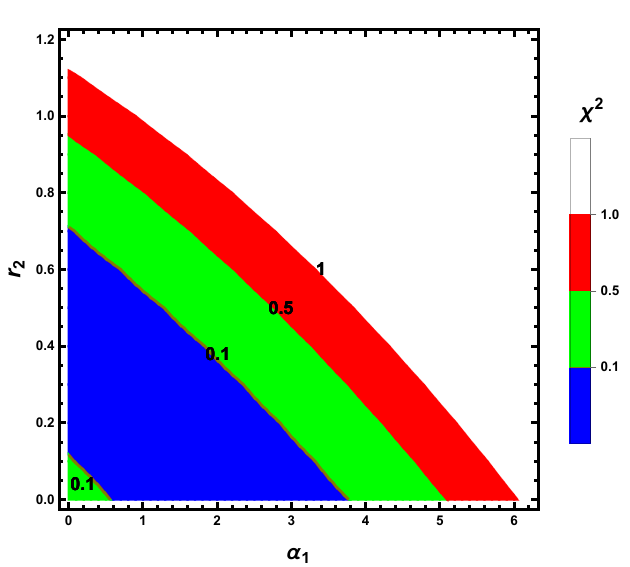}
        \caption{\label{profile1_4.297}$M=4. 297\times10^6 M_{\odot}$ and $D=8277$pc}
    \end{subfigure}
    \caption{\label{contour plots for Sgr A* considering profile 1}The figures represent  contour plots of $\chi^2$ (calculated using \Cref{chi-square} corresponding to $a_{min}$) for Keck (\Cref{profile1_3.951,profile1_3.975}) and GRAVITY collaboration (\Cref{profile1_4.261,profile1_4.297}) mass and distance measurements  for Sgr A* surrounded by plasma profile 1. In the figures, the red color region of $\alpha_1- r_2$ plane corresponds to  $0.5\leq\chi^2\leq1$, the region with $0.1\leq\chi^2\leq0.5$ is colored green and the blue region represents $0\leq\chi^2\leq0.1$. For all figures the angle of inclination $\theta_i=46^\circ$.}
\label{Fig15}
\end{figure}

    \item In \Cref{profile1_4.261,profile1_4.297}, $\alpha_1\leq6$ and $r_2\leq1.1$ for the region $\chi^2\leq1$. For the region $\chi^2\leq0.5$  we obtain the bounds, $\alpha_1\lesssim5$ and $r_2\lesssim0.9$. For the most observationally favored  region $\chi^2\leq0.1$,  we obtain $0.4\leq\alpha_1\leq3.6$ and $0.1\leq r_2\leq0.7$. % The bounds on $r_2$ and $\alpha_1$ indicates  for $\alpha_1<0.$4  the allowed range of $r_2$ excludes Kerr scenario in the most observationally favored region. 
This implies that for $\Delta\Theta_{th}$ evaluated by using the mass $M\text{ and distance }D$ reported by the GRAVITY collaboration, $r_2=0$ is not most observationally favored. This is further supported by our finding that, for mass and distance measurement  $M=4.261\times10^6 M_{\odot}\text{, } D=8246.7$pc, $\alpha_1=0.9,r_2=0.2$ and $a=0.81$  gives the lowest $\chi^2$. Also when mass and distance measurement   $M=4.297\times10^6M_{\odot}\text{, }D=8277$pc  is considered, $\alpha_1=1.6,r_2=0.2$ and $a=0.27$ gives the lowest $\chi^2$.  
\item Furthermore for  the mass and distance measurements, $(M=4.261\times10^6 M_{\odot}\text{, } D=8246.7\text{pc})$ we get  $0.1\leq a\leq1$ and  for $(M=4.297\times10^6 M_{\odot}\text{, } D=8277\text{pc})$ we get $0\leq a\leq1$  as the allowed spin  within $1-\sigma$.  
\end{itemize}
We next consider the case of profile 2. The contours for $\chi^2$ for different mass and distance measurements of Sgr A* are shown in \Cref{profile2_3.951,profile2_3.975,profile2_4.261,profile2_4.297}.

\begin{itemize}

 \item  Similar to the case of profile 1, the contour plots in \Cref{contour plots for M87* considering profile 2} and \Cref{contour plots for Sgr A* considering profile 2} are similar in nature.   For a given contour,  as $\alpha_2$ increases in the allowed range the upper bound of $r_2$ decreases and vice-versa.
 
\item In \Cref{profile2_3.951,profile2_3.975}, $\alpha_2\leq4.4$ and $r_2\lesssim1$ for the region $\chi^2\leq1$. For the region $\chi^2\leq0.5$, we obtain the bounds, $\alpha_2\leq3.6$ and $r_2\leq0.8$. For the most observationally favored  region $\chi^2\leq0.1$, $\alpha_2\leq2.4$ and $r_2\leq0.5$. Furthermore, $\alpha_1>4.4$ and $r_2>1$ are ruled outside $1-\sigma$. 
\item For the Keck team the mass and distance estimate $M=3.951\times10^6 M_{\odot},D=7935$pc,  $\alpha_2=0.4,r_2=0.1$ and $a=0.76$ gives the lowest $\chi^2$ and for the mass and distance estimate $M=3.975\times10^{6} M_{\odot}\text{ and }D=7959$pc, $\alpha_2=1.3,r_2=0$ and $a=0.3$ gives the lowest $\chi^2$.

\item In \Cref{profile2_4.261,profile2_4.297}, $\alpha_2\lesssim5.2$ and $r_2\leq1.1$ for the region $\chi^2\leq1$. For region $\chi^2\leq0.5$  we obtain the bounds, $\alpha_2\lesssim4.4$ and $r_2\lesssim0.95$. For the most observationally favored  region $\chi^2\leq0.1$,  we obtain $0.4\leq\alpha_2\leq3.2$ and $0.1\leq r_2\leq0.7$.  In the case of profile 2 also we observe that the Kerr scenario is not most observationally favored although it is allowed within $1-\sigma$.   Furthermore, $\alpha_1>5.2$ and $r_2>1.1$ are ruled outside  $1-\sigma$.   
\item For the GRAVITY collaboration  estimates $M=4.261\times10^6 M_{\odot},D=8246.7$pc,  $\alpha_2=1.2,r_2=0.2$ and $a=0.27$ gives the lowest $\chi^2$ and for $M=4.297\times10^{6} M_{\odot},D=8277$pc, $\alpha_2=0,r_2=0.5$ and $a=0.075$ gives the lowest $\chi^2$.
\item For all the Keck team and GRAVITY collaboration mass measurements the allowed range of spin within $\chi^2\leq1$ is $0\leq a\leq1$.
\end{itemize}

From the constraints obtained for Sgr A* on the plasma parameters $\alpha_1$ and $\alpha_2$ and the dilaton parameter $r_2$ considering profile 1 and profile 2 we note that very high plasma densities are not favored in both cases. Interestingly, for the case of Sgr A* we find in general, the bounds of $\alpha_2$ (\Cref{contour plots for Sgr A* considering profile 2}) are less than the  bounds on $\alpha_1$ (\Cref{contour plots for Sgr A* considering profile 1}). This, indicates profile 2 has stronger contracting effect on the shadow than profile 1 which is contrary to what we observed in case of M87* (refer \Cref{Constraining dilaton charge and plasma environment from M87* shadow,p1,p2}). Thus, the contracting effects of profile 1 and profile 2 are sensitive to mass $M$ and distance $D$. 
\begin{figure}[H]
\begin{subfigure}{0.5\textwidth}
        \centering
        \includegraphics[width=\linewidth]{ 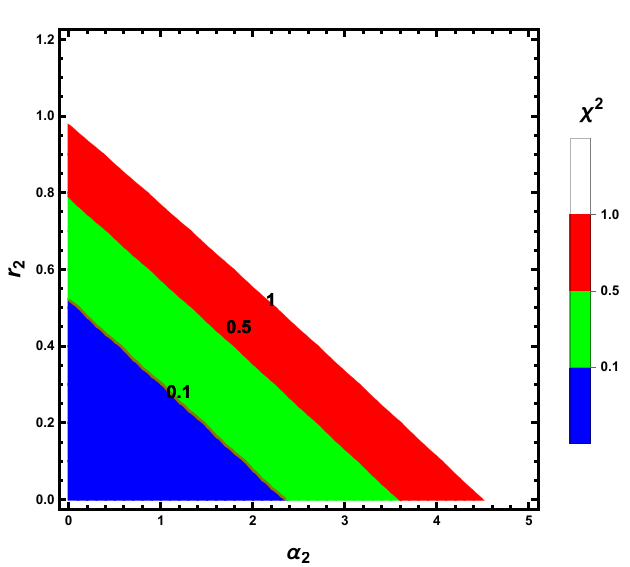}
        \caption{\label{profile2_3.951}$M=3. 951\times10^6$ and $D=7935$pc}
    \end{subfigure}\hfill
    \begin{subfigure}{0.5\textwidth}
        \centering
        \includegraphics[width=\linewidth]{ 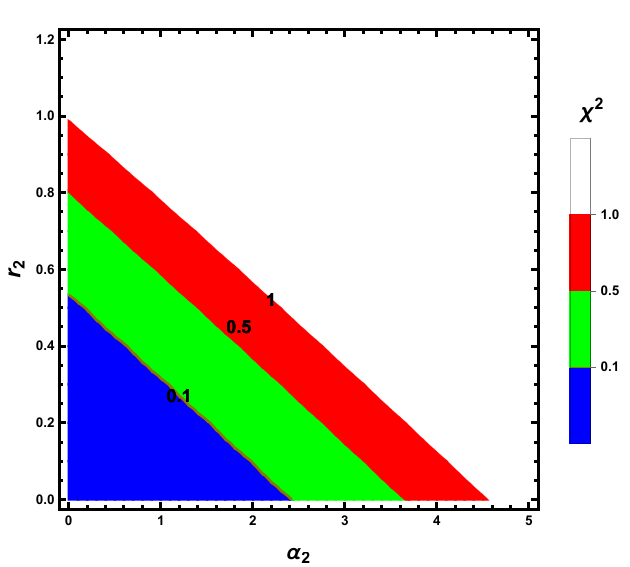}
        \caption{\label{profile2_3.975}$M=3. 975\times10^6$ and $D=7959$pc}
     \end{subfigure}\hfill
    \begin{subfigure}{0.5\textwidth}
        \centering
        \includegraphics[width=\linewidth]{ 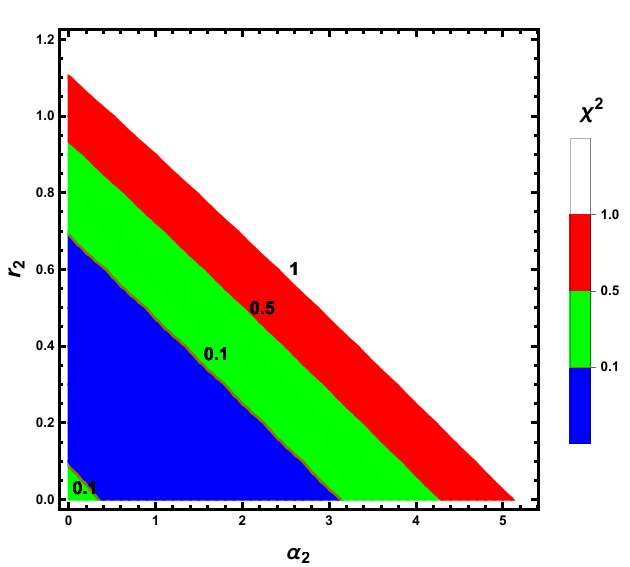}
        \caption{\label{profile2_4.261}$M=4. 261\times10^6$ and $D=8246. 7$pc}
    \end{subfigure}\hfill
    \begin{subfigure}{0.5\textwidth}
        \centering
        \includegraphics[width=\linewidth]{ 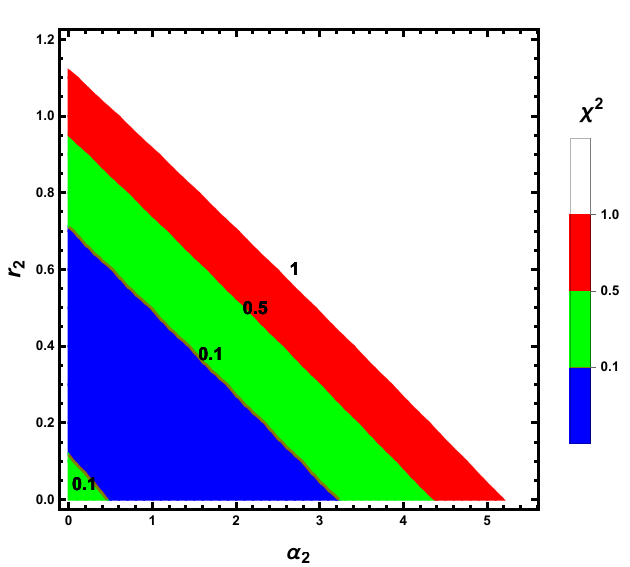}
        \caption{\label{profile2_4.297}$M=4. 297\times10^6$ and $D=8277$pc}
    \end{subfigure}
    \caption{\label{contour plots for Sgr A* considering profile 2}The figures represent  contour plots of $\chi^2$ (calculated using \Cref{chi-square} corresponding to $a_{min}$) for Keck (\Cref{profile2_3.951,profile2_3.975}) and GRAVITY collaboration (\Cref{profile2_4.261,profile2_4.297}) mass and distance measurements  for Sgr A* surrounded by plasma profile 2. In the figures, the red color region of $\alpha_1- r_2$ plane corressponds to  $0.5\leq\chi^2\leq1$, the region with $0.1\leq\chi^2\leq0.5$ is colored green and the blue region represents $0\leq\chi^2\leq0.1$. For all figures the angle of inclination $\theta_i=46^\circ$.}
\label{Fig16}
\end{figure}

The constraints on the plasma parameter and the dilaton charge corresponding  to two sets of  distance and mass measurements reported by the Keck team ($r_2>1$ and $\alpha_1>5.2$ ruled outside $1-\sigma$ for profile 1 and $r_2>1$ and $\alpha_1>4.4$ ruled outside $1-\sigma$ for profile 2) are  nearly same because of small variation in these measurements. This is also the case for the constraints obtained from the mass and distance measurements reported by the GRAVITY collaboration  ($r_2>1.1$ and $\alpha_1>6$ for profile 1 and $r_2>1.1$ and $\alpha_1>5.2$ for profile 2). More importantly using mass and distance measurements by GRAVITY collaboration, we find for the plasma profiles 1 and 2 an allowed range of plasma parameters ($\alpha_1\lesssim0.45$ and $\alpha_2\lesssim0.4$) for which the most observationally favored region excludes the Kerr scenario. This is not observed when  Keck team's mass and distance measurements are used. It is worth mentioning when the dilaton charge was constrained using  $\Delta\Theta_{obs}=48.7\mu as$  by the EHT collaboration and $M$ and $D$ reported by the GRAVITY collaboration, a non zero dilaton $r_2$ was favored even without plasma \cite{Sahoo:2023czj}. The allowed spin range within $\chi^2\leq1$ using both mass and distance estimates by the Keck team and the GRAVITY collaboration is $0\lesssim a\lesssim1$ for both plasma profiles 1 and 2.  Thus, the shadow observation fails to provide strong constraints on spin of Sgr A*. There are various spin estimates of Sgr A*  reported previouly in literature.  The spin of Sgr A* was estimated to be $a\sim0.92$ in \cite{2009ApJ...706..497M} , $a\sim0.5$ in  \cite{2012ApJ...755..133S}, $a\lesssim0.1$ in\cite{Fragione:2020khu}, $a\sim0.22$ in \cite{Belanger:2006gm}, $a>0.4$\cite{Meyer:2006fd}, $a\sim0.52$ in \cite{Genzel:2003as} and $a=0.9\pm0.06$ in \cite{Daly:2023axh}. Clearly, these previous spin estimates  are not  consistent.   
Lastly, one important finding for both cases of plasma profiles 1 and 2 is that, for plasma environments with higher density, present shadow observations disfavor higher dilaton charges. This was also observed in the case of M87*( refer \Cref{Constraining dilaton charge and plasma environment from M87* shadow}).    

\begin{figure}[t!]
\begin{subfigure}{0.5\textwidth}
        \centering
        \includegraphics[width=0.8\linewidth]{ 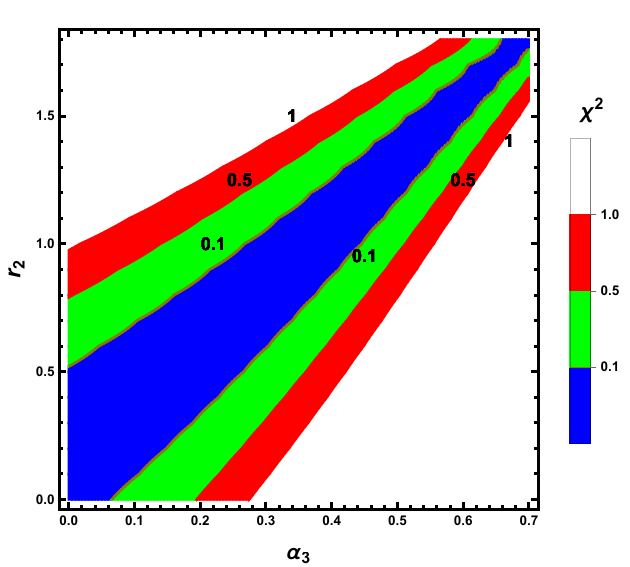}
        \caption{\label{profile3_3.951}$M=3.951\times10^6$ and $D=7935$pc}
    \end{subfigure}\hfill
    \begin{subfigure}{0.5\textwidth}
        \centering
        \includegraphics[width=0.8\linewidth]{ 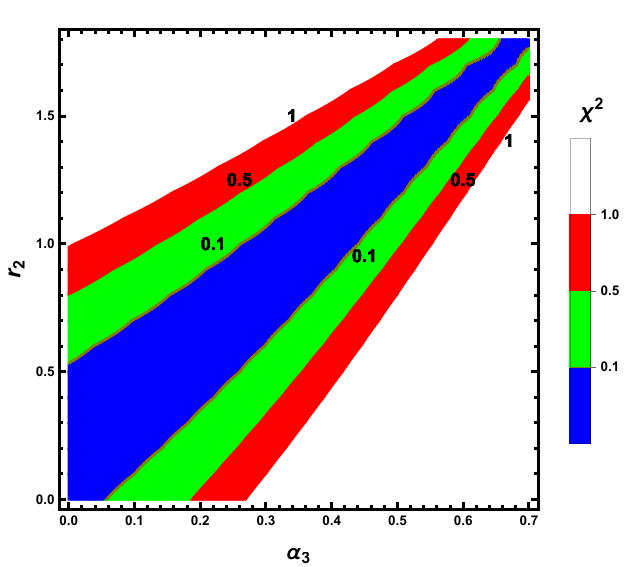}
        \caption{\label{profile3_3.975}$M=3.975\times10^6$ and $D=7959$pc}
     \end{subfigure}\hfill
    \begin{subfigure}{0.5\textwidth}
        \centering
        \includegraphics[width=0.8\linewidth]{ 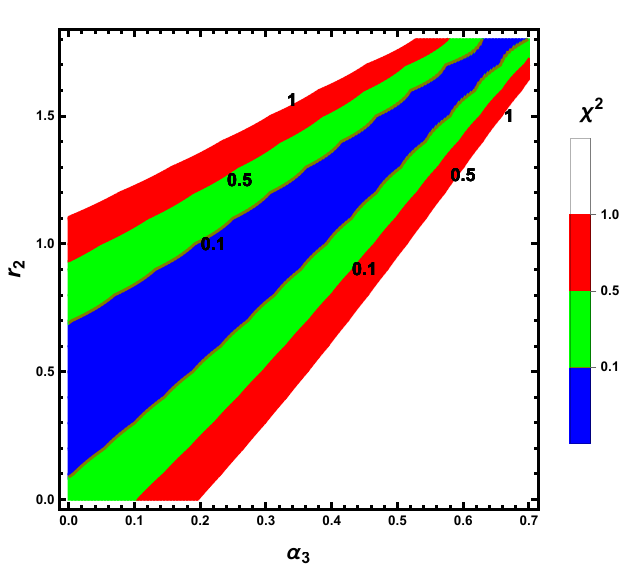}
        \caption{\label{profile3_4.261}$M=4.261\times10^6$ and $D=8246. 7$pc}
    \end{subfigure}\hfill
    \begin{subfigure}{0.5\textwidth}
        \centering
        \includegraphics[width=0.8\linewidth]{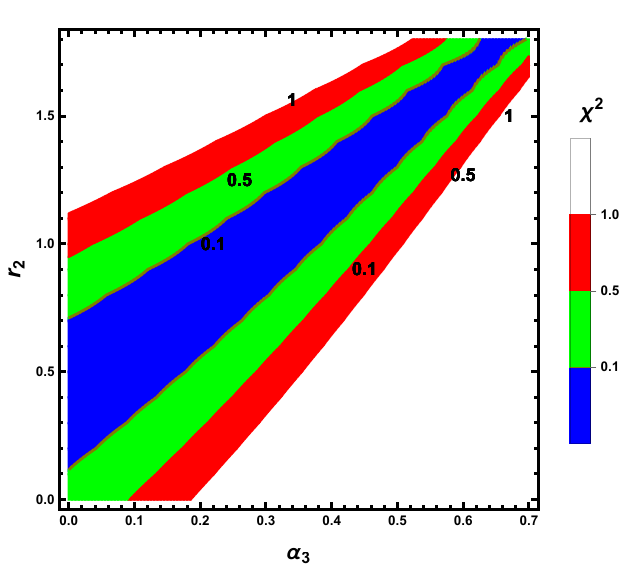}
        \caption{\label{profile3_4.297}$M=4.297\times10^6$ and $D=8277$pc}
    \end{subfigure}
    \caption{\label{contour plots for Sgr A* considering profile 3}The figures represent  contour plots of $\chi^2$ (calculated using \Cref{chi-square} corresponding to $a_{min}$) for Keck and GRAVITY collaboration mass and distance measurements  for Sgr A* surrounded by plasma profile 3. In the figures, the red color region of $\alpha_1- r_2$ plane corressponds to  $0.5\leq\chi^2\leq1$, the region with $0.1\leq\chi^2\leq0.5$ is colored green and the blue region represents $0\leq\chi^2\leq0.1$. For all figures the angle of inclination $\theta_i=46^\circ$.}
\label{Fig17-1}
\end{figure}

We now proceed to discuss the case of homogeneous plasma. In \Cref{profile3_3.951,profile3_3.975,profile3_4.261,profile3_4.297} we observe the following 
\begin{itemize}

\item Because of the expansive effect of homogeneous plasma on the shadow (refer \Cref{h1,h2,h3}) we can reproduce both the upper $1-\sigma$  and lower $1-\sigma$  values of $\Delta\Theta_{th}$. Hence, we get two $\chi^2=1,0.5 \text{ and }0.1$ contours.
\item For a given $r_2$, lower $1-\sigma$ of $\Delta\Theta_{obs}$ requires a smaller $\alpha_3$ compared to upper $1-\sigma$ value of $\Delta\Theta_{obs}$ (refer \Cref{profile3_3.951,profile3_3.975,profile3_4.261,profile3_4.297}).
\item Black hole with larger dilaton charge requires denser homogeneous plasma environment to explain the data.
\item For any given $r_2$ we always have some $\alpha_3$ (where $0\leq\alpha_3\leq1$) which can explain the data. Hence. for homogeneous plasma we cannot rule out any parameter space of $r_2\text{ and }\alpha_3$.

\item The above observations are true for mass and distance measurements reported by the Keck and GRAVITY collaboration. 

\end{itemize}

\subsection{Constrains on the plasma parameter from the electron number density and accretion rate estimates near M87* and Sgr A*}
{  The plasma parameter $\alpha$ relates the plasma frequency to the frequency at which the observations are done such that,
\begin{align}
    \label{12}
   \frac{ \omega_{P}(r, \theta)^2}{\omega_{0}^2}=\alpha \left(\frac{f(r)+g(\theta)}{\rho}\right)
\end{align}
where, $\alpha=\frac{ \omega_b^2}{\omega_{0}^2}$ and $\rho=r(r+r_2) + a^2 \cos^2\theta$. Also, the plasma frequency $\omega_P^2$ is related to the electron number density $\mathcal{N}$ by,
\begin{align}
\omega_P^2=\frac{4 \pi e^2}{m_e}\mathcal{N}(x^\mu)
\label{13}
\end{align}
Comparing Equations \ref{12} and \ref{13} we get,
\begin{align}
\frac{ \omega_{P}(r, \theta)^2}{\omega_{0}^2}=\alpha \left(\frac{f(r)+g(\theta)}{\rho}\right)=\frac{4 \pi e^2}{m_e \omega_{0}^2}\mathcal{N}(x^\mu)
\label{14}
\end{align}
where, $\omega_0$ is the frequency of light measured by an observer at rest at infinity. This is because, the frequency $\omega(x)$ of photon which reaches the observer at $r$ is given by,
\begin{align}
\omega(x^\mu)=\frac{\omega_0}{\sqrt{-g_{tt}}}>\omega_P (x^\mu)
\label{15}
\end{align}
In the limit $r\to \infty$, $g_{tt}\to -1$ for the Kerr-Sen metric and hence from Equation \ref{15} we obtain $\omega_\infty=\omega_0$. The BH shadow related observations made by the EHT are at $\nu_0=230$ GHz which implies, that $\omega_0=2\pi \nu_0$.

To proceed further, we now consider different plasma profiles given in Equations \ref{eqn:p1}, \ref{eqn:p2} and \ref{eqn:p3} which can be used in Equation \ref{14}.
To constrain $\alpha_1$,  $\alpha_2$ and  $\alpha_3$ we also need information about the electron number density $\mathcal{N}$. We first discuss the constrains on the plasma parameters from the available number density/accretion rate estimates of M87* and then for Sgr A*.

\paragraph{M87*:}
The EHT team has reported that the electron number density $\mathcal{N}$ around M87* is $\sim 10^4-10^7 ~ \rm {cm^{-3}}$ \cite{EventHorizonTelescope:2021srq}. Such a number density has been estimated in the emission region $r\simeq 5 r_g$, based on a one-zone isothermal sphere model \cite{EventHorizonTelescope:2021srq}. Using this in Equation \ref{14} along with $\theta=\pi/2$, we summarize below the results for the three plasma profiles:
\begin{table}[h!]
\centering
\begin{tabular}{|c|c|c|c|c|c|}
\hline
\textbf{$r_2$} & \textbf{$\mathcal{N}$} &  {$r$ (in $R_g$)}  & \textbf{$\alpha_1$} & \textbf{$\alpha_2$} & \textbf{$\alpha_3$}\\
\hline
0 & $\rm 10^7 cm^{-3}$  &  5 & $1.7\times 10^{-7}$ & $3.81 \times 10^{-7}$  & $1.52 \times 10^{-8}$\\
\hline
0 & $\rm 10^4 cm^{-3}$  &  5 & $1.7\times 10^{-10}$ & $3.81 \times 10^{-10}$  & $1.52 \times 10^{-11}$\\
\hline
2 & $\rm 10^7 cm^{-3}$  &  5 & $2.37 \times 10^{-7}$ & $ 5.33\times 10^{-7}$  & $ 1.52\times 10^{-8}$\\
\hline
2 & $\rm 10^4 cm^{-3}$  &  5 & $2.37 \times 10^{-10}$ & $ 5.33 \times 10^{-10}$  & $ 1.52\times 10^{-11}$\\
\hline
\end{tabular}
\caption{Estimates of the plasma parameters $\alpha_1$, $\alpha_2$ and $\alpha_3$ from the electron number density data of M87*.}
\label{m87}
\end{table}
%%%%%%%%%%%%%%%%%%%%%%%%%%%%%%%%%%%%%%%%%%%%%%%%%%%%%%%%%%%%%%%%%

Assuming radial free fall of the plasma particles from rest at infinity (profile 1) \cite{shapiro1974accretion}, Perlick et al.\cite{Perlick:2015vta} has given a relation between $\alpha_1$ and the mass accretion rate, such that,
\begin{align}
\alpha_1=\frac{e^2\dot{M}c^3}{\sqrt{2} m_e m_p \omega_0^2 G^2 M^2}
\label{16}
\end{align}
where $e$ is the electron's charge, $m_e$ is the electron's mass, $m_p$ is the mass of the proton, $c$ is the speed of light, $M$ and $\dot{M}$ are respectively the black hole mass and accretion rate. 
Although this was derived for a Schwarzschild BH, it holds equally well for the Kerr or the Kerr-Sen BH \cite{shapiro1974accretion} (see Section \ref{p1}). 
The EHT Team reports the mass accretion rate of M87* to be ${\rm (3-20)\times 10^{-4} M_\odot yr^{-1}}$ \cite{EventHorizonTelescope:2021srq}. Using, $\dot{M}\simeq {\rm 20\times 10^{-4} M_\odot yr^{-1} }$ in Equation \ref{16}, we get $\alpha_1\simeq 2.58 \times 10^{-10}$ while $\dot{M}\simeq {\rm 3\times 10^{-4} M_\odot yr^{-1} }$ in Equation \ref{16} yields $\alpha_1\simeq 3.87 \times 10^{-11}$. A more recent study \cite{Drew:2025euq} reports that the accretion rate of M87* lies in the range: ${\rm 4 \times 10^{-5} M_\odot yr^{-1} } - {\rm 4 \times 10^{-1} M_\odot yr^{-1} }$.
If we assume M87* to be a Kerr BH with $\dot{M}\simeq {\rm 4 \times 10^{-1} M_\odot yr^{-1} }$ in Equation \ref{16}, then $\alpha_1\simeq 5.15\times 10^{-8}$ is obtained.
The above discussion elucidates that the electron number density and the accretion rate estimates of M87* indicate a very small plasma parameter, irrespective of the choice of the plasma profile. These estimates are not very precise and there have been other estimates on the accretion rate of M87* \cite{Kuo:2014pqa,Feng:2016gif}, but all of them indicate a very low plasma density.

In Section \ref{Sec5-1} we have discussed constrains on the plasma parameter from the angular diameter of the shadow of M87*. This data is more precise than the number density estimate. From Fig. \ref{contour plots for M87* 6.2 considering profile 1}, Fig. \ref{angular_6.2} and Fig. \ref{homogeneous_6.2} we note the following observationally allowed ranges for the plasma parameters: $0\lesssim \alpha_1 \lesssim 2.5$, $0\lesssim \alpha_2 \lesssim 3.7$ and $0\lesssim \alpha_3 \lesssim 0.3$ (assuming $r_2=0$ and $M\sim 6.2\times 10^9 M_\odot$), which are consistent with the findings based on the number density/accretion rate estimates. We also note from the aforesaid figures that for the inhomogeneous profiles (profile 1 and profile 2), the presence of a dilaton charge decreases the allowed range of $\alpha$. Now, if we additionally take into account the previous electron number density or the accretion rate estimates, then we should consider $\alpha_1, \alpha_2\approx 0$ (as discussed above) which allows larger values of $r_2$ within the observed 1-$\sigma$ (Fig. \ref{Fig12-1} and Fig. \ref{Fig13-1}) and $r_{2,max}\simeq 0.48$. Moreover, even with $\alpha\approx 0$, the Kerr scenario or mildly charged dilaton BHs are more favored compared to BHs with high dilaton charges (blue shaded region in Figs. \ref{Fig12-1} and \ref{Fig13-1}). 

\paragraph{Sgr A*:} For Sgr A*, the electron number density has not been precisely estimated, rather, there exist estimates on the accretion rate \cite{Quataert:1999ng,Marrone:2006vu,Bower:2018wsw,Yoon:2020yew}. Knowing $\dot{M}$ and using Equation \ref{16} we can determine $\alpha_1$. In the table below we provide estimates of $\alpha_1$:
\begin{table}[h!]
\centering
\begin{tabular}{|c|c|}
\hline
\textbf{$\dot{M} $} & \textbf{$\alpha_1$} \\
\hline
$ {\rm 10^{-7} M_\odot yr^{-1}-10^{-9} M_\odot yr^{-1}}$ \cite{Yoon:2020yew} & $2.95\times 10^{-8} -2.95\times 10^{-10}$  \\
\hline
$\sim {\rm 10^{-8} M_\odot yr^{-1}} $ \cite{Bower:2018wsw} &  $2.95\times 10^{-9}$ \\
\hline
$ {\rm 2\times 10^{-7} M_\odot yr^{-1}-2\times 10^{-9} M_\odot yr^{-1}}$ \cite{Marrone:2006vu}  & $5.89\times 10^{-8} -5.89\times 10^{-10}$ \\ \hline
$\lesssim {\rm 10^{-5} M_\odot yr^{-1}}$ \cite{Quataert:1999ng} & $\lesssim 2.95\times 10^{-6}$  \\
\hline
\end{tabular}
\caption{Estimates of the plasma parameter $\alpha_1$ from the accretion rate of Sgr A*.}
\label{SgrA}
\end{table}
%%%%%%%%%%%%%%%%%%%%%%%%%%%%%%%%%%%%%%%%%%%%%%%%%%%%%%%%%%%%%%%%%

Table \ref{SgrA} reveal that $\alpha_1$ is indeed very small. Similarly, $\alpha_2$ and $\alpha_3$ are also expected to be small as in the case of M87*. However, the shadow angular diameter of Sgr A* allows $0\lesssim \alpha_1 \lesssim 6$ (Fig. \ref{Fig15}) consistent with the $\alpha_1$ results obtained from the accretion rate estimates. When one takes into account both the shadow angular diameter and the accretion rate estimate, then Fig. \ref{Fig15} reveals that for $\alpha_1\approx 0$ the dilaton parameter as high as $r_2\sim 1.1$ is allowed within the observed 1-$\sigma$ interval. The figure also reveals that the Kerr scenario or mild/moderately charged dilaton BHs are favored equally by the present observations.

From the above discussion we note that the Kerr-Sen scenario could have been completely ruled out if we had a high density plasma surounding M87* and Sgr A*. Since this is not the case, the degeneracy between the Kerr scenario and the Kerr-Sen scenario with small $r_2$ cannot be lifted from the observed shadow angular diameters of M87* and Sgr A* with the current level of precision. Our results reveal that the background geometry has a stronger effect on the shadow size than the surrounding plasma, as also mentioned in \cite{EventHorizonTelescope:2025dua}.

We now give an order of magnitude estimate of the plasma densities near M87* and Sgr A* (assuming that they are Kerr BHs) which can have an observable effect on the shadow at 230 GHz.
From Figs. \ref{Fig12-1}, \ref{Fig13-1}, \ref{Fig15} and \ref{Fig16} of the revised version (the inhomogeneous plasma profiles) we note that if the plasma parameter $\approx 0.8$, then it has an observable effect on the shadow. For the homogeneous plasma profile, we note that, if the plasma parameter $\alpha_3\approx 0.1$ (Figs. \ref{Fig14} and \ref{Fig17-1}), then it has an observable effect on the shadow. Assuming these values of the plasma parameters and considering Equation \ref{14}, we now give an estimate of the electron number density and the plasma density at the equatorial plane near the BH (e.g. $r\sim 5 r_g$).  

\begin{table}[h!]
\centering
\begin{tabular}{|c|c|c|c|}
\hline
Plasma parameter  & {$r$ (in $R_g$)} &  $\mathcal{N}$ (in ${\rm cm^{-3}}$)  & $n=m_p n_e$ (in gm ${\rm cm^{-3}}$) \\
\hline
$\alpha_1\approx 0.8$ & 5  & $4.69\times 10^{13}$  & $7.83\times 10^{-11}$\\
\hline
$\alpha_2\approx 0.8$ & 5  & $6.29\times 10^{13}$  &  $1.05\times 10^{10}$\\
\hline
$\alpha_3\approx 0.1$ &  5 & $6.56\times 10^{13}$  &  $1.1\times 10^{10}$\\
\hline
\end{tabular}
\caption{Estimates of the electron number density and plasma density near the BH which can have an observable effect on the shadow.}
\label{T1}
\end{table}
Table \ref{T1} reveals that $\alpha\approx 1$ results in very high electron number densities which are far larger than the estimates reported in the literature \cite{EventHorizonTelescope:2021srq}. The above $\mathcal{N}$ may decrease by an order if we consider a non-zero dilaton charge which is still several order of magnitude larger than the existing estimates \cite{EventHorizonTelescope:2021srq}. Following \cite{Perlick:2015vta} we can provide estimates of the mass accretion rate from $\mathcal{N}$ reported in Table \ref{T1}. 
Assuming profile 1, it can be shown that \cite{Perlick:2015vta},
\begin{align}
\mathcal{N}(r)=\frac{\dot{M}}{4\sqrt{2r_g}\pi m_p c }\frac{1}{r^{3/2}}
\label{17-1}
\end{align}
Using Equation \ref{17-1} and $r\sim 5 r_g$ we get the following $\dot{M}$ for M87* and Sgr A* (Table \ref{T2}), 
\begin{table}[h!]
\centering
\begin{tabular}{|c|c|c|}
\hline
Source  &  $\mathcal{N}$ (in ${\rm cm^{-3}}$)  & $\dot{M}$ (in $M_\odot{\rm yr^{-1}}$) \\
\hline
M87*   & $4.69\times 10^{13}$  & $6.19\times 10^6$\\
\hline
Sgr A*  & $4.69\times 10^{13}$  &  $2.58 $\\
\hline
\end{tabular}
\caption{Accretion rate estimates for M87* and Sgr A* assuming profile 1 and $\alpha_1\approx 0.8$.}
\label{T2}
\end{table}
which are far larger than previous estimates \cite{EventHorizonTelescope:2021srq,Yoon:2020yew,Bower:2018wsw,Marrone:2006vu,Quataert:1999ng}.
%%%%%%%%%%%%%%%%%%%%%%%%%%%%%%%%%%%%%%%%%%%%%%%%%%%%%%%%%%%%%%%%%

Thus, we note that the light propagation condition $\omega(x)>\omega_P(x)$ provides an upper bound on the plasma parameters (Figs. \ref{Fig1} and \ref{Fig5}) which is further constrained from the shadow angular diameter data and even more more constrained when the electron number density or accretion rate estimates are taken into account. This indicates that the background metric has a much stronger effect on the shadow than the surrounding plasma environment.

}
%{  Moreover, it is however important to note that we considered a maximum offset of 10\% between the emission ring and the shadow diameter \cite{EventHorizonTelescope:2019dse,EventHorizonTelescope:2019ggy} which gave us $\Delta\Theta_{obs}\simeq 37.8 \pm 3 \mu as$. If we consider a smaller offset (e.g. 5\%) then the upper bound on $r_2$, i.e. $r_{2,max}$ will decrease and the Kerr scenario will be more preferred.}

\section{Conclusion\label{conclusion}}

In this work, we  have studied the influence of plasma on the shadow characteristics of the Kerr-Sen black hole,  a black hole solution \cite{Sen:1992ua} derived from Einstein-Maxwell Dilaton-Axion (EMDA) gravity.  By considering different plasma profiles,  we  derived constraints on the dilaton charge and plasma parameters which enable us to understand the interplay between the background metric and the plasma environment in explaining the observed images of M87*\cite{EventHorizonTelescope:2019dse} and Sgr A*\cite{EventHorizonTelescope:2022apq}. Astrophysical black holes  are not isolated systems and are in general surrounded by accretion disks consisting of plasma. Thus, the present work is a continuation of our previous work\cite{Sahoo:2023czj} where the effect of plasma was not considered. Further, this would also contribute to the understanding of black hole properties in alternative gravity models. We have calculated the equations describing the shape of the shadow of a Kerr Sen black hole surrounded by non-magnetized, pressureless plasma and studied the variation of the shadow with dilaton charge $r_2$, plasma parameter $\alpha_i$  $(i=1,2,3 \text{ for profile 1, 2 and 3 respectively} )$, spin $a$ and angle of inclination $\theta_i$. We observed that:

\begin{itemize}
    \item With increase in the dilaton charge $r_2$ the   shadow size decreases irrespective of the spin $a$ of the black hole, the angle of inclination of the observer $\theta_i$ and the plasma environment considered. This indicates that the contracting effect of $r_2$ on the shadow size is generic.
    \item The distortion of the shadow from its circular shape  increases as both the spin $a$ and inclination angle $\theta_i$ are increased. The shift in the  geometric center of the shadow increases with increase in the spin $a$.   
    \item The increase in plasma parameters $\alpha_1$ (for profile 1 represented by \Cref{eqn:p1}) and $\alpha_2$ (for profile 2 represented by \Cref{eqn:p2})  decrease the  shadow size irrespective of the choice of $r_2$, $a$ and $\theta_i$.  Thus, the non-homogeneous plasma environments   have a contracting effect on the shadow. 
    \item However, the degree of contraction of the shadow size due to the plasma profiles 1 and 2 depend on the mass $M$ and the distance $D$ of the black hole from the observer. 
    \item  Furthermore,  as the plasma parameters ($\alpha_1\text{ and }\alpha_2$) were increased the shape of the shadow became more circular even for high spin $a$ and inclination angle $\theta_i$ but the shift in geometric center persisted. This indicates that the distortion of the shadow from circular shape lessens in presence of plasma, i.e. inhomogeneous plasma environments may obliterate the effects of high spin and inclination.
    \item   A light ray propagating in the Kerr Sen spacetime can reach an observer only when plasma parameters $\alpha_1$ and $\alpha_2$ are less than a certain upper bound. These bounds, which are directly related to the density of  the plasma (refer \Cref{sec:4}),  strongly  depend on the dilaton charge of the Kerr Sen black hole (refer \Cref{p1,p2}). As $r_2$ increases, the upper bounds on both $\alpha_1$ and $\alpha_2$ decrease (refer \Cref{F1,F2}). Thus Kerr Sen black holes surrounded by  plasma environments with density above the corresponding theoretical upper bounds  may not be detectable by electromagnetic observations.  
    \item The homogeneous plasma (plasma profile 3 represented by \Cref{eqn:p3}) on the other hand, has an expanding effect on the shadow size and 
unlike the inhomogeneous plasma environments, the shadow continues to be non-circular due to the combined effect of spin $a$ and inclination $\theta_i$ in the presence of moderately dense homogeneous plasma environments.
    \end{itemize}
In order to decipher the observationally favored dilaton charge and plasma parameters we have used the EHT  observations related to the shadows of M87* and Sgr A*. The EHT collaboration reported the   angular diameter of the primary ring of M87* to be $(42\pm3)\mu as$ with a maximum offset of $10\%$ between the   primary ring and the  shadow \cite{EventHorizonTelescope:2019ggy}. We have taken into account the maximum offset, thus the shadow angular diameter of M87* is taken to be $\Delta\Theta_{obs}=(37.8\pm3)\mu as$. For Sgr A*, the EHT collaboration reported the angular diameter of the shadow to be $\Delta\Theta_{obs}=(48.7\pm 7)\mu as$.   We have used the methodology described towards the end of \Cref{eht} to obtain the constrains. We have used previously reported mass $M$, distance $D$ and inclination angle $\theta_i$ measurements  for M87* and Sgr A* to calculate their theoretical angular diameter.

For M87* the distance $D=16.8$Mpc\cite{Blakeslee:2009tc,Bird:2010rd,Cantiello:2018ffy}, inclination angle $\theta_i=17^\circ$\cite{Tamburini:2019vrf} and mass estimates from gas dynamics studies\cite{EventHorizonTelescope:2019dse,Walsh:2013uua}, and stellar dynamics studies\cite{EventHorizonTelescope:2019dse,Gebhardt:2000fk,Gebhardt:2009cr} are used to calculate the theoretical shadow angular diameter $\Delta\Theta_{th}$. For purpose of completeness and comparison, we also constrained the dilaton charge and the plasma parameter using the mass estimated by the EHT team \cite{EventHorizonTelescope:2019dse,EventHorizonTelescope:2019ggy}. We report the following important results:
\begin{itemize}
    \item  The allowed range of dilaton charge within the observed $1-\sigma$ interval depends on the plasma environment. In the case of non-homogeneous plasma (which are also more realistic), as the plasma density is increased (i.e., $\alpha_1$ and $\alpha_2$ are enhanced in the allowed range), the corresponding upper bounds of $r_2$ required to reproduce the observed image of M87*  within $1-\sigma$ decreases. This is due to the contracting effect of the shadow size due to $\alpha_1$, $\alpha_2$ and $r_2$.
    \item  However, the presence of homogeneous plasma (which has an expanding effect on shadow size) enhances the upper bounds of $r_2$ such that the observed shadow is reproduced. When homogeneous plasma is considered, we find that the entire allowed range of $\alpha_3$, (i.e., $0\leq\alpha_3\leq 1$) lies within the observed $1-\sigma$, hence no constrains are obtained.
    \item For the non-homogeneous  profiles 1 and  2 when mass estimate of M87* from gas dynamics studies ($M\sim 3.5\times10^9M_{\odot}$) is used to calculate $\Delta\Theta_{th}$, the observed anuglar diameter $\Delta\Theta_{obs}$ of M87* could not be reproduced even within 1-$\sigma$ for any combination of $r_2$, $a$, $\alpha_1$ and $\alpha_2$. This is consistent with our previous work \cite{Sahoo:2023czj} where the effects of plasma were not taken into account. However, with homogeneous plasma $\Delta\Theta_{obs}$ can be reproduced if $\alpha_3\gtrsim 0.72$  is considered. Thus, the observed shadow of M87* rules out homogeneous plasma parameter $\alpha_3\lesssim 0.72$ while inhomogeneous plasma environments  rule out $M\sim 3.5\times10^9M_{\odot}$ which is also inconsistent with the mass estimate of M87* by the EHT team, i.e.,  $M= 6.5\pm 0.7\times10^9M_{\odot}$.
    \item When the mass of M87* based on stellar dynamics studies ($M=6.2\times10^9 M_{\odot}$) is used to calculate $\Delta\Theta_{th}$, we observe that  $r_2>0.48$ fails to reproduce the image of M87* within the observed $1-\sigma$, irrespective of the inhomogeneous plasma environment considered. Further, the shadow of M87* rules out $\alpha_1>2.5$ and  $\alpha_2>3.8$. 

\item {  Independent constraints on the plasma parameter can be obtained from the accretion rate and the electron number density estimates near M87* \cite{EventHorizonTelescope:2021srq} which indicate that the magnitude of the plasma parameters is indeed very small, e.g., $10^{-11}\lesssim \alpha_1, \alpha_2 \lesssim 10^{-7}$. This is also true for the homogeneous plasma parameter $\alpha_3$, which in turn implies that the plasma density near M87* is extremely low. 
Thus, if we take into account both the shadow angular diameter data of M87* and the electron number density estimate near M87*, then the plasma parameter is $\approx 0$ which allows larger values of $r_2$ within the observed 1-$\sigma$, e.g. if the plasma density is described by the inhomogeneous profiles, then $0\lesssim r_{2}\lesssim 0.48$ is possible. Therefore, with the current level of precision of the shadow angular diameter data of M87*, one cannot distinguish between the Kerr scenario and the mildly charged Kerr-Sen scenario.}

%The lowest value of $\chi^2$ considering plasma profile 1 is obtained for $\alpha_1\sim0.1$, $r_2\sim0$ and $a\sim0.2$. When plasma profile 2 is considered  $\alpha_2\sim0.2$, $r_2\sim0$ and $a\sim0.1$ minimizes the $\chi^2$. Thus, the Kerr scenario is more favored by the EHT observations of M87* for both the inhomogeneous plasma profiles considered here, in agreement with our previous work \cite{Sahoo:2023czj}. The silhouette of M87* fails to constrain its dilaton charge if a homogeneous plasma environment is considered.
    \item The allowed spin range for M87* using the stellar dynamics mass estimate within the observed $1-\sigma$ interval is found to be  $0\lesssim a\lesssim0.3$. The spin constraint is the nearly the same for both the inhomogeneous plasma profiles. The previously reported spin estimates of M87*\cite{Nemmen:2019idv,Tamburini:2019vrf} are higher than the bounds obtained here which may be due to the  weak dependence of angular diameter of the shadow on the spin at low inclination angles (which for M87* is $\theta_i\approx 17^\circ$) \cite{Johannsen:2010ru,Takahashi:2004xh,Nemmen:2019idv}. 
\begin{comment}

    \item For the sake of completeness, we also considered the mass of M87*  estimated by the EHT collaboration (i.e., $M=6.5\times10^9 M_{\odot}$) assuming M87* as a Kerr black hole. For plasma profile 1, $\alpha_1>3.6$ ( refer \Cref{contour plots for M87* 6.5 considering profile 1})  and for plasma profile 2,  $\alpha_2>5.4$ (\Cref{angular_6.5}) are ruled out outside $1-\sigma$. The constraints on the dilaton charge $r_2$ in this case should not be considered for testing the Kerr Sen scenario as the mass has been estimated already assuming the Kerr scenario. The allowed range of spin of M87* obtained considering this mass for both the profiles turn out to be $0\lesssim a\lesssim0.9$.
\end{comment}

\end{itemize}

Next we summarize our constraints on the dilaton charge $r_2$ and the plasma parameters using the EHT observations of Sgr A*. In case of Sgr A* also, we used previously reported  measurements of mass, distance and inclination angle to calculate the theoretical angular diameter $\Delta\Theta_{th}$ of the shadow.
The inclination angle  $\theta_i\approx46^\circ$ \cite{abuter2019geometric} while the mass and distance estimates are provided by the Keck team and the GRAVITY collaboration.  

%used in calculating the theoretical angular diameter of the shadow of Sgr A*  

%using results reported by the  Keck collaboration keeping the red shift parameter free$(M=3.975\times10^6 M_{\odot},D=7959)$pc \cite{Do:2019txf} and another set by keeping it fixed to unity$(M=3.951\times10^6 M_{\odot},D=7935)$pc\cite{Do:2019txf}. The mass and distance measurements using astrometry observations reported by the GRAVITY collaboration without considering optical aberrations$(M=4.261\times10^6 M_{\odot},D=8246.7)$pc \cite{GRAVITY:2020gka,GRAVITY:2021xju} and taking into account effects of optical aberrations$(M=4.297\times10^6 M_{\odot},D=8277)$pc \cite{GRAVITY:2021xju,GRAVITY:2020gka,1976ApAvni}.
 \begin{itemize}
     \item For the two sets of Keck collaboration mass and distance measurements, the constraints on the dilaton charge and the plasma environment obtained are nearly the same because the two sets of mass and distance measurements are not very different. When $\Delta\Theta_{th}$ is calculated with the aforesaid mass and distance estimates, we find that $r_2>1$ is ruled out outside the observed $1-\sigma$ interval irrespective of the inhomogeneous plasma environment considered. Further, the shadow of Sgr A* disfavors $\alpha_1>5.2$ and $\alpha_2>4.4$. 
     \item  When the mass and distance reported by the Keck collaboration that fixes the redshift parameter to unity is used to calculate $\Delta\Theta_{th}$, $\alpha_1\sim0.3$, $r_2\sim0.2$ and $a\sim0.45$ minimizes the $\chi^2$ for plasma profile 1 while $\alpha_2\sim0.4$, $r_2\sim0.1$ and $a\sim0.76$ minimizes the $\chi^2$ for plasma profile 2, indicating a marginal preference towards the EMDA scenario. When mass and distance reported by the Keck team by keeping redshift parameter free is considered, the lowest value of $\chi^2$ is obtained for $\alpha_1\sim1.2, r_2\sim 0$ and $a\sim0.7$ (for profile 1) while $\alpha_2\sim1.3, r_2\sim 0$ and $a\sim0.3$ minimizes the $\chi^2$ (for plasma profile 2), thereby favoring the Kerr scenario. Thus, either GR or Kerr-Sen black holes with mild dilaton  charges is favored.
     \item When the two sets of  mass and distance measurements of Sgr A* by the GRAVITY collaboration  are considered to calculate $\Delta\Theta_{th}$,   $r_2>1.1$ is ruled out outside $1-\sigma$ irrespective of the choice of the inhomogeneous plasma environments, while $\alpha_1>6$ and  $\alpha_2>5.2$ are ruled out by the image of Sgr A*.
    \item When $\Delta\Theta_{th}$ is calculated using mass and distance estimates of Sgr A* by the GRAVITY collaboration without considering optical aberration, $\alpha_1\sim0.9$, $r_2\sim0.2$ and $a\sim0.81$ minimizes the $\chi^2$ (for plasma profile 1) while $\alpha_2\sim1.2$, $r_2\sim0.2$ and $a\sim0.27$ minimizes the $\chi^2$ (for plasma profile 2). When mass and distance assuming optical aberration is considered, the lowest value of $\chi^2$ is obtained for $\alpha_1\sim1.6,r_2\sim 0.2$ and $a\sim0.27$ for profile 1 and $\alpha_2\sim0,r_2\sim 0.5$ and $a\sim0.075$ for profile 2. Thus, in this case the EMDA scenario is favored compared to GR, although the Kerr scenario is allowed within the 1-$\sigma$ and this result holds good for both the inhomogeneous plasma enviromments.
    \item {  Independent constrains on the plasma parameter can be obtained from the accretion rate estimates of Sgr A*, which indicates that $10^{-10}\lesssim \alpha\lesssim 10^{-8}$ thereby revealing that the plasma density near Sgr A* is too low to have an observable effect on the shadow. This in turn allows dilaton charge in the range $0\lesssim r_2\lesssim 1.1$ to address the shadow angular diameter within the observed 1-$\sigma$. Thus, with the current precision, the Kerr scenario or mild/moderately charged dilaton BHs are favored equally by the shadow related observations of Sgr A*.}
     \item  The present analysis could not constrain the spin of Sgr A*.  Interestingly, previous estimates assuming Sgr A* to be a Kerr BH yield diverse results covering the entire allowed range \cite{2009ApJ...706..497M,2012ApJ...755..133S,Fragione:2020khu,Belanger:2006gm,Meyer:2006fd,Genzel:2003as,Daly:2023axh} which are often not mutually consistent. This probably indicates revisiting the spin estimate of Sgr A*. 

     \end{itemize}
{  The above discussion elucidates that even with the current precision of the data, shadow related observations rule out very large values of dilaton charge for M87* and Sgr A*
in presence of inhomogeneous plasma environments (which seem to be more realistic). Moreover, the shadow of M87* and Sgr A* rule out very dense inhomogeneous plasma environments surrounding these objects. Further constraints on the plasma density comes from the  
accretion rate or the electron number density estimate near these objects which indicate that the plasma density in their vicinity is too low to have an observable effect on the shadow.   
In fact, the Kerr-Sen scenario could have been completely ruled out if we had a high density plasma surounding M87* and Sgr A*. 
Since this is not the case, within the observed 1-$\sigma$ interval, the present data cannot distinguish between the Kerr and the Kerr-Sen black holes with mild/moderate dilaton charges. This also indicates that that the background geometry has a stronger effect on the shadow size than the surrounding plasma which has been further supported by the persistent shadow of M87* reported by the EHT team \cite{EventHorizonTelescope:2024dhe}. Thus, the shadow provides a cleaner probe to the background spacetime compared to other electromagnetic observations, e.g. the continuum spectrum or the Fe-line. Hence, the finding that M87* and Sgr A* do not harbour large dilaton charges seem to be quite robust. The present findings are consistent with earlier results from observations related to the continuum spectrum \cite{Banerjee:2020qmi}, the quasi-periodic oscillations \cite{Dasgupta:2025fuh} and the observed jet-power \cite{Banerjee:2020ubc}, although using different BH samples.
%However, shadow related observations fail to provide strong constraints on the spin of M87* and Sgr A*, which may be attributed to the low inclination angle of these objects.  
With the availability of more precise data, the error bars on the observed angular diameters are expected to reduce which in turn will enable us to establish stronger constrains on the background spacetime.}
%While the shadow of M87* marginally favors the Kerr scenario, the shadow of Sgr A* exhibits a slight preference towards the Kerr-Sen scenario. 
%In this regard, it may be worthwhile to mention that the shadow provides a cleaner probe to the background spacetime compared to other electromagnetic observations, e.g. the continuum spectrum or the Fe-line, which has been further supported by the persistent shadow of M87* reported by the EHT team \cite{EventHorizonTelescope:2024dhe}. Hence, the finding that BHs with large dilaton charges are disfavored by shadow related observations seem to be quite robust. The present findings are consistent with earlier results from observations related to the continuum spectrum \cite{Banerjee:2020qmi}, the quasi-periodic oscillations \cite{Dasgupta:2025fuh} and the observed jet-power \cite{Banerjee:2020ubc}, although using different BH samples.
%Shadow related observations fail to provide strong constraints on the spin of M87* and Sgr A*, which may be attributed to the low inclination angle of these objects. With the availability of more precise data, the error bars on the observed angular diameters are expected to reduce which in turn will enable us to establish stronger constrains on the background spacetime.}

\section*{Acknowledgements}
Research of I.B. is funded by the Start-Up
Research Grant from SERB, DST, Government of India
(Reg. No. SRG/2021/000418).\\

\bibliography{references.bib,reference-2.bib}

\providecommand{\href}[2]{#2}\begingroup\raggedright\begin{thebibliography}{100}

\bibitem{LIGOScientific:2016aoc}
{\bfseries LIGO Scientific, Virgo} Collaboration, B.~P. Abbott {\em et~al.},
  ``{Observation of Gravitational Waves from a Binary Black Hole Merger},''
  \href{http://dx.doi.org/10.1103/PhysRevLett.116.061102}{{\em Phys. Rev.
  Lett.} {\bfseries 116} no.~6, (2016) 061102},
  \href{http://arxiv.org/abs/1602.03837}{{\ttfamily arXiv:1602.03837 [gr-qc]}}.

\bibitem{LIGOScientific:2017vwq}
{\bfseries LIGO Scientific, Virgo} Collaboration, B.~P. Abbott {\em et~al.},
  ``{GW170817: Observation of Gravitational Waves from a Binary Neutron Star
  Inspiral},'' \href{http://dx.doi.org/10.1103/PhysRevLett.119.161101}{{\em
  Phys. Rev. Lett.} {\bfseries 119} no.~16, (2017) 161101},
  \href{http://arxiv.org/abs/1710.05832}{{\ttfamily arXiv:1710.05832 [gr-qc]}}.

\bibitem{LIGOScientific:2018mvr}
{\bfseries LIGO Scientific, Virgo} Collaboration, B.~P. Abbott {\em et~al.},
  ``{GWTC-1: A Gravitational-Wave Transient Catalog of Compact Binary Mergers
  Observed by LIGO and Virgo during the First and Second Observing Runs},''
  \href{http://dx.doi.org/10.1103/PhysRevX.9.031040}{{\em Phys. Rev. X}
  {\bfseries 9} no.~3, (2019) 031040},
  \href{http://arxiv.org/abs/1811.12907}{{\ttfamily arXiv:1811.12907
  [astro-ph.HE]}}.

\bibitem{EventHorizonTelescope:2019dse}
{\bfseries Event Horizon Telescope} Collaboration, K.~Akiyama {\em et~al.},
  ``{First M87 Event Horizon Telescope Results. I. The Shadow of the
  Supermassive Black Hole},''
  \href{http://dx.doi.org/10.3847/2041-8213/ab0ec7}{{\em Astrophys. J. Lett.}
  {\bfseries 875} (2019) L1}, \href{http://arxiv.org/abs/1906.11238}{{\ttfamily
  arXiv:1906.11238 [astro-ph.GA]}}.

\bibitem{EventHorizonTelescope:2019ggy}
{\bfseries Event Horizon Telescope} Collaboration, K.~Akiyama {\em et~al.},
  ``{First M87 Event Horizon Telescope Results. VI. The Shadow and Mass of the
  Central Black Hole},'' \href{http://dx.doi.org/10.3847/2041-8213/ab1141}{{\em
  Astrophys. J. Lett.} {\bfseries 875} no.~1, (2019) L6},
  \href{http://arxiv.org/abs/1906.11243}{{\ttfamily arXiv:1906.11243
  [astro-ph.GA]}}.

\bibitem{EventHorizonTelescope:2019jan}
{\bfseries Event Horizon Telescope} Collaboration, K.~Akiyama {\em et~al.},
  ``{First M87 Event Horizon Telescope Results. III. Data Processing and
  Calibration},'' \href{http://dx.doi.org/10.3847/2041-8213/ab0c57}{{\em
  Astrophys. J. Lett.} {\bfseries 875} no.~1, (2019) L3},
  \href{http://arxiv.org/abs/1906.11240}{{\ttfamily arXiv:1906.11240
  [astro-ph.GA]}}.

\bibitem{EventHorizonTelescope:2019pgp}
{\bfseries Event Horizon Telescope} Collaboration, K.~Akiyama {\em et~al.},
  ``{First M87 Event Horizon Telescope Results. V. Physical Origin of the
  Asymmetric Ring},'' \href{http://dx.doi.org/10.3847/2041-8213/ab0f43}{{\em
  Astrophys. J. Lett.} {\bfseries 875} no.~1, (2019) L5},
  \href{http://arxiv.org/abs/1906.11242}{{\ttfamily arXiv:1906.11242
  [astro-ph.GA]}}.

\bibitem{EventHorizonTelescope:2019ths}
{\bfseries Event Horizon Telescope} Collaboration, K.~Akiyama {\em et~al.},
  ``{First M87 Event Horizon Telescope Results. IV. Imaging the Central
  Supermassive Black Hole},''
  \href{http://dx.doi.org/10.3847/2041-8213/ab0e85}{{\em Astrophys. J. Lett.}
  {\bfseries 875} no.~1, (2019) L4},
  \href{http://arxiv.org/abs/1906.11241}{{\ttfamily arXiv:1906.11241
  [astro-ph.GA]}}.

\bibitem{EventHorizonTelescope:2019uob}
{\bfseries Event Horizon Telescope} Collaboration, K.~Akiyama {\em et~al.},
  ``{First M87 Event Horizon Telescope Results. II. Array and
  Instrumentation},'' \href{http://dx.doi.org/10.3847/2041-8213/ab0c96}{{\em
  Astrophys. J. Lett.} {\bfseries 875} no.~1, (2019) L2},
  \href{http://arxiv.org/abs/1906.11239}{{\ttfamily arXiv:1906.11239
  [astro-ph.IM]}}.

\bibitem{EventHorizonTelescope:2021dqv}
{\bfseries Event Horizon Telescope} Collaboration, P.~Kocherlakota {\em
  et~al.}, ``{Constraints on black-hole charges with the 2017 EHT observations
  of M87*},'' \href{http://dx.doi.org/10.1103/PhysRevD.103.104047}{{\em Phys.
  Rev. D} {\bfseries 103} no.~10, (2021) 104047},
  \href{http://arxiv.org/abs/2105.09343}{{\ttfamily arXiv:2105.09343 [gr-qc]}}.

\bibitem{EventHorizonTelescope:2022apq}
{\bfseries Event Horizon Telescope} Collaboration, K.~Akiyama {\em et~al.},
  ``{First Sagittarius A* Event Horizon Telescope Results. II. EHT and
  Multiwavelength Observations, Data Processing, and Calibration},''
  \href{http://dx.doi.org/10.3847/2041-8213/ac6675}{{\em Astrophys. J. Lett.}
  {\bfseries 930} no.~2, (2022) L13}.

\bibitem{EventHorizonTelescope:2022exc}
{\bfseries Event Horizon Telescope} Collaboration, K.~Akiyama {\em et~al.},
  ``{First Sagittarius A* Event Horizon Telescope Results. IV. Variability,
  Morphology, and Black Hole Mass},''
  \href{http://dx.doi.org/10.3847/2041-8213/ac6736}{{\em Astrophys. J. Lett.}
  {\bfseries 930} no.~2, (2022) L15}.

\bibitem{EventHorizonTelescope:2022urf}
{\bfseries Event Horizon Telescope} Collaboration, K.~Akiyama {\em et~al.},
  ``{First Sagittarius A* Event Horizon Telescope Results. V. Testing
  Astrophysical Models of the Galactic Center Black Hole},''
  \href{http://dx.doi.org/10.3847/2041-8213/ac6672}{{\em Astrophys. J. Lett.}
  {\bfseries 930} no.~2, (2022) L16}.

\bibitem{EventHorizonTelescope:2022wkp}
{\bfseries Event Horizon Telescope} Collaboration, K.~Akiyama {\em et~al.},
  ``{First Sagittarius A* Event Horizon Telescope Results. I. The Shadow of the
  Supermassive Black Hole in the Center of the Milky Way},''
  \href{http://dx.doi.org/10.3847/2041-8213/ac6674}{{\em Astrophys. J. Lett.}
  {\bfseries 930} no.~2, (2022) L12}.

\bibitem{EventHorizonTelescope:2022wok}
{\bfseries Event Horizon Telescope} Collaboration, K.~Akiyama {\em et~al.},
  ``{First Sagittarius A* Event Horizon Telescope Results. III. Imaging of the
  Galactic Center Supermassive Black Hole},''
  \href{http://dx.doi.org/10.3847/2041-8213/ac6429}{{\em Astrophys. J. Lett.}
  {\bfseries 930} no.~2, (2022) L14}.

\bibitem{EventHorizonTelescope:2022xqj}
{\bfseries Event Horizon Telescope} Collaboration, K.~Akiyama {\em et~al.},
  ``{First Sagittarius A* Event Horizon Telescope Results. VI. Testing the
  Black Hole Metric},'' \href{http://dx.doi.org/10.3847/2041-8213/ac6756}{{\em
  Astrophys. J. Lett.} {\bfseries 930} no.~2, (2022) L17}.

\bibitem{Will:2014kxa}
C.~M. Will, ``{The Confrontation between General Relativity and Experiment},''
  \href{http://dx.doi.org/10.12942/lrr-2014-4}{{\em Living Rev. Rel.}
  {\bfseries 17} (2014) 4}, \href{http://arxiv.org/abs/1403.7377}{{\ttfamily
  arXiv:1403.7377 [gr-qc]}}.

\bibitem{Wright:2024mco}
M.~Wright, J.~Janquart, and N.~K. Johnson-McDaniel, ``{Effect of Deviations
  from General Relativity on Searches for Gravitational-wave Microlensing and
  Type II Strong Lensing},''
  \href{http://dx.doi.org/10.3847/1538-4357/ad9d3e}{{\em Astrophys. J.}
  {\bfseries 981} no.~2, (2025) 133},
  \href{http://arxiv.org/abs/2403.08957}{{\ttfamily arXiv:2403.08957 [gr-qc]}}.

\bibitem{LIGOScientific:2016lio}
{\bfseries LIGO Scientific, Virgo} Collaboration, B.~P. Abbott {\em et~al.},
  ``{Tests of general relativity with GW150914},''
  \href{http://dx.doi.org/10.1103/PhysRevLett.116.221101}{{\em Phys. Rev.
  Lett.} {\bfseries 116} no.~22, (2016) 221101},
  \href{http://arxiv.org/abs/1602.03841}{{\ttfamily arXiv:1602.03841 [gr-qc]}}.
  [Erratum: Phys.Rev.Lett. 121, 129902 (2018)].

\bibitem{Berti:2015itd}
E.~Berti {\em et~al.}, ``{Testing General Relativity with Present and Future
  Astrophysical Observations},''
  \href{http://dx.doi.org/10.1088/0264-9381/32/24/243001}{{\em Class. Quant.
  Grav.} {\bfseries 32} (2015) 243001},
  \href{http://arxiv.org/abs/1501.07274}{{\ttfamily arXiv:1501.07274 [gr-qc]}}.

\bibitem{Martin:2012bt}
J.~Martin, ``{Everything You Always Wanted To Know About The Cosmological
  Constant Problem (But Were Afraid To Ask)},''
  \href{http://dx.doi.org/10.1016/j.crhy.2012.04.008}{{\em Comptes Rendus
  Physique} {\bfseries 13} (2012) 566--665},
  \href{http://arxiv.org/abs/1205.3365}{{\ttfamily arXiv:1205.3365
  [astro-ph.CO]}}.

\bibitem{Weinberg:1988cp}
S.~Weinberg, ``{The Cosmological Constant Problem},''
  \href{http://dx.doi.org/10.1103/RevModPhys.61.1}{{\em Rev. Mod. Phys.}
  {\bfseries 61} (1989) 1--23}.

\bibitem{SupernovaSearchTeam:1998fmf}
{\bfseries Supernova Search Team} Collaboration, A.~G. Riess {\em et~al.},
  ``{Observational evidence from supernovae for an accelerating universe and a
  cosmological constant},'' \href{http://dx.doi.org/10.1086/300499}{{\em
  Astron. J.} {\bfseries 116} (1998) 1009--1038},
  \href{http://arxiv.org/abs/astro-ph/9805201}{{\ttfamily
  arXiv:astro-ph/9805201}}.

\bibitem{SupernovaCosmologyProject:1998vns}
{\bfseries Supernova Cosmology Project} Collaboration, S.~Perlmutter {\em
  et~al.}, ``{Measurements of $\Omega$ and $\Lambda$ from 42 high redshift
  supernovae},'' \href{http://dx.doi.org/10.1086/307221}{{\em Astrophys. J.}
  {\bfseries 517} (1999) 565--586},
  \href{http://arxiv.org/abs/astro-ph/9812133}{{\ttfamily
  arXiv:astro-ph/9812133}}.

\bibitem{Penrose:1964wq}
R.~Penrose, ``{Gravitational collapse and space-time singularities},''
  \href{http://dx.doi.org/10.1103/PhysRevLett.14.57}{{\em Phys. Rev. Lett.}
  {\bfseries 14} (1965) 57--59}.

\bibitem{Hawking:1970zqf}
S.~W. Hawking and R.~Penrose, ``{The Singularities of gravitational collapse
  and cosmology},'' \href{http://dx.doi.org/10.1098/rspa.1970.0021}{{\em Proc.
  Roy. Soc. Lond. A} {\bfseries 314} (1970) 529--548}.

\bibitem{Hawking:1966vg}
S.~W. Hawking, ``{Singularities in the universe},''
  \href{http://dx.doi.org/10.1103/PhysRevLett.17.444}{{\em Phys. Rev. Lett.}
  {\bfseries 17} (1966) 444--445}.

\bibitem{Page:2004xp}
D.~N. Page, ``{Hawking radiation and black hole thermodynamics},''
  \href{http://dx.doi.org/10.1088/1367-2630/7/1/203}{{\em New J. Phys.}
  {\bfseries 7} (2005) 203},
  \href{http://arxiv.org/abs/hep-th/0409024}{{\ttfamily arXiv:hep-th/0409024}}.

\bibitem{Bambi:2017khi}
C.~Bambi, \href{http://dx.doi.org/10.1007/978-981-10-4524-0}{{\em {Black Holes:
  A Laboratory for Testing Strong Gravity}}}.
\newblock Springer, 2017.

\bibitem{Gralla:2019xty}
S.~E. Gralla, D.~E. Holz, and R.~M. Wald, ``{Black Hole Shadows, Photon Rings,
  and Lensing Rings},''
  \href{http://dx.doi.org/10.1103/PhysRevD.100.024018}{{\em Phys. Rev. D}
  {\bfseries 100} no.~2, (2019) 024018},
  \href{http://arxiv.org/abs/1906.00873}{{\ttfamily arXiv:1906.00873
  [astro-ph.HE]}}.

\bibitem{Teo:2003ltt}
E.~Teo, ``{Spherical Photon Orbits Around a Kerr Black Hole},''
  \href{http://dx.doi.org/10.1023/A:1026286607562}{{\em Gen. Rel. Grav.}
  {\bfseries 35} no.~11, (2003) 1909--1926}.

\bibitem{perlick2000ray}
V.~Perlick, {\em Ray optics, Fermat’s principle, and applications to general
  relativity}, vol.~61.
\newblock Springer Science \& Business Media, 2000.

\bibitem{Perlick:2004tq}
V.~Perlick, ``{Gravitational lensing from a spacetime perspective},'' {\em
  Living Rev. Rel.} {\bfseries 7} (2004) 9. \url{10.12942/lrr-2004-9}.

\bibitem{Synge:1966okc}
J.~L. Synge, ``{The Escape of Photons from Gravitationally Intense Stars},''
  \href{http://dx.doi.org/10.1093/mnras/131.3.463}{{\em Mon. Not. Roy. Astron.
  Soc.} {\bfseries 131} no.~3, (1966) 463--466}.

\bibitem{1974IAUS...64..132B}
J.~M. {Bardeen}, ``{Properties of Black Holes Relevant to Their Observation
  (invited Paper)},'' in {\em Gravitational Radiation and Gravitational
  Collapse}, C.~{Dewitt-Morette}, ed., vol.~64 of {\em IAU Symposium}, p.~132.
\newblock Jan., 1974.

\bibitem{Perlick:2015vta}
V.~Perlick, O.~Y. Tsupko, and G.~S. Bisnovatyi-Kogan, ``{Influence of a plasma
  on the shadow of a spherically symmetric black hole},''
  \href{http://dx.doi.org/10.1103/PhysRevD.92.104031}{{\em Phys. Rev. D}
  {\bfseries 92} no.~10, (2015) 104031},
  \href{http://arxiv.org/abs/1507.04217}{{\ttfamily arXiv:1507.04217 [gr-qc]}}.

\bibitem{Perlick:2017fio}
V.~Perlick and O.~Y. Tsupko, ``{Light propagation in a plasma on Kerr
  spacetime: Separation of the Hamilton-Jacobi equation and calculation of the
  shadow},'' \href{http://dx.doi.org/10.1103/PhysRevD.95.104003}{{\em Phys.
  Rev. D} {\bfseries 95} no.~10, (2017) 104003},
  \href{http://arxiv.org/abs/1702.08768}{{\ttfamily arXiv:1702.08768 [gr-qc]}}.

\bibitem{Rogatko:2002qe}
M.~Rogatko, ``{Positivity of energy in Einstein-Maxwell axion dilaton
  gravity},'' \href{http://dx.doi.org/10.1088/0264-9381/19/20/303}{{\em Class.
  Quant. Grav.} {\bfseries 19} (2002) 5063--5072},
  \href{http://arxiv.org/abs/hep-th/0209126}{{\ttfamily arXiv:hep-th/0209126}}.

\bibitem{Sen:1992ua}
A.~Sen, ``{Rotating charged black hole solution in heterotic string theory},''
  \href{http://dx.doi.org/10.1103/PhysRevLett.69.1006}{{\em Phys. Rev. Lett.}
  {\bfseries 69} (1992) 1006--1009},
  \href{http://arxiv.org/abs/hep-th/9204046}{{\ttfamily arXiv:hep-th/9204046}}.

\bibitem{Gyulchev:2006zg}
G.~N. Gyulchev and S.~S. Yazadjiev, ``{Kerr-Sen dilaton-axion black hole
  lensing in the strong deflection limit},''
  \href{http://dx.doi.org/10.1103/PhysRevD.75.023006}{{\em Phys. Rev. D}
  {\bfseries 75} (2007) 023006},
  \href{http://arxiv.org/abs/gr-qc/0611110}{{\ttfamily arXiv:gr-qc/0611110}}.

\bibitem{An:2017hby}
J.~An, J.~Peng, Y.~Liu, and X.-H. Feng, ``{Kerr-Sen Black Hole as Accelerator
  for Spinning Particles},''
  \href{http://dx.doi.org/10.1103/PhysRevD.97.024003}{{\em Phys. Rev. D}
  {\bfseries 97} no.~2, (2018) 024003},
  \href{http://arxiv.org/abs/1710.08630}{{\ttfamily arXiv:1710.08630 [gr-qc]}}.

\bibitem{Younsi:2016azx}
Z.~Younsi, A.~Zhidenko, L.~Rezzolla, R.~Konoplya, and Y.~Mizuno, ``{New method
  for shadow calculations: Application to parametrized axisymmetric black
  holes},'' \href{http://dx.doi.org/10.1103/PhysRevD.94.084025}{{\em Phys. Rev.
  D} {\bfseries 94} no.~8, (2016) 084025},
  \href{http://arxiv.org/abs/1607.05767}{{\ttfamily arXiv:1607.05767 [gr-qc]}}.

\bibitem{Hioki:2008zw}
K.~Hioki and U.~Miyamoto, ``{Hidden symmetries, null geodesics, and photon
  capture in the Sen black hole},''
  \href{http://dx.doi.org/10.1103/PhysRevD.78.044007}{{\em Phys. Rev. D}
  {\bfseries 78} (2008) 044007},
  \href{http://arxiv.org/abs/0805.3146}{{\ttfamily arXiv:0805.3146 [gr-qc]}}.

\bibitem{Narang:2020bgo}
A.~Narang, S.~Mohanty, and A.~Kumar, ``{Test of Kerr-Sen metric with black hole
  observations},'' \href{http://arxiv.org/abs/2002.12786}{{\ttfamily
  arXiv:2002.12786 [gr-qc]}}.

\bibitem{Jana:2023sil}
S.~Jana and S.~Kar, ``{Shadows in dyonic Kerr-Sen black holes},''
  \href{http://arxiv.org/abs/2303.14513}{{\ttfamily arXiv:2303.14513 [gr-qc]}}.

\bibitem{Sahoo:2023czj}
S.~K. Sahoo, N.~Yadav, and I.~Banerjee, ``{Imprints of
  Einstein-Maxwell-dilaton-axion gravity in the observed shadows of Sgr A* and
  M87*},'' \href{http://dx.doi.org/10.1103/PhysRevD.109.044008}{{\em Phys. Rev.
  D} {\bfseries 109} no.~4, (2024) 044008},
  \href{http://arxiv.org/abs/2305.14870}{{\ttfamily arXiv:2305.14870 [gr-qc]}}.

\bibitem{Banerjee:2020qmi}
I.~Banerjee, B.~Mandal, and S.~SenGupta, ``{Implications of
  Einstein\textendash{}Maxwell dilaton\textendash{}axion gravity from the black
  hole continuum spectrum},''
  \href{http://dx.doi.org/10.1093/mnras/staa3232}{{\em Mon. Not. Roy. Astron.
  Soc.} {\bfseries 500} no.~1, (2020) 481--492},
  \href{http://arxiv.org/abs/2007.13980}{{\ttfamily arXiv:2007.13980 [gr-qc]}}.

\bibitem{Tripathi:2021rwb}
A.~Tripathi, B.~Zhou, A.~B. Abdikamalov, D.~Ayzenberg, and C.~Bambi,
  ``{Constraints on Einstein-Maxwell dilaton-axion gravity from X-ray
  reflection spectroscopy},''
  \href{http://dx.doi.org/10.1088/1475-7516/2021/07/002}{{\em JCAP} {\bfseries
  07} (2021) 002}, \href{http://arxiv.org/abs/2103.07593}{{\ttfamily
  arXiv:2103.07593 [astro-ph.HE]}}.

\bibitem{Dasgupta:2025fuh}
A.~Dasgupta, N.~Tiwari, and I.~Banerjee, ``{Signatures of Einstein-Maxwell
  dilaton-axion gravity from the observed quasi-periodic oscillations in black
  holes},'' \href{http://arxiv.org/abs/2503.02708}{{\ttfamily arXiv:2503.02708
  [gr-qc]}}.

\bibitem{Banerjee:2020ubc}
I.~Banerjee, B.~Mandal, and S.~SenGupta, ``{Signatures of Einstein-Maxwell
  dilaton-axion gravity from the observed jet power and the radiative
  efficiency},'' \href{http://dx.doi.org/10.1103/PhysRevD.103.044046}{{\em
  Phys. Rev. D} {\bfseries 103} no.~4, (2021) 044046},
  \href{http://arxiv.org/abs/2007.03947}{{\ttfamily arXiv:2007.03947 [gr-qc]}}.

\bibitem{Abramowicz:2011xu}
M.~A. Abramowicz and P.~C. Fragile, ``{Foundations of Black Hole Accretion Disk
  Theory},'' \href{http://dx.doi.org/10.12942/lrr-2013-1}{{\em Living Rev.
  Rel.} {\bfseries 16} (2013) 1},
  \href{http://arxiv.org/abs/1104.5499}{{\ttfamily arXiv:1104.5499
  [astro-ph.HE]}}.

\bibitem{PhysRevLett.17.455}
D.~O. Muhleman and I.~D. Johnston, ``Radio propagation in the solar
  gravitational field,''
  \href{http://dx.doi.org/10.1103/PhysRevLett.17.455}{{\em Phys. Rev. Lett.}
  {\bfseries 17} (Aug, 1966) 455--458}.
  \url{https://link.aps.org/doi/10.1103/PhysRevLett.17.455}.

\bibitem{Muhleman:1970zz}
D.~O. Muhleman, R.~D. Ekers, and E.~B. Fomalont, ``{Radio Interferometric Test
  of the General Relativistic Light Bending Near the Sun},''
  \href{http://dx.doi.org/10.1103/PhysRevLett.24.1377}{{\em Phys. Rev. Lett.}
  {\bfseries 24} (1970) 1377--1380}.

\bibitem{2000rofp.book.....P}
V.~{Perlick}, {\em {Ray Optics, Fermat's Principle, and Applications to General
  Relativity}}, vol.~61.
\newblock 2000.

\bibitem{Bisnovatyi-Kogan:2010flt}
G.~S. Bisnovatyi-Kogan and O.~Y. Tsupko, ``{Gravitational lensing in a
  non-uniform plasma},''
  \href{http://dx.doi.org/10.1111/j.1365-2966.2010.16290.x}{{\em Mon. Not. Roy.
  Astron. Soc.} {\bfseries 404} (2010) 1790--1800},
  \href{http://arxiv.org/abs/1006.2321}{{\ttfamily arXiv:1006.2321
  [astro-ph.CO]}}.

\bibitem{Tsupko:2013cqa}
O.~Y. Tsupko and G.~S. Bisnovatyi-Kogan, ``{Gravitational lensing in plasma:
  Relativistic images at homogeneous plasma},''
  \href{http://dx.doi.org/10.1103/PhysRevD.87.124009}{{\em Phys. Rev. D}
  {\bfseries 87} no.~12, (2013) 124009},
  \href{http://arxiv.org/abs/1305.7032}{{\ttfamily arXiv:1305.7032
  [astro-ph.CO]}}.

\bibitem{2013ApSS.346..513M}
V.~S. {Morozova}, B.~J. {Ahmedov}, and A.~A. {Tursunov}, ``{Gravitational
  lensing by a rotating massive object in a plasma},''
  \href{http://dx.doi.org/10.1007/s10509-013-1458-6}{{\em ApSS} {\bfseries 346}
  no.~2, (Aug., 2013) 513--520}.

\bibitem{Crisnejo:2018uyn}
G.~Crisnejo and E.~Gallo, ``{Weak lensing in a plasma medium and gravitational
  deflection of massive particles using the Gauss-Bonnet theorem. A unified
  treatment},'' \href{http://dx.doi.org/10.1103/PhysRevD.97.124016}{{\em Phys.
  Rev. D} {\bfseries 97} no.~12, (2018) 124016},
  \href{http://arxiv.org/abs/1804.05473}{{\ttfamily arXiv:1804.05473 [gr-qc]}}.

\bibitem{Crisnejo:2018ppm}
G.~Crisnejo, E.~Gallo, and A.~Rogers, ``{Finite distance corrections to the
  light deflection in a gravitational field with a plasma medium},''
  \href{http://dx.doi.org/10.1103/PhysRevD.99.124001}{{\em Phys. Rev. D}
  {\bfseries 99} no.~12, (2019) 124001},
  \href{http://arxiv.org/abs/1807.00724}{{\ttfamily arXiv:1807.00724 [gr-qc]}}.

\bibitem{Crisnejo:2019xtp}
G.~Crisnejo, E.~Gallo, and J.~R. Villanueva, ``{Gravitational lensing in
  dispersive media and deflection angle of charged massive particles in terms
  of curvature scalars and energy-momentum tensor},''
  \href{http://dx.doi.org/10.1103/PhysRevD.100.044006}{{\em Phys. Rev. D}
  {\bfseries 100} no.~4, (2019) 044006},
  \href{http://arxiv.org/abs/1905.02125}{{\ttfamily arXiv:1905.02125 [gr-qc]}}.

\bibitem{Crisnejo:2019ril}
G.~Crisnejo, E.~Gallo, and K.~Jusufi, ``{Higher order corrections to deflection
  angle of massive particles and light rays in plasma media for stationary
  spacetimes using the Gauss-Bonnet theorem},''
  \href{http://dx.doi.org/10.1103/PhysRevD.100.104045}{{\em Phys. Rev. D}
  {\bfseries 100} no.~10, (2019) 104045},
  \href{http://arxiv.org/abs/1910.02030}{{\ttfamily arXiv:1910.02030 [gr-qc]}}.

\bibitem{Liu:2016eju}
C.-Q. Liu, C.-K. Ding, and J.-L. Jing, ``{Effects of Homogeneous Plasma on
  Strong Gravitational Lensing of Kerr Black Holes},''
  \href{http://dx.doi.org/10.1088/0256-307X/34/9/090401}{{\em Chin. Phys.
  Lett.} {\bfseries 34} no.~9, (2017) 090401},
  \href{http://arxiv.org/abs/1610.02128}{{\ttfamily arXiv:1610.02128 [gr-qc]}}.

\bibitem{Rogers:2015dla}
A.~Rogers, ``{Frequency-dependent effects of gravitational lensing within
  plasma},'' \href{http://dx.doi.org/10.1093/mnras/stv903}{{\em Mon. Not. Roy.
  Astron. Soc.} {\bfseries 451} no.~1, (2015) 17--25},
  \href{http://arxiv.org/abs/1505.06790}{{\ttfamily arXiv:1505.06790 [gr-qc]}}.

\bibitem{Rogers:2016xcc}
A.~Rogers, ``{Escape and Trapping of Low-Frequency Gravitationally Lensed Rays
  by Compact Objects within Plasma},''
  \href{http://dx.doi.org/10.1093/mnras/stw2829}{{\em Mon. Not. Roy. Astron.
  Soc.} {\bfseries 465} no.~2, (2017) 2151--2159},
  \href{http://arxiv.org/abs/1611.01269}{{\ttfamily arXiv:1611.01269 [gr-qc]}}.

\bibitem{Er:2013efa}
X.~Er and S.~Mao, ``{Effects of plasma on gravitational lensing},''
  \href{http://dx.doi.org/10.1093/mnras/stt2043}{{\em Mon. Not. Roy. Astron.
  Soc.} {\bfseries 437} no.~3, (2014) 2180--2186},
  \href{http://arxiv.org/abs/1310.5825}{{\ttfamily arXiv:1310.5825
  [astro-ph.CO]}}.

\bibitem{Tsukamoto:2014tja}
N.~Tsukamoto, Z.~Li, and C.~Bambi, ``{Constraining the spin and the deformation
  parameters from the black hole shadow},''
  \href{http://dx.doi.org/10.1088/1475-7516/2014/06/043}{{\em JCAP} {\bfseries
  06} (2014) 043}, \href{http://arxiv.org/abs/1403.0371}{{\ttfamily
  arXiv:1403.0371 [gr-qc]}}.

\bibitem{Tsukamoto:2017fxq}
N.~Tsukamoto, ``{Black hole shadow in an asymptotically-flat, stationary, and
  axisymmetric spacetime: The Kerr-Newman and rotating regular black holes},''
  \href{http://dx.doi.org/10.1103/PhysRevD.97.064021}{{\em Phys. Rev. D}
  {\bfseries 97} no.~6, (2018) 064021},
  \href{http://arxiv.org/abs/1708.07427}{{\ttfamily arXiv:1708.07427 [gr-qc]}}.

\bibitem{Kumar:2022zky}
P.~Kumar and P.~Beniamini, ``{Gravitational lensing in the presence of plasma
  scattering with application to Fast Radio Bursts},''
  \href{http://dx.doi.org/10.1093/mnras/stad160}{{\em Mon. Not. Roy. Astron.
  Soc.} {\bfseries 520} no.~1, (2023) 247--258},
  \href{http://arxiv.org/abs/2208.03332}{{\ttfamily arXiv:2208.03332
  [astro-ph.HE]}}.

\bibitem{Crisnejo:2022qlv}
G.~Crisnejo, E.~Gallo, E.~F. Boero, and O.~M. Moreschi, ``{Perturbative and
  numerical approach to plasma strong lensing},''
  \href{http://dx.doi.org/10.1103/PhysRevD.107.084041}{{\em Phys. Rev. D}
  {\bfseries 107} no.~8, (2023) 084041},
  \href{http://arxiv.org/abs/2212.14297}{{\ttfamily arXiv:2212.14297 [gr-qc]}}.

\bibitem{Bisnovatyi-Kogan:2022yzj}
G.~S. Bisnovatyi-Kogan and O.~Y. Tsupko, ``{Time delay induced by plasma in
  strong lens systems},'' \href{http://dx.doi.org/10.1093/mnras/stad2030}{{\em
  Mon. Not. Roy. Astron. Soc.} {\bfseries 524} no.~2, (2023) 3060--3067},
  \href{http://arxiv.org/abs/2301.00053}{{\ttfamily arXiv:2301.00053 [gr-qc]}}.

\bibitem{Tsupko:2019axo}
O.~Y. Tsupko and G.~S. Bisnovatyi-Kogan, ``{Hills and holes in the microlensing
  light curve due to plasma environment around gravitational lens},''
  \href{http://dx.doi.org/10.1093/mnras/stz3365}{{\em Mon. Not. Roy. Astron.
  Soc.} {\bfseries 491} no.~4, (2020) 5636--5649},
  \href{http://arxiv.org/abs/1910.03457}{{\ttfamily arXiv:1910.03457 [gr-qc]}}.

\bibitem{Sun:2022ujt}
J.~Sun, X.~Er, and O.~Y. Tsupko, ``{Binary microlensing with plasma environment
  \textendash{} star and planet},''
  \href{http://dx.doi.org/10.1093/mnras/stad200}{{\em Mon. Not. Roy. Astron.
  Soc.} {\bfseries 520} no.~1, (2023) 994--1004},
  \href{http://arxiv.org/abs/2211.13442}{{\ttfamily arXiv:2211.13442
  [astro-ph.SR]}}.

\bibitem{Er_2020}
X.~Er, Y.-P. Yang, and A.~Rogers, ``The effects of plasma lensing on the
  inferred dispersion measures of fast radiobursts,''
  \href{http://dx.doi.org/10.3847/1538-4357/ab66b1}{{\em The Astrophysical
  Journal} {\bfseries 889} no.~2, (Feb, 2020) 158}.
  \url{https://dx.doi.org/10.3847/1538-4357/ab66b1}.

\bibitem{Kala:2024fvg}
S.~Kala, H.~Nandan, A.~Abebe, and S.~Roy, ``{Gravitational lensing around a
  dual-charged stringy black hole in plasma background},''
  \href{http://dx.doi.org/10.1140/epjc/s10052-024-13362-9}{{\em Eur. Phys. J.
  C} {\bfseries 84} no.~10, (2024) 1089}.

\bibitem{Kala:2022uog}
S.~Kala, H.~Nandan, and P.~Sharma, ``{Shadow and weak gravitational lensing of
  a rotating regular black hole in a non-minimally coupled Einstein-Yang-Mills
  theory in the presence of plasma},''
  \href{http://dx.doi.org/10.1140/epjp/s13360-022-02634-6}{{\em Eur. Phys. J.
  Plus} {\bfseries 137} no.~4, (2022) 457},
  \href{http://arxiv.org/abs/2207.10717}{{\ttfamily arXiv:2207.10717 [gr-qc]}}.

\bibitem{Yan:2019etp}
H.~Yan, ``{Influence of a plasma on the observational signature of a high-spin
  Kerr black hole},'' \href{http://dx.doi.org/10.1103/PhysRevD.99.084050}{{\em
  Phys. Rev. D} {\bfseries 99} no.~8, (2019) 084050},
  \href{http://arxiv.org/abs/1903.04382}{{\ttfamily arXiv:1903.04382 [gr-qc]}}.

\bibitem{Chowdhuri:2020ipb}
A.~Chowdhuri and A.~Bhattacharyya, ``{Shadow analysis for rotating black holes
  in the presence of plasma for an expanding universe},''
  \href{http://dx.doi.org/10.1103/PhysRevD.104.064039}{{\em Phys. Rev. D}
  {\bfseries 104} no.~6, (2021) 064039},
  \href{http://arxiv.org/abs/2012.12914}{{\ttfamily arXiv:2012.12914 [gr-qc]}}.

\bibitem{Badia:2021kpk}
J.~Bad\'\i{}a and E.~F. Eiroa, ``{Shadow of axisymmetric, stationary, and
  asymptotically flat black holes in the presence of plasma},''
  \href{http://dx.doi.org/10.1103/PhysRevD.104.084055}{{\em Phys. Rev. D}
  {\bfseries 104} no.~8, (2021) 084055},
  \href{http://arxiv.org/abs/2106.07601}{{\ttfamily arXiv:2106.07601 [gr-qc]}}.

\bibitem{Badia:2022phg}
J.~Bad\'\i{}a and E.~F. Eiroa, ``{Shadows of rotating Einstein-Maxwell-dilaton
  black holes surrounded by a plasma},''
  \href{http://dx.doi.org/10.1103/PhysRevD.107.124028}{{\em Phys. Rev. D}
  {\bfseries 107} no.~12, (2023) 124028},
  \href{http://arxiv.org/abs/2210.03081}{{\ttfamily arXiv:2210.03081 [gr-qc]}}.

\bibitem{Bezdekova:2022gib}
B.~Bezdekova, V.~Perlick, and J.~Bicak, ``{Light propagation in a plasma on an
  axially symmetric and stationary spacetime: Separability of the
  Hamilton\textendash{}Jacobi equation and shadow},''
  \href{http://dx.doi.org/10.1063/5.0106433}{{\em J. Math. Phys.} {\bfseries
  63} no.~9, (2022) 092501}, \href{http://arxiv.org/abs/2204.05593}{{\ttfamily
  arXiv:2204.05593 [gr-qc]}}.

\bibitem{Briozzo:2022mgg}
G.~Briozzo, E.~Gallo, and T.~M\"adler, ``{Shadows of rotating black holes in
  plasma environments with aberration effects},''
  \href{http://dx.doi.org/10.1103/PhysRevD.107.124004}{{\em Phys. Rev. D}
  {\bfseries 107} no.~12, (2023) 124004},
  \href{http://arxiv.org/abs/2211.05620}{{\ttfamily arXiv:2211.05620 [gr-qc]}}.

\bibitem{Perlick:2023znh}
V.~Perlick and O.~Y. Tsupko, ``{Light propagation in a plasma on Kerr
  spacetime. II. Plasma imprint on photon orbits},''
  \href{http://dx.doi.org/10.1103/PhysRevD.109.064063}{{\em Phys. Rev. D}
  {\bfseries 109} no.~6, (2024) 064063},
  \href{http://arxiv.org/abs/2311.10615}{{\ttfamily arXiv:2311.10615 [gr-qc]}}.

\bibitem{EMDAGarfinkle}
D.~Garfinkle, G.~T. Horowitz, and A.~Strominger, ``Charged black holes in
  string theory,'' \href{http://dx.doi.org/10.1103/PhysRevD.43.3140}{{\em Phys.
  Rev. D} {\bfseries 43} (May, 1991) 3140--3143}.
  \url{https://link.aps.org/doi/10.1103/PhysRevD.43.3140}.

\bibitem{EMDAGarfinkle2}
D.~Garfinkle, G.~T. Horowitz, and A.~Strominger, ``Erratum: Charged black holes
  in string theory,'' \href{http://dx.doi.org/10.1103/PhysRevD.45.3888}{{\em
  Phys. Rev. D} {\bfseries 45} (May, 1992) 3888--3888}.
  \url{https://link.aps.org/doi/10.1103/PhysRevD.45.3888}.

\bibitem{Ganguly:2014pwa}
C.~Ganguly and S.~SenGupta, ``{Penrose process in a charged
  axion\textendash{}dilaton coupled black hole},''
  \href{http://dx.doi.org/10.1140/epjc/s10052-016-4058-0}{{\em Eur. Phys. J. C}
  {\bfseries 76} no.~4, (2016) 213},
  \href{http://arxiv.org/abs/1401.6826}{{\ttfamily arXiv:1401.6826 [hep-th]}}.

\bibitem{Boyer:1966qh}
R.~H. Boyer and R.~W. Lindquist, ``{Maximal analytic extension of the Kerr
  metric},'' \href{http://dx.doi.org/10.1063/1.1705193}{{\em J. Math. Phys.}
  {\bfseries 8} (1967) 265}.

\bibitem{Pal:2021nul}
K.~Pal, K.~Pal, R.~Shaikh, and T.~Sarkar, ``{Shadows in conformally related
  gravity theories},''
  \href{http://dx.doi.org/10.1016/j.physletb.2022.137109}{{\em Phys. Lett. B}
  {\bfseries 829} (2022) 137109},
  \href{http://arxiv.org/abs/2110.13723}{{\ttfamily arXiv:2110.13723 [gr-qc]}}.

\bibitem{Piotr}
P.~T. Chru\`{s}ciel, ``{Lectures on Energy in General Relativity},''.

\bibitem{Gebhardt:2011yw}
K.~Gebhardt, J.~Adams, D.~Richstone, T.~R. Lauer, S.~M. Faber, K.~Gultekin,
  J.~Murphy, and S.~Tremaine, ``{The Black-Hole Mass in M87 from Gemini/NIFS
  Adaptive Optics Observations},''
  \href{http://dx.doi.org/10.1088/0004-637X/729/2/119}{{\em Astrophys. J.}
  {\bfseries 729} (2011) 119}, \href{http://arxiv.org/abs/1101.1954}{{\ttfamily
  arXiv:1101.1954 [astro-ph.CO]}}.

\bibitem{Abuter:2018uum}
R.~Abuter {\em et~al.}, ``{Detection of orbital motions near the last stable
  circular orbit of the massive black hole SgrA*},''
  \href{http://dx.doi.org/10.1051/0004-6361/201834294}{{\em Astron. Astrophys.}
  {\bfseries 618} (2018) }, \href{http://arxiv.org/abs/1810.12641}{{\ttfamily
  arXiv:1810.12641 [astro-ph.GA]}}.

\bibitem{KumarWalia:2024omf}
R.~Kumar~Walia, P.~Kocherlakota, D.~O. Chang, and K.~Salehi, ``{Spacetime
  measurements with the photon ring},''
  \href{http://dx.doi.org/10.1103/PhysRevD.111.104074}{{\em Phys. Rev. D}
  {\bfseries 111} no.~10, (2025) 104074},
  \href{http://arxiv.org/abs/2411.15119}{{\ttfamily arXiv:2411.15119 [gr-qc]}}.

\bibitem{Vagnozzi:2022moj}
S.~Vagnozzi {\em et~al.}, ``{Horizon-scale tests of gravity theories and
  fundamental physics from the Event Horizon Telescope image of Sagittarius
  A},'' \href{http://dx.doi.org/10.1088/1361-6382/acd97b}{{\em Class. Quant.
  Grav.} {\bfseries 40} no.~16, (2023) 165007},
  \href{http://arxiv.org/abs/2205.07787}{{\ttfamily arXiv:2205.07787 [gr-qc]}}.

\bibitem{breuer1980propagation}
R.~A. Breuer and J.~Ehlers, ``Propagation of high-frequency electromagnetic
  waves through a magnetized plasma in curved space-time. i,'' {\em Proceedings
  of the Royal Society of London. A. Mathematical and Physical Sciences}
  {\bfseries 370} no.~1742, (1980) 389--406.

\bibitem{breuer1981propagation}
R.~A. Breuer and J.~Ehlers, ``Propagation of high-frequency electromagnetic
  waves through a magnetized plasma in curved space-time. ii. application of
  the asymptotic approximation,'' {\em Proceedings of the Royal Society of
  London. A. Mathematical and Physical Sciences} {\bfseries 374} no.~1756,
  (1981) 65--86.

\bibitem{synge1960relativity}
J.~Synge, {\em Relativity: The General Theory}.
\newblock No.~v. 1 in North-Holland series in physics. North-Holland Publishing
  Company, 1960.
\newblock \url{https://books.google.co.in/books?id=CqoNAQAAIAAJ}.

\bibitem{Carter:1968rr}
B.~Carter, ``{Global structure of the Kerr family of gravitational fields},''
  \href{http://dx.doi.org/10.1103/PhysRev.174.1559}{{\em Phys. Rev.} {\bfseries
  174} (1968) 1559--1571}.

\bibitem{Banerjee:2021nza}
I.~Banerjee, V.~S. Chawan, B.~Mandal, S.~K. Sahoo, and S.~SenGupta, ``{Quasar
  continuum spectrum disfavors black holes with a magnetic monopole charge},''
  \href{http://dx.doi.org/10.1103/PhysRevD.105.064073}{{\em Phys. Rev. D}
  {\bfseries 105} no.~6, (2022) 064073},
  \href{http://arxiv.org/abs/2112.05385}{{\ttfamily arXiv:2112.05385 [gr-qc]}}.

\bibitem{Grenzebach:2014fha}
A.~Grenzebach, V.~Perlick, and C.~L\"ammerzahl, ``{Photon Regions and Shadows
  of Kerr-Newman-NUT Black Holes with a Cosmological Constant},''
  \href{http://dx.doi.org/10.1103/PhysRevD.89.124004}{{\em Phys. Rev. D}
  {\bfseries 89} no.~12, (2014) 124004},
  \href{http://arxiv.org/abs/1403.5234}{{\ttfamily arXiv:1403.5234 [gr-qc]}}.

\bibitem{Grenzebach:2015oea}
A.~Grenzebach, V.~Perlick, and C.~L\"ammerzahl, ``{Photon Regions and Shadows
  of Accelerated Black Holes},''
  \href{http://dx.doi.org/10.1142/S0218271815420249}{{\em Int. J. Mod. Phys. D}
  {\bfseries 24} no.~09, (2015) 1542024},
  \href{http://arxiv.org/abs/1503.03036}{{\ttfamily arXiv:1503.03036 [gr-qc]}}.

\bibitem{shapiro1974accretion}
S.~L. Shapiro, ``Accretion onto black holes: The emergent radiation spectrum.
  iii. rotating (kerr) black holes,'' {\em Astrophysical Journal, Vol. 189, pp.
  343-352 (1974)} {\bfseries 189} (1974) 343--352.

\bibitem{Mosallanezhad:2013gre}
A.~Mosallanezhad, S.~Abbassi, and N.~Beiranvand, ``{Structure of
  advection-dominated accretion discs with outflows: the role of toroidal
  magnetic fields},'' \href{http://dx.doi.org/10.1093/mnras/stt2048}{{\em Mon.
  Not. Roy. Astron. Soc.} {\bfseries 437} no.~4, (2014) 3112--3123},
  \href{http://arxiv.org/abs/1310.6318}{{\ttfamily arXiv:1310.6318
  [astro-ph.HE]}}.

\bibitem{Rees:1982pe}
M.~J. Rees, E.~S. Phinney, M.~C. Begelman, and R.~D. Blandford, ``{Ion
  supported tori and the origin of radio jets},''
  \href{http://dx.doi.org/10.1038/295017a0}{{\em Nature} {\bfseries 295} (1982)
  17--21}.

\bibitem{Komissarov:2006nz}
S.~S. Komissarov, ``{Magnetized Tori around Kerr Black Holes: Analytic
  Solutions with a Toroidal Magnetic Field},''
  \href{http://dx.doi.org/10.1111/j.1365-2966.2006.10183.x}{{\em Mon. Not. Roy.
  Astron. Soc.} {\bfseries 368} (2006) 993--1000},
  \href{http://arxiv.org/abs/astro-ph/0601678}{{\ttfamily
  arXiv:astro-ph/0601678}}.

\bibitem{Kulsrud:1991jt}
R.~Kulsrud and A.~Loeb, ``{Dynamics and gravitational interaction of waves in
  nonuniform media},'' \href{http://dx.doi.org/10.1103/PhysRevD.45.525}{{\em
  Phys. Rev. D} {\bfseries 45} (1992) 525--531}.

\bibitem{Gebhardt:2009cr}
K.~Gebhardt and J.~Thomas, ``{The Black Hole Mass, Stellar M/L, and Dark Halo
  in M87},'' \href{http://dx.doi.org/10.1088/0004-637X/700/2/1690}{{\em
  Astrophys. J.} {\bfseries 700} (2009) 1690--1701},
  \href{http://arxiv.org/abs/0906.1492}{{\ttfamily arXiv:0906.1492
  [astro-ph.CO]}}.

\bibitem{Gebhardt:2000fk}
K.~Gebhardt {\em et~al.}, ``{A Relationship between nuclear black hole mass and
  galaxy velocity dispersion},'' \href{http://dx.doi.org/10.1086/312840}{{\em
  Astrophys. J. Lett.} {\bfseries 539} (2000) L13},
  \href{http://arxiv.org/abs/astro-ph/0006289}{{\ttfamily arXMartin J. Comptes
  Rendus Physique 13:566 (2012)iv:astro-ph/0006289}}.

\bibitem{Walsh:2013uua}
J.~L. Walsh, A.~J. Barth, L.~C. Ho, and M.~Sarzi, ``{The M87 Black Hole Mass
  from Gas-dynamical Models of Space Telescope Imaging Spectrograph
  Observations},'' \href{http://dx.doi.org/10.1088/0004-637X/770/2/86}{{\em
  Astrophys. J.} {\bfseries 770} (2013) 86},
  \href{http://arxiv.org/abs/1304.7273}{{\ttfamily arXiv:1304.7273
  [astro-ph.CO]}}.

\bibitem{Blakeslee:2009tc}
J.~P. Blakeslee, A.~Jordan, S.~Mei, P.~Cote, L.~Ferrarese, L.~Infante, E.~W.
  Peng, J.~L. Tonry, and M.~J. West, ``{The ACS Fornax Cluster Survey. V.
  Measurement and Recalibration of Surface Brightness Fluctuations and a
  Precise Value of the Fornax--Virgo Relative Distance},''
  \href{http://dx.doi.org/10.1088/0004-637X/694/1/556}{{\em Astrophys. J.}
  {\bfseries 694} (2009) 556--572},
  \href{http://arxiv.org/abs/0901.1138}{{\ttfamily arXiv:0901.1138
  [astro-ph.CO]}}.

\bibitem{Bird:2010rd}
S.~Bird, W.~E. Harris, J.~P. Blakeslee, and C.~Flynn, ``{The Inner Halo of M87:
  A First Direct View of the Red-Giant Population},''
  \href{http://dx.doi.org/10.1051/0004-6361/201014876}{{\em Astron. Astrophys.}
  {\bfseries 524} (2010) A71}, \href{http://arxiv.org/abs/1009.3202}{{\ttfamily
  arXiv:1009.3202 [astro-ph.GA]}}.

\bibitem{Tamburini:2019vrf}
F.~Tamburini, B.~Thid\'e, and M.~Della~Valle, ``{Measurement of the spin of the
  M87 black hole from its observed twisted light},''
  \href{http://dx.doi.org/10.1093/mnrasl/slz176}{{\em Mon. Not. Roy. Astron.
  Soc.} {\bfseries 492} no.~1, (2020) L22--L27},
  \href{http://arxiv.org/abs/1904.07923}{{\ttfamily arXiv:1904.07923
  [astro-ph.HE]}}.

\bibitem{Do:2019txf}
T.~Do {\em et~al.}, ``{Relativistic redshift of the star S0-2 orbiting the
  Galactic center supermassive black hole},''
  \href{http://dx.doi.org/10.1126/science.aav8137}{{\em Science} {\bfseries
  365} no.~6454, (2019) 664--668},
  \href{http://arxiv.org/abs/1907.10731}{{\ttfamily arXiv:1907.10731
  [astro-ph.GA]}}.

\bibitem{GRAVITY:2020gka}
{\bfseries GRAVITY} Collaboration, R.~Abuter {\em et~al.}, ``{Detection of the
  Schwarzschild precession in the orbit of the star S2 near the Galactic centre
  massive black hole},''
  \href{http://dx.doi.org/10.1051/0004-6361/202037813}{{\em Astron. Astrophys.}
  {\bfseries 636} (2020) L5}, \href{http://arxiv.org/abs/2004.07187}{{\ttfamily
  arXiv:2004.07187 [astro-ph.GA]}}.

\bibitem{GRAVITY:2021xju}
{\bfseries GRAVITY} Collaboration, R.~Abuter {\em et~al.}, ``{Mass distribution
  in the Galactic Center based on interferometric astrometry of multiple
  stellar orbits},'' \href{http://dx.doi.org/10.1051/0004-6361/202142465}{{\em
  Astron. Astrophys.} {\bfseries 657} (2022) L12},
  \href{http://arxiv.org/abs/2112.07478}{{\ttfamily arXiv:2112.07478
  [astro-ph.GA]}}.

\bibitem{abuter2019geometric}
R.~Abuter, A.~Amorim, M.~Baub{\"o}ck, J.~Berger, H.~Bonnet, W.~Brandner,
  Y.~Cl{\'e}net, V.~C. Du~Foresto, P.~De~Zeeuw, J.~Dexter, {\em et~al.}, ``A
  geometric distance measurement to the galactic center black hole with 0.3\%
  uncertainty,'' {\em Astronomy \& Astrophysics} {\bfseries 625} (2019) L10.

\bibitem{Will:2007pp}
C.~M. Will, ``{Testing the general relativistic no-hair theorems using the
  Galactic center black hole SgrA*},''
  \href{http://dx.doi.org/10.1086/528847}{{\em Astrophys. J. Lett.} {\bfseries
  674} (2008) L25--L28}, \href{http://arxiv.org/abs/0711.1677}{{\ttfamily
  arXiv:0711.1677 [astro-ph]}}.

\bibitem{Will:2016pgm}
C.~M. Will and M.~Maitra, ``{Relativistic orbits around spinning supermassive
  black holes. Secular evolution to 4.5 post-Newtonian order},''
  \href{http://dx.doi.org/10.1103/PhysRevD.95.064003}{{\em Phys. Rev. D}
  {\bfseries 95} no.~6, (2017) 064003},
  \href{http://arxiv.org/abs/1611.06931}{{\ttfamily arXiv:1611.06931 [gr-qc]}}.

\bibitem{Fernandez:2023kro}
R.~F. Fern{\'a}ndez, R.~Della~Monica, and I.~de~Martino, ``{Constraining an
  Einstein-Maxwell-dilaton-axion black hole at the Galactic Center with the
  orbit of the S2 star},''
  \href{http://dx.doi.org/10.1088/1475-7516/2023/08/039}{{\em JCAP} {\bfseries
  08} (2023) 039}, \href{http://arxiv.org/abs/2306.06937}{{\ttfamily
  arXiv:2306.06937 [gr-qc]}}.

\bibitem{GRAVITY:2018ofz}
{\bfseries GRAVITY} Collaboration, R.~Abuter {\em et~al.}, ``{Detection of the
  gravitational redshift in the orbit of the star S2 near the Galactic centre
  massive black hole},''
  \href{http://dx.doi.org/10.1051/0004-6361/201833718}{{\em Astron. Astrophys.}
  {\bfseries 615} (2018) L15},
  \href{http://arxiv.org/abs/1807.09409}{{\ttfamily arXiv:1807.09409
  [astro-ph.GA]}}.

\bibitem{DellaMonica:2021fdr}
R.~Della~Monica and I.~de~Martino, ``{Unveiling the nature of SgrA* with the
  geodesic motion of S-stars},''
  \href{http://dx.doi.org/10.1088/1475-7516/2022/03/007}{{\em JCAP} {\bfseries
  03} no.~03, (2022) 007}, \href{http://arxiv.org/abs/2112.01888}{{\ttfamily
  arXiv:2112.01888 [astro-ph.GA]}}.

\bibitem{Bambhaniya:2022xbz}
P.~Bambhaniya, A.~B. Joshi, D.~Dey, P.~S. Joshi, A.~Mazumdar, T.~Harada, and
  K.-i. Nakao, ``{Relativistic orbits of S2 star in the presence of scalar
  field},'' \href{http://dx.doi.org/10.1140/epjc/s10052-024-12477-3}{{\em Eur.
  Phys. J. C} {\bfseries 84} no.~2, (2024) 124},
  \href{http://arxiv.org/abs/2209.12610}{{\ttfamily arXiv:2209.12610 [gr-qc]}}.

\bibitem{Shaikh:2021yux}
R.~Shaikh, K.~Pal, K.~Pal, and T.~Sarkar, ``{Constraining alternatives to the
  Kerr black hole},'' \href{http://dx.doi.org/10.1093/mnras/stab1779}{{\em Mon.
  Not. Roy. Astron. Soc.} {\bfseries 506} no.~1, (2021) 1229--1236},
  \href{http://arxiv.org/abs/2102.04299}{{\ttfamily arXiv:2102.04299 [gr-qc]}}.

\bibitem{Shaikh:2022ivr}
R.~Shaikh, ``{Testing black hole mimickers with the Event Horizon Telescope
  image of Sagittarius A*},''
  \href{http://dx.doi.org/10.1093/mnras/stad1383}{{\em Mon. Not. Roy. Astron.
  Soc.} {\bfseries 523} no.~1, (2023) 375--384},
  \href{http://arxiv.org/abs/2208.01995}{{\ttfamily arXiv:2208.01995 [gr-qc]}}.

\bibitem{Pal:2023wqg}
K.~Pal, K.~Pal, R.~Shaikh, and T.~Sarkar, ``{A rotating modified JNW spacetime
  as a Kerr black hole mimicker},''
  \href{http://dx.doi.org/10.1088/1475-7516/2023/11/060}{{\em JCAP} {\bfseries
  11} (2023) 060}, \href{http://arxiv.org/abs/2305.07518}{{\ttfamily
  arXiv:2305.07518 [gr-qc]}}.

\bibitem{Islam:2024sph}
S.~U. Islam, S.~G. Ghosh, and S.~D. Maharaj, ``{Investigating rotating black
  holes in bumblebee gravity: insights from EHT observations},''
  \href{http://dx.doi.org/10.1088/1475-7516/2024/12/047}{{\em JCAP} {\bfseries
  12} (2024) 047}, \href{http://arxiv.org/abs/2410.05395}{{\ttfamily
  arXiv:2410.05395 [gr-qc]}}.

\bibitem{Islam:2022wck}
S.~U. Islam, J.~Kumar, R.~Kumar~Walia, and S.~G. Ghosh, ``{Investigating Loop
  Quantum Gravity with Event Horizon Telescope Observations of the Effects of
  Rotating Black Holes},''
  \href{http://dx.doi.org/10.3847/1538-4357/aca411}{{\em Astrophys. J.}
  {\bfseries 943} no.~1, (2023) 22},
  \href{http://arxiv.org/abs/2211.06653}{{\ttfamily arXiv:2211.06653 [gr-qc]}}.

\bibitem{Kumar:2018ple}
R.~Kumar and S.~G. Ghosh, ``{Black Hole Parameter Estimation from Its
  Shadow},'' \href{http://dx.doi.org/10.3847/1538-4357/ab77b0}{{\em Astrophys.
  J.} {\bfseries 892} (2020) 78},
  \href{http://arxiv.org/abs/1811.01260}{{\ttfamily arXiv:1811.01260 [gr-qc]}}.

\bibitem{Afrin:2021imp}
M.~Afrin, R.~Kumar, and S.~G. Ghosh, ``{Parameter estimation of hairy Kerr
  black holes from its shadow and constraints from M87*},''
  \href{http://dx.doi.org/10.1093/mnras/stab1260}{{\em Mon. Not. Roy. Astron.
  Soc.} {\bfseries 504} no.~4, (2021) 5927--5940},
  \href{http://arxiv.org/abs/2103.11417}{{\ttfamily arXiv:2103.11417 [gr-qc]}}.

\bibitem{Raza:2024zkp}
M.~A. Raza, M.~Zubair, and E.~Maqsood, ``{Influence of plasma on the optical
  appearance of spinning black hole in Kalb-Ramond gravity and its Existence
  around M87* and Sgr A*},''
  \href{http://dx.doi.org/10.1088/1475-7516/2024/05/047}{{\em JCAP} {\bfseries
  05} (2024) 047}, \href{http://arxiv.org/abs/2401.04779}{{\ttfamily
  arXiv:2401.04779 [gr-qc]}}.

\bibitem{1976ApAvni}
Y.~{Avni}, ``{Energy spectra of X-ray clusters of galaxies.},''
  \href{http://dx.doi.org/10.1086/154870}{{\em {Astrophys. J.}} {\bfseries
  {210}} (Dec, 1976) {642--646}}.

\bibitem{EventHorizonTelescope:2024dhe}
{\bfseries Event Horizon Telescope} Collaboration, K.~Akiyama {\em et~al.},
  ``{The persistent shadow of the supermassive black hole of M~87. I.
  Observations, calibration, imaging, and analysis},''
  \href{http://dx.doi.org/10.1051/0004-6361/202347932}{{\em Astron. Astrophys.}
  {\bfseries 681} (2024) A79}.

\bibitem{Nemmen:2019idv}
R.~Nemmen, ``{The Spin of M87*},''
  \href{http://dx.doi.org/10.3847/2041-8213/ab2fd3}{{\em Astrophys. J. Lett.}
  {\bfseries 880} no.~2, (2019) L26},
  \href{http://arxiv.org/abs/1905.02143}{{\ttfamily arXiv:1905.02143
  [astro-ph.HE]}}.

\bibitem{Takahashi:2004xh}
R.~Takahashi, ``{Shapes and positions of black hole shadows in accretion disks
  and spin parameters of black holes},''
  \href{http://dx.doi.org/10.1086/422403}{{\em J. Korean Phys. Soc.} {\bfseries
  45} (2004) S1808--S1812},
  \href{http://arxiv.org/abs/astro-ph/0405099}{{\ttfamily
  arXiv:astro-ph/0405099}}.

\bibitem{Johannsen:2010ru}
T.~Johannsen and D.~Psaltis, ``{Testing the No-Hair Theorem with Observations
  in the Electromagnetic Spectrum: II. Black-Hole Images},''
  \href{http://dx.doi.org/10.1088/0004-637X/718/1/446}{{\em Astrophys. J.}
  {\bfseries 718} (2010) 446--454},
  \href{http://arxiv.org/abs/1005.1931}{{\ttfamily arXiv:1005.1931
  [astro-ph.HE]}}.

\bibitem{2009ApJ...706..497M}
M.~{Mo{\'s}cibrodzka}, C.~F. {Gammie}, J.~C. {Dolence}, H.~{Shiokawa}, and
  P.~K. {Leung}, ``{Radiative Models of SGR A* from GRMHD Simulations},''
  \href{http://dx.doi.org/10.1088/0004-637X/706/1/497}{{\em "{Astrophys. J.}"}
  {\bfseries 706} no.~1, (Nov., 2009) 497--507},
  \href{http://arxiv.org/abs/0909.5431}{{\ttfamily arXiv:0909.5431
  [astro-ph.HE]}}.

\bibitem{2012ApJ...755..133S}
R.~V. {Shcherbakov}, R.~F. {Penna}, and J.~C. {McKinney}, ``{Sagittarius A*
  Accretion Flow and Black Hole Parameters from General Relativistic Dynamical
  and Polarized Radiative Modeling},''
  \href{http://dx.doi.org/10.1088/0004-637X/755/2/133}{{\em "{Astrophys. J.}"}
  {\bfseries 755} no.~2, (Aug., 2012) 133},
  \href{http://arxiv.org/abs/1007.4832}{{\ttfamily arXiv:1007.4832
  [astro-ph.HE]}}.

\bibitem{Fragione:2020khu}
G.~Fragione and A.~Loeb, ``{An upper limit on the spin of SgrA$^*$ based on
  stellar orbits in its vicinity},''
  \href{http://dx.doi.org/10.3847/2041-8213/abb9b4}{{\em Astrophys. J. Lett.}
  {\bfseries 901} no.~2, (2020) L32},
  \href{http://arxiv.org/abs/2008.11734}{{\ttfamily arXiv:2008.11734
  [astro-ph.GA]}}.

\bibitem{Belanger:2006gm}
G.~Belanger, R.~Terrier, O.~C. De~Jager, A.~Goldwurm, and F.~Melia, ``{Periodic
  Modulations in an X-ray Flare from Sagittarius A*},''
  \href{http://dx.doi.org/10.1088/1742-6596/54/1/066}{{\em J. Phys. Conf. Ser.}
  {\bfseries 54} (2006) 420--426},
  \href{http://arxiv.org/abs/astro-ph/0604337}{{\ttfamily
  arXiv:astro-ph/0604337}}.

\bibitem{Meyer:2006fd}
L.~Meyer, A.~Eckart, R.~Schoedel, W.~J. Duschl, K.~Muzic, M.~Dovciak, and
  V.~Karas, ``{Near-infrared polarimetry setting constraints on the orbiting
  spot model for Sgr A* flares},''
  \href{http://dx.doi.org/10.1051/0004-6361:20065925}{{\em Astron. Astrophys.}
  {\bfseries 460} (2006) 15},
  \href{http://arxiv.org/abs/astro-ph/0610104}{{\ttfamily
  arXiv:astro-ph/0610104}}.

\bibitem{Genzel:2003as}
R.~Genzel, R.~Schodel, T.~Ott, A.~Eckart, T.~Alexander, F.~Lacombe, D.~Rouan,
  and B.~Aschenbach, ``{Near-infrared flares from accreting gas around the
  supermassive black hole at the galactic centre},''
  \href{http://dx.doi.org/10.1038/nature02065}{{\em Nature} {\bfseries 425}
  (2003) 934--937}, \href{http://arxiv.org/abs/astro-ph/0310821}{{\ttfamily
  arXiv:astro-ph/0310821}}.

\bibitem{Daly:2023axh}
R.~A. Daly, M.~Donahue, C.~P. O'Dea, B.~Sebastian, D.~Haggard, and A.~Lu,
  ``{New black hole spin values for Sagittarius A* obtained with the outflow
  method},'' \href{http://dx.doi.org/10.1093/mnras/stad3228}{{\em Mon. Not.
  Roy. Astron. Soc.} {\bfseries 527} no.~1, (2023) 428--436},
  \href{http://arxiv.org/abs/2310.12108}{{\ttfamily arXiv:2310.12108
  [astro-ph.GA]}}.

\bibitem{EventHorizonTelescope:2021srq}
{\bfseries Event Horizon Telescope} Collaboration, K.~Akiyama {\em et~al.},
  ``{First M87 Event Horizon Telescope Results. VIII. Magnetic Field Structure
  near The Event Horizon},''
  \href{http://dx.doi.org/10.3847/2041-8213/abe4de}{{\em Astrophys. J. Lett.}
  {\bfseries 910} no.~1, (2021) L13},
  \href{http://arxiv.org/abs/2105.01173}{{\ttfamily arXiv:2105.01173
  [astro-ph.HE]}}.

\bibitem{Drew:2025euq}
M.~Drew, J.~S. Stanway, B.~A. Patterson, T.~J. Walton, and D.~Ward-Thompson,
  ``{New Estimates of the Spin and Accretion Rate of the Black Hole M87*},''
  \href{http://dx.doi.org/10.3847/2041-8213/adc90e}{{\em Astrophys. J. Lett.}
  {\bfseries 984} no.~1, (2025) L31},
  \href{http://arxiv.org/abs/2505.17035}{{\ttfamily arXiv:2505.17035
  [astro-ph.HE]}}.

\bibitem{Kuo:2014pqa}
C.~Y. Kuo {\em et~al.}, ``{Measuring Mass Accretion Rate onto the Supermassive
  Black Hole in M87 Using Faraday Rotation Measure with the Submillimeter
  Array},'' \href{http://dx.doi.org/10.1088/2041-8205/783/2/L33}{{\em
  Astrophys. J. Lett.} {\bfseries 783} (2014) L33},
  \href{http://arxiv.org/abs/1402.5238}{{\ttfamily arXiv:1402.5238
  [astro-ph.GA]}}.

\bibitem{Feng:2016gif}
J.~Feng, Q.~Wu, and R.-S. Lu, ``{An accretion-jet model for M87: interpreting
  the spectral energy distribution and Faraday rotation measure},''
  \href{http://dx.doi.org/10.3847/0004-637X/830/1/6}{{\em Astrophys. J.}
  {\bfseries 830} no.~1, (2016) 6},
  \href{http://arxiv.org/abs/1607.08054}{{\ttfamily arXiv:1607.08054
  [astro-ph.HE]}}.

\bibitem{Quataert:1999ng}
E.~Quataert, R.~Narayan, and M.~Reid, ``{What is the accretion rate in sgr
  a*?},'' \href{http://dx.doi.org/10.1086/312035}{{\em Astrophys. J. Lett.}
  {\bfseries 517} (1999) L101},
  \href{http://arxiv.org/abs/astro-ph/9903412}{{\ttfamily
  arXiv:astro-ph/9903412}}.

\bibitem{Marrone:2006vu}
D.~P. Marrone, J.~M. Moran, J.~H. Zhao, and R.~Rao, ``{An Unambiguous Detection
  of Faraday Rotation in Sagittarius A*},''
  \href{http://dx.doi.org/10.1086/510850}{{\em Astrophys. J. Lett.} {\bfseries
  654} (2006) L57--L60},
  \href{http://arxiv.org/abs/astro-ph/0611791}{{\ttfamily
  arXiv:astro-ph/0611791}}.

\bibitem{Bower:2018wsw}
G.~C. Bower {\em et~al.}, ``{ALMA Polarimetry of Sgr A*: Probing the Accretion
  Flow from the Event Horizon to the Bondi Radius},''
  \href{http://dx.doi.org/10.3847/1538-4357/aae983}{{\em Astrophys. J.}
  {\bfseries 868} no.~2, (2018) 101},
  \href{http://arxiv.org/abs/1810.07317}{{\ttfamily arXiv:1810.07317
  [astro-ph.HE]}}.

\bibitem{Yoon:2020yew}
D.~Yoon, K.~Chatterjee, S.~Markoff, D.~van Eijnatten, Z.~Younsi, M.~Liska, and
  A.~Tchekhovskoy, ``{Spectral and imaging properties of Sgr A* from
  high-resolution 3D GRMHD simulations with radiative cooling},''
  \href{http://dx.doi.org/10.1093/mnras/staa3031}{{\em Mon. Not. Roy. Astron.
  Soc.} {\bfseries 499} no.~3, (2020) 3178--3192},
  \href{http://arxiv.org/abs/2009.14227}{{\ttfamily arXiv:2009.14227
  [astro-ph.HE]}}.

\bibitem{EventHorizonTelescope:2025dua}
{\bfseries Event Horizon Telescope} Collaboration, K.~Akiyama {\em et~al.},
  ``{The persistent shadow of the supermassive black hole of M87. II. Model
  comparisons and theoretical interpretations},''
  \href{http://dx.doi.org/10.1051/0004-6361/202451296}{{\em Astron. Astrophys.}
  {\bfseries 693} (2025) A265}.

\bibitem{Cantiello:2018ffy}
M.~Cantiello {\em et~al.}, ``{A Precise Distance to the Host Galaxy of the
  Binary Neutron Star Merger GW170817 Using Surface Brightness Fluctuations},''
  \href{http://dx.doi.org/10.3847/2041-8213/aaad64}{{\em Astrophys. J. Lett.}
  {\bfseries 854} no.~2, (2018) L31},
  \href{http://arxiv.org/abs/1801.06080}{{\ttfamily arXiv:1801.06080
  [astro-ph.GA]}}.

\end{thebibliography}\endgroup
\bibliographystyle{utphys1}
\end{document}